\def\asca{{\it ASCA\/}}

\def\axaf{{\it AXAF\/}}
\def\chandra{{\it Chandra\/}}

\def\heao1{{\it HEAO-1\/}}
\def\hst{{\it {\it HST}\/}}

\def\iso{{\it ISO\/}}

\def\rosat{{\it ROSAT\/}}

\def\ltsima{$\; \buildrel < \over \sim \;$}
\def\simlt{\lower.5ex\hbox{\ltsima}}
\def\gtsima{$\; \buildrel > \over \sim \;$}
\def\simgt{\lower.5ex\hbox{\gtsima}}

\documentclass[preprint]{aastex}

\begin{document}


\title{The Chandra Deep Survey of the Hubble Deep Field North Area. II. 
  Results from the Caltech Faint Field Galaxy Redshift Survey Area$^{1}$ }


\author{A.E.~Hornschemeier,$^{2}$ ~
W.N.~Brandt,$^2$
G.P.~Garmire,$^2$
D.P.~Schneider,$^2$
A.J.~Barger,$^{3,4,5,6}$~
P.S.~Broos,$^2$
L.L.~Cowie,$^{3,6}$
L.K.~Townsley,$^2$
M.W.~Bautz,$^7$
D.N.~Burrows,$^2$
G.~Chartas,$^2$
E.D.~Feigelson,$^2$ 
R.E.~Griffiths,$^8$~
D.~Lumb,$^9$ ~  
J.A.~Nousek,$^2$ ~ 
L.W.~Ramsey,$^{2}$~ and
W.L.W.~Sargent$^{10}$ 
}

\footnotetext[1]{Based on
observations obtained at the W. M. Keck Observatory which is operated jointly
by the California Institute of Technology and the University of California.  Based  
on observations obtained by the Hobby-Eberly
Telescope, which is a joint project of The University of Texas at Austin,
The Pennsylvania State University, Stanford University,
Ludwig-Maximillians-Universit\"at M\"unchen, and Georg-August-Universit\"at
G\"ottingen.}

\footnotetext[2]{Department of Astronomy \& Astrophysics, 525 Davey Laboratory, 
The Pennsylvania State University, University Park, PA 16802}

\footnotetext[3]{Institute for Astronomy, University of Hawaii, 
2680 Woodlawn Drive, Honolulu, HI 96822 }

\footnotetext[4]{Department of Astronomy, University of Wisconsin-Madison,
475 N. Charter Street, Madison, WI 53706}

\footnotetext[5]{Hubble Fellow and Chandra Fellow at Large}

\footnotetext[6]{Visiting Astronomer, W.M. Keck Observatory, jointly operated by the
 California Institute of Technology and the University of California}

\footnotetext[7]{Massachusetts Institute of Technology, Center for Space Research, 
70 Vassar Street, Building 37, Cambridge, MA 02139}

\footnotetext[8]{Department of Physics, Carnegie Mellon University, Pittsburgh, PA 15213}

\footnotetext[9]{Astrophysics Division, ESTEC, Keplerlaan~1, 2200~AG Noordwijk, The Netherlands}

\footnotetext[10]{Palomar Observatory, California Institute of Technology, Pasadena, CA 91125}


\begin{abstract}

A deep X-ray survey of the Hubble Deep Field 
North (\hbox{HDF-N}) and its environs is performed 
using data collected by the Advanced CCD 
Imaging Spectrometer (ACIS) on board the {\it Chandra X-ray Observatory\/}. 
Currently a 221.9~ks exposure is available, the deepest ever presented, 
and here we give results on X-ray sources located 
in the $8.6^{\prime}\times 8.7^{\prime}$ area covered by the 
Caltech Faint Field Galaxy Redshift Survey (the ``Caltech area''). This 
area has (1) deep photometric coverage in several optical and near-infrared 
bands, (2) extensive coverage at radio, submillimeter and mid-infrared 
wavelengths, and (3) some of the deepest and most complete 
spectroscopic coverage ever obtained. It is also where the X-ray
data have the greatest sensitivity; the minimum detectable fluxes in 
the 0.5--2~keV (soft) and 2--8~keV (hard) bands are 
$\approx 1.3\times 10^{-16}$~erg~cm$^{-2}$~s$^{-1}$ and 
$\approx 6.5\times 10^{-16}$~erg~cm$^{-2}$~s$^{-1}$, respectively.
More than $\approx 80$\% of the extragalactic X-ray
background in the hard band is resolved.

The 82 \chandra\ sources detected in the Caltech area are correlated 
with more than 25 multiwavelength source catalogs, and the 
results of these correlations as well as spectroscopic follow-up 
results obtained with the Keck and Hobby-Eberly Telescopes are presented. All 
but nine of the \chandra\ sources are detected optically with 
$R \simlt 26.5$. Redshifts are available for 39\% of the \chandra\ sources, 
including 96\% of the sources with $R<23$; the redshift range is 
0.1--3.5, with most sources having $z < 1.5$.  Eight of the X-ray 
sources are located in the \hbox{HDF-N} itself, including two 
not previously reported. A population of
X-ray faint, optically bright, nearby galaxies emerges at soft-band  
fluxes of $\simlt 3 \times 10^{-16}$~erg~cm$^{-2}$~s$^{-1}$. 

Our multiwavelength correlations have set the tightest constraints to 
date on the X-ray emission properties of $\mu$Jy radio sources, 
mid-infrared sources detected by \iso, and very red (${\cal R}-K_{\rm s}>5.0$)
objects. 
Sixteen of the 67 1.4~GHz $\mu$Jy sources in the Caltech area are
detected in the X-ray band, and the detection rates for starburst-type 
and AGN-candidate $\mu$Jy sources are comparable. Only  
two of the 17 red, optically faint ($I>25$) $\mu$Jy sources are 
detected in X-rays. While 
many of the starburst-type $\mu$Jy sources appear to contain 
obscured AGN, the \chandra\ data are consistent with the majority of the 
$\mu$Jy radio sources being powered by star formation. 
Eleven of the $\approx 100$ \iso\ mid-infrared sources
found in and near the \hbox{HDF-N} are detected in X-rays. 
In the \hbox{HDF-N} itself, where both the 
infrared and the X-ray coverage are deepest, it is notable that six 
of the eight \chandra\ sources are detected by \iso; most of these are 
known to be AGN where the X-ray and infrared detections reveal both 
the direct and indirect accretion power being generated. The high 
X-ray to infrared matching rate bodes well for future sensitive 
infrared observations of faint X-ray sources. 

Four of the 33 very red objects that have been 
identified in the Caltech area by Hogg et~al. (2000a) are detected in X-rays; 
these four are among our hardest 
\chandra\ sources, and we argue that they contain moderately
luminous obscured AGN. Overall, however, the small \chandra\ 
detection fraction suggests a relatively small AGN content 
in the optically selected very red object population.
A stacking analysis of the very red objects not detected
individually by \chandra\ yields a soft-band detection with
an average soft-band X-ray flux of  
$\approx 1.9\times 10^{-17}$~erg~cm$^{-2}$~s$^{-1}$; the 
observed emission may be associated with the hot interstellar 
media of moderate redshift elliptical galaxies.

Constraints on AGN candidates, extended X-ray sources, 
and Galactic objects in the Caltech area are also presented.

\end{abstract}


\keywords{
diffuse radiation~--
surveys~--
cosmology: observations~--
galaxies: active~--
X-rays: galaxies~--
X-rays: general.}


\section{Introduction \label{intro}}

Observations with the \chandra\ {\it X-ray Observatory} are in the process 
of revolutionizing studies of the X-ray background and the sources
that comprise it. Not only has most of the 2--8~keV background now
been resolved into point sources (e.g., Brandt et~al. 2000; 
Mushotzky et~al. 2000; Giacconi et~al. 2001, hereafter G01),  
 but the high-quality
\chandra\ positions also allow these sources to be matched 
unambiguously to (often faint) multiwavelength counterparts. 
Surveys over a broad X-ray band have finally reached the depths 
needed to complement the most sensitive surveys in the radio, 
submillimeter and infrared bands. In these bands, star formation 
and lower luminosity active nuclei in moderate-to-high 
redshift galaxies are thought to dominate the energy production
(e.g., Richards et~al. 1998, hereafter R98; Aussel et~al. 1999,
hereafter A99;  Barger, Cowie, \& Richards 2000, hereafter BCR00).

We have undertaken a deep X-ray survey with the \chandra\ {\it X-ray 
Observatory} (hereafter \chandra; Weisskopf, O'Dell, \& van~Speybroeck 1996)
centered on the most intensively studied extragalactic patch of sky,
the Hubble Deep Field North (\hbox{HDF-N}; Williams et~al. 1996, hereafter W96; 
Ferguson, Dickinson, \& Williams 2000). The Advanced CCD Imaging Spectrometer 
(ACIS; G.P. Garmire et~al., in preparation) has achieved 221.9~ks 
of exposure in four spatially staggered segments over an 
$\approx 17^{\prime}\times 19^{\prime}$ region (see Figure~\ref{emap}).
In the area of deepest coverage, the minimum detectable fluxes in the 
0.5--2~keV (soft) and 2--8~keV (hard) bands are 
$\approx 1.3\times 10^{-16}$~erg~cm$^{-2}$~s$^{-1}$ and 
$\approx 6.5\times 10^{-16}$~erg~cm$^{-2}$~s$^{-1}$, respectively.

In Hornschemeier et~al. (2000, hereafter H00), we presented first
results from the survey based on 164.4~ks of data; our focus was almost
entirely on the \hbox{HDF-N} itself, although we also presented tight
constraints on the X-ray emission properties of submillimeter sources
found outside the \hbox{HDF-N}. In this paper, we present results
derived from more data over a significantly larger area: that studied by the
Caltech Faint Field Galaxy Redshift Survey (hereafter the ``Caltech
area''; e.g., Hogg et~al. 2000a, hereafter Hogg00; Cohen et~al. 2000,
hereafter C00).  This area covers $8.6^{\prime}\times 8.7^{\prime}$ and
has (1) deep photometric coverage in several optical and near-infrared
bands, (2) extensive coverage at radio, submillimeter and mid-infrared
wavelengths, and (3) some of the deepest and most complete
spectroscopic coverage ever obtained. This is also the region where our
X-ray data have the greatest sensitivity (see Figure~\ref{emap}).  The
optical and near-infrared photometric coverage includes the HDF-N
itself (W96), the eight \hst\ flanking fields adjacent to the
\hbox{HDF-N} (W96; these reach $I\approx 25$), and the deep
ground-based imaging presented by Hogg00 which reaches
$U_{\rm n}=25$, $G=26$, ${\cal R}=25.5$ and $K_{\rm s}=20.5$.
We also have deep $I$ and $V$ band images covering the field
(Barger et~al. 1999) as well as deep Keck $R$ (BCR00)  images covering
portions of the field.
The multiwavelength coverage includes radio surveys and maps (e.g.,
R98; Muxlow et~al. 1999; Garrett et~al. 2000; Richards 2000, hereafter R00; and
references therein), submillimeter surveys (e.g., Hughes et~al. 1998;
BCR00; Borys et~al. 2001; Chapman et~al. 2001), and infrared surveys (e.g., A99; Thompson
et~al. 1999; Dickinson et~al. 2000; Hogg et~al. 2000b).
Finally, more than 670 redshifts have been obtained in this area, 
primarily with the Keck Telescope (C00 and references therein). 

We have detected 82 \chandra\ sources in the Caltech area, and in this
paper we present the X-ray source catalog and correlate it with the multiwavelength
source catalogs described above. Our goals are (1) to understand the broad-band
emission and nature of the sources producing most of the X-ray
background in the 0.5--8 keV band and (2) to investigate the X-ray
emission properties of physically interesting sources identified at
other wavelengths. We present optical spectra for 15 of these sources,
using data obtained by the Keck and Hobby-Eberly Telescopes.
A companion paper, G.P. Garmire et~al., in preparation (Paper III), 
presents the number counts
derived from this field and further details on the X-ray spectral 
properties of these sources. 

The Galactic column density along this line of sight
is $(1.6\pm 0.4)\times 10^{20}$~cm$^{-2}$ (Stark et~al. 1992), and
$H_0=70$~km~s$^{-1}$ Mpc$^{-1}$ and $q_0=0.1$ are adopted throughout this paper.
Coordinates throughout this paper are J2000.


\section{Chandra ACIS Observations \label{ACISobs}}

The field containing the \hbox{HDF-N} was observed with the
\chandra\ ACIS for a total exposure time of 221.9~ks in four separate
observations between 1999 Nov and 2000 Feb; a journal of
observations is presented in Table~\ref{chandralog}.
ACIS consists of ten CCDs designed for efficient X-ray detection and
spectroscopy. Four of the CCDs (ACIS-I; CCDs I0--I3) are arranged in a
$2\times 2$ array with each CCD tipped slightly to approximate the
curved focal surface of the \chandra\ High Resolution Mirror Assembly
(HRMA).  The remaining six CCDs (ACIS-S; CCDs S0--S5) are set in a
linear array and are tipped to approximate the Rowland circle of the
objective gratings that can be inserted behind the HRMA. The CCD which
lies on-axis in ACIS-I is I3.  Each CCD subtends \hbox{an
$8.3^{\prime}\times 8.3^{\prime}$} square on the sky.
The \hbox{HDF-N} was placed near the ``aim point" of the ACIS-I array
during all four observations.
 While observing with ACIS-I, two CCDs from ACIS-S, typically S2 and
 S3, can be operated.  CCD S3 was turned off during the \hbox{HDF-N}
observations due to the higher background level of this device; this
property may cause telemetry saturation during background flares. CCD
S2 was operated during our observations, but since it is far off-axis,
it is outside the Caltech area under study in this paper.

The full ACIS-I field of view is 
$16.9^{\prime}\times 16.9^{\prime}$,\footnote{There are 
small gaps between the CCDs; see Figure~\ref{emap}.} 
although the region of full exposure (221.9~ks) is somewhat reduced due
to the different pointings of the four separate observations.  These
offsets were necessary to satisfy the roll constraints of
\chandra\ while keeping the \hbox{HDF-N} itself near the area of best
focus and away from the gaps between the four ACIS-I CCDs  (see
Figure~\ref{emap}).  The on-axis image quality is~$\approx
0.5^{\prime\prime}$ FWHM increasing to~$\approx 3.0^{\prime\prime}$
FWHM at $\approx 4.0^{\prime}$ off-axis.\footnote{Each ACIS pixel
is~$0.49^{\prime \prime}$ on a side.}  The focal-plane temperature,
which governs several characteristics of the CCD behavior, in
particular the Charge Transfer Inefficiency (CTI), was $-110^\circ$C
during the first three observations and $-120^\circ$C during the
fourth.  The version numbers of the \axaf \footnote{{\it Advanced X-ray
Astrophysics Facility\/}, the name of the \chandra\ mission before
launch.} Science Center Data System software used for the processing of
these data were R4CU4UPD2 for the first three observations and
R4CU5UPD2 for the fourth.  The data processing dates were 1999 Nov 26,
1999 Nov 27, 1999 Nov 28,
 and 2000 Apr 19.

The data were corrected for the radiation damage the CCDs suffered
during the first few weeks of the mission (Hill et~al. 2000; Prigozhin
et~al. 2000) following the procedure of Townsley et~al.  (2000), which
corrects simultaneously for both position-dependent gain shifts and
event grade changes.    There is still some ($\approx 11$\%) residual 
quantum efficiency (QE) degradation in the $-110^\circ$C data at the
center of the ACIS-I array at 5.9~keV;\footnote{The effect of CTI is most 
severe at higher energies and farthest from the read-out nodes.  The center
of the ACIS-I array is far from the read-out nodes, so these values are the
worst-case scenario.} but this is improved over the 
 $\approx 19$\% loss incurred without CTI correction (Townsley et~al. 2000). 
  At $-120^\circ$C,  the 5.9~keV QE
loss from CTI is $\approx$ 10\% at the center of the ACIS-I array 
before CTI correction, and it is
$\approx5$\% with correction.  For a $\Gamma=1.4$ power law 
($N = A E^{-\Gamma}$, 
 where $N$ is the number of photons per second per
cm$^{-2}$ per keV, $A$ is a normalization constant, and $\Gamma$ is the
photon index), 5\% of the hard-band energy flux is lost due
to this residual QE deficiency, so the incurred error is generally
small compared to measurement error.   We do not correct for this
residual QE loss, as it is dependent on the spectral shape of
the source, and the X-ray photon statistics are generally too limited
to constrain the detailed spectral properties of the sources.  For
extremely hard sources, where most of the photon flux occurs at
observed energies greater than 5~keV, it is possible that the hard-band
fluxes will have been underestimated by a slightly larger factor.
However, the HRMA effective area is quite small above 5~keV, so few
such extremely hard sources would be detected in the current sample regardless
of the effects of CTI.

The background was stable during the entire 221.9~ks of observation.
The 0.5--8~keV light curves have been inspected for all of the
observations, and they are free from strong flaring due to ``space
weather." Over the
four observations, only seven 3.3-s frames were ``dropped" from the
telemetry, which amounts to a data loss of less than 0.01\% (frames are
dropped when background flaring occurs, causing the telemetry to be
saturated). During any observation the background is stable to within 
$\approx 20$\%, and the mean background levels from the four observations
agree to within 10\%.  The average background over the fully exposed portion
of the Caltech area is 0.053~count~pixel$^{-1}$ in the 0.5--8~keV band.
 Thus, even with a total observation length of 221.9~ks, 
the observation is still photon limited. \chandra\ ACIS should remain photon
limited near the optical axis and for point source detection 
in the 0.5--8 keV band out to exposure lengths of $\approx 4$~Ms.

\section{Chandra Source Searching and Analysis}

The X-ray data analysis performed in this paper was carried out using
the \chandra\ X-ray Center's (CXC's) \chandra\ Interactive Analysis of
Observations ({\sc ciao}) software as well as the Interactive Data 
Language ({\sc idl}) based Tools
for ACIS Real-Time Analysis ({\sc tara}) software developed at Penn
State (publicly available; see Broos et~al. 2000).

\subsection{X-ray Image Creation \label{xrimagecreate}}

The four observations were co-added after cross-registration using 12
bright X-ray sources detected in each of the individual observations
near the optical axis (see Figure~\ref{register}).
 All observations were reassigned to the coordinate system of
observation 966 (see Table~\ref{chandralog}), requiring shifts of
$\approx 1^{\prime \prime}$ and very small rotational corrections 
(0.01--$0.20^{\circ}$).  After cross-registration, the offsets of the other
three observations from the coordinate system of 966 were zero within plausible 
errors with $0.22^{\prime \prime}$ root mean square (RMS) dispersion in right ascension and
$0.19^{\prime \prime}$ RMS dispersion in declination.  Thus internal
cross-registration of the four observation datasets is accurate 
to within $0.3^{\prime\prime}$.

Absolute X-ray source positions were obtained by matching bright ($>$
25 counts) 0.5--8~keV X-ray sources to either $R < 23$ sources in the
Hogg00 catalog or any radio source in the R00
 1.4 GHz catalog.   Using the astrometry as determined on board
\chandra\, before any astrometric corrections were made, the 22 matched
sources had a mean offset from the X-ray sources of 1.88$\pm
0.70^{\prime \prime}$.  After shifting and slightly rotating the
positions of the X-ray sources so that the mean offset  
was zero, the RMS scatter in the offsets 
was $0.46^{\prime \prime}$.  The dispersion in positional offsets for these 22 sources is
shown in Figure~\ref{register}, along with an ellipse showing the RMS 
 dispersion of offsets for this group.  We adopt
$0.5^{\prime\prime}$ as our systematic positional uncertainty but note
that for X-ray sources with $\simlt$10 counts and/or larger off-axis
angles ($\simgt 3.0^{\prime}$), the positional uncertainty will be
dominated by photon statistics and will be $\simgt 1.0^{\prime \prime}$.

Images were created from 0.5--8~keV (full band), 0.5--2~keV (soft band)
and 2--8~keV (hard band) using the {\sc ciao datamodel}
 software.  \asca\ event grades 0, 2, 3, 4 and 6 were used in all
analyses, as this is the grade distribution currently being supported
by the CXC.  In the future, as support is developed for different grade
distributions, the effective background may be reduced by choosing a
subset of these grades (compare with Brandt et~al. 2000).  Neglecting
the 8--10~keV data improves the signal-to-noise ratio in the hard-band,
as the effective area of the mirror is steeply decreasing while the
overall background is increasing with energy over this range.  The raw
images are shown in Figure~3 for both the soft and hard bands.  Figure
\ref{imagesmooth} displays the same data adaptively smoothed at the
$3\sigma$ level using the code of Ebeling, White, \& Rangarajan (2001).
We did not use the adaptively smoothed images for source searching, but
they do show most of the detected X-ray sources more clearly than the
raw data.  Note that some of the sources we detect do not appear in
Figure~\ref{imagesmooth} because they fall below the 3$\sigma$ level of
the adaptive smoothing.

\subsection{X-ray Exposure Map \label{xrexpmapcreate}}

\begin{figure}
\vspace{-0.5truein}
\epsscale{0.8}
\figurenum{1}
\plotone{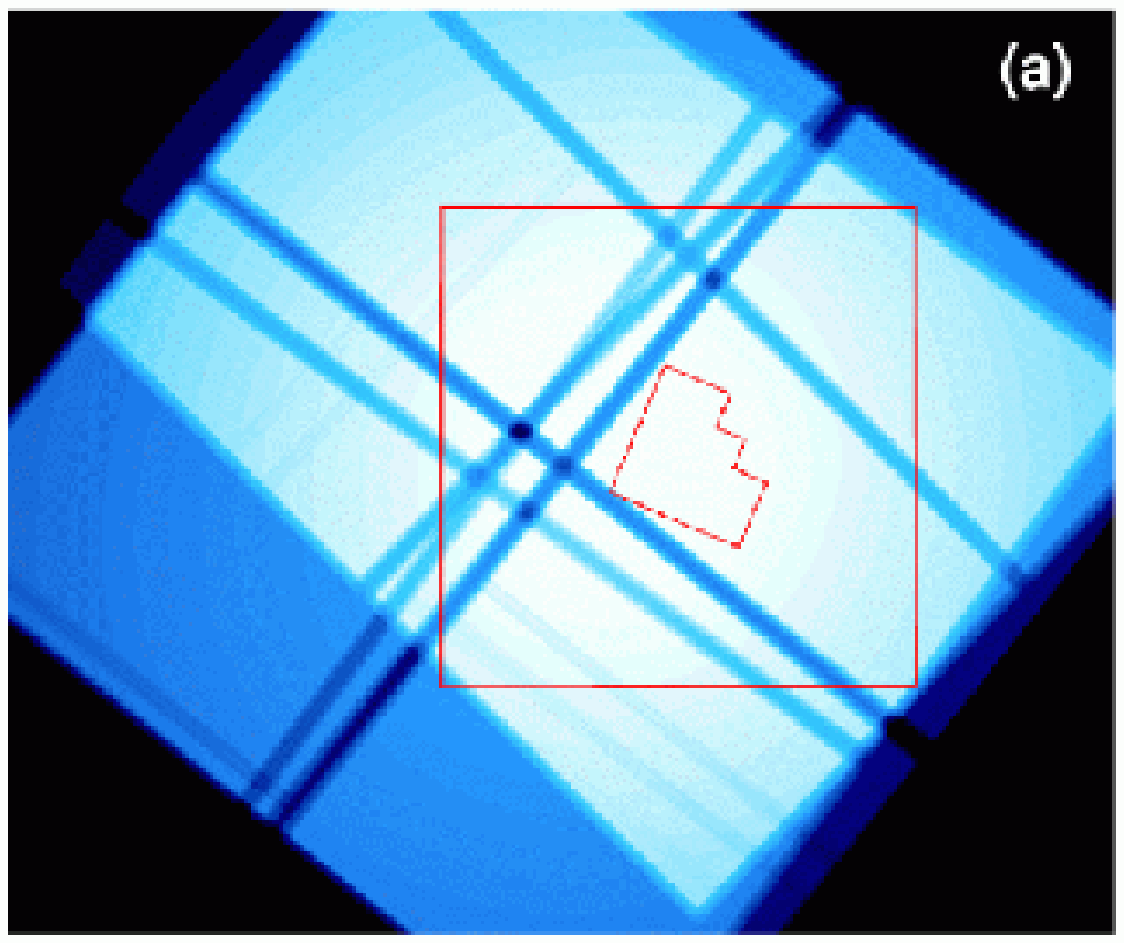}
\epsscale{1.0}
\plottwo{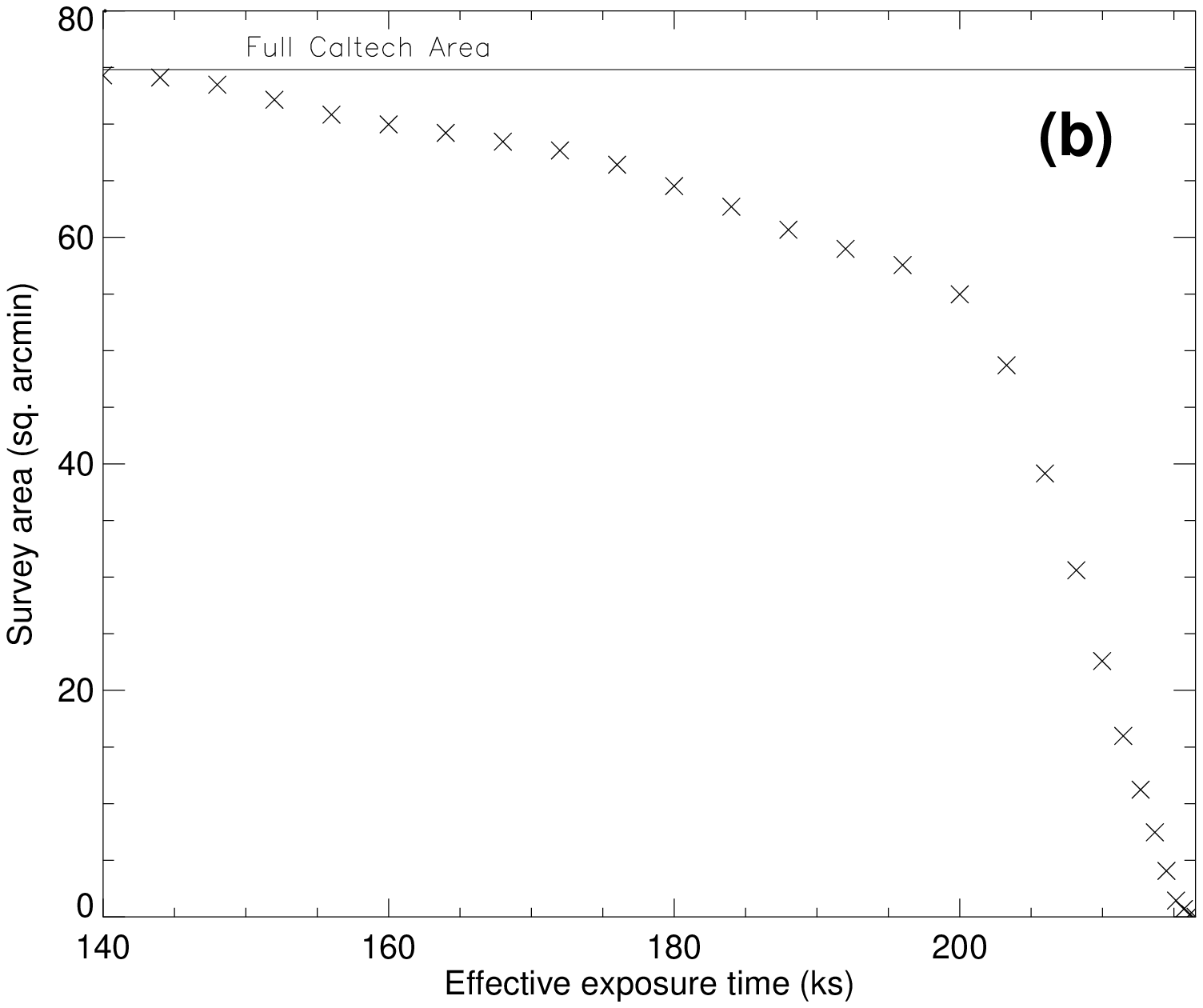}{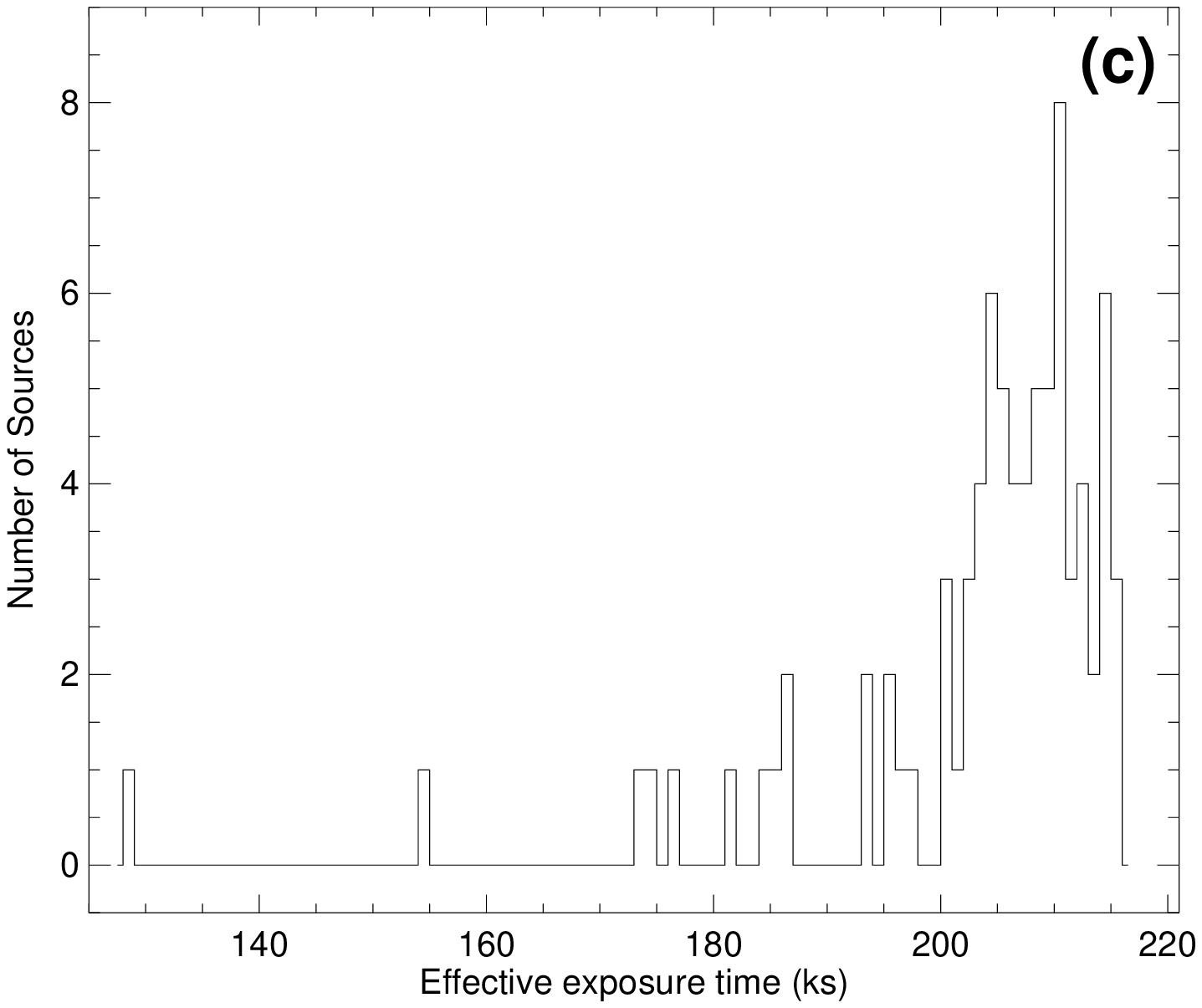}

\vspace{-0.1truein}
\caption{(a) The combined exposure map for the four
ACIS-I observations.  The whitest regions correspond to the highest
exposure times.  The dark grooves running through the exposure map correspond
to the gaps between the ACIS-I CCDs. North is up,  and East is to the left.
 The large square indicates the Caltech area, which is
$8.6^{\prime} \times 8.7^{\prime}$.  The
small polygon indicates the \hbox{HDF-N}.  (b) The observed
survey area for the Caltech area as a function of effective exposure time in that area.
Approximately 75\% of the Caltech area has an effective exposure time of $> 200$~ks, which
corresponds to the whitest regions in the exposure map.   (c) The distribution
of effective exposure times for the 82 X-ray sources detected in the Caltech area.
\label{emap}}
\end{figure}

\begin{figure}
\vspace{-0.25in}
\epsscale{0.6}
\figurenum{2}
\plotone{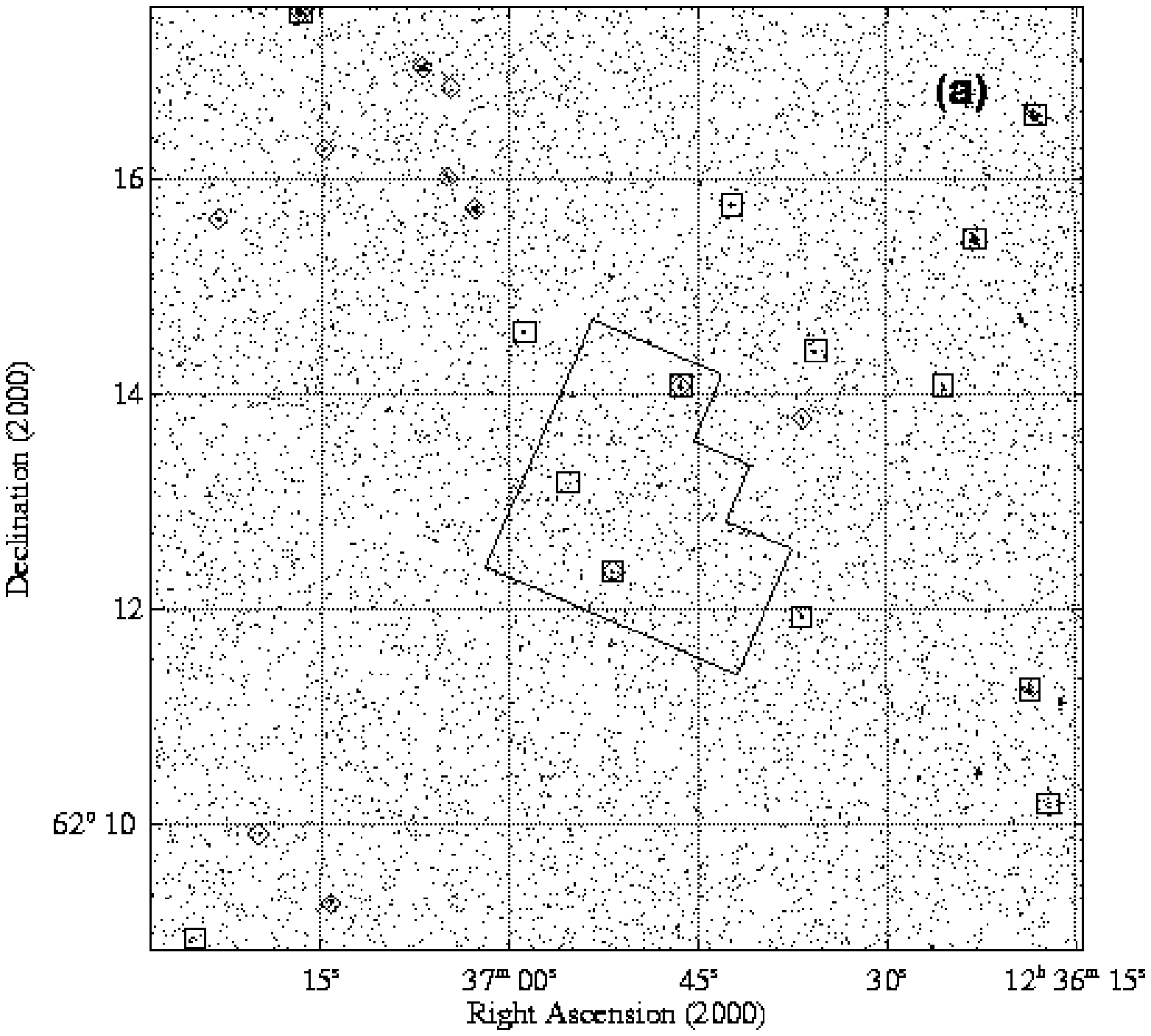}
\epsscale{0.5}
\plotone{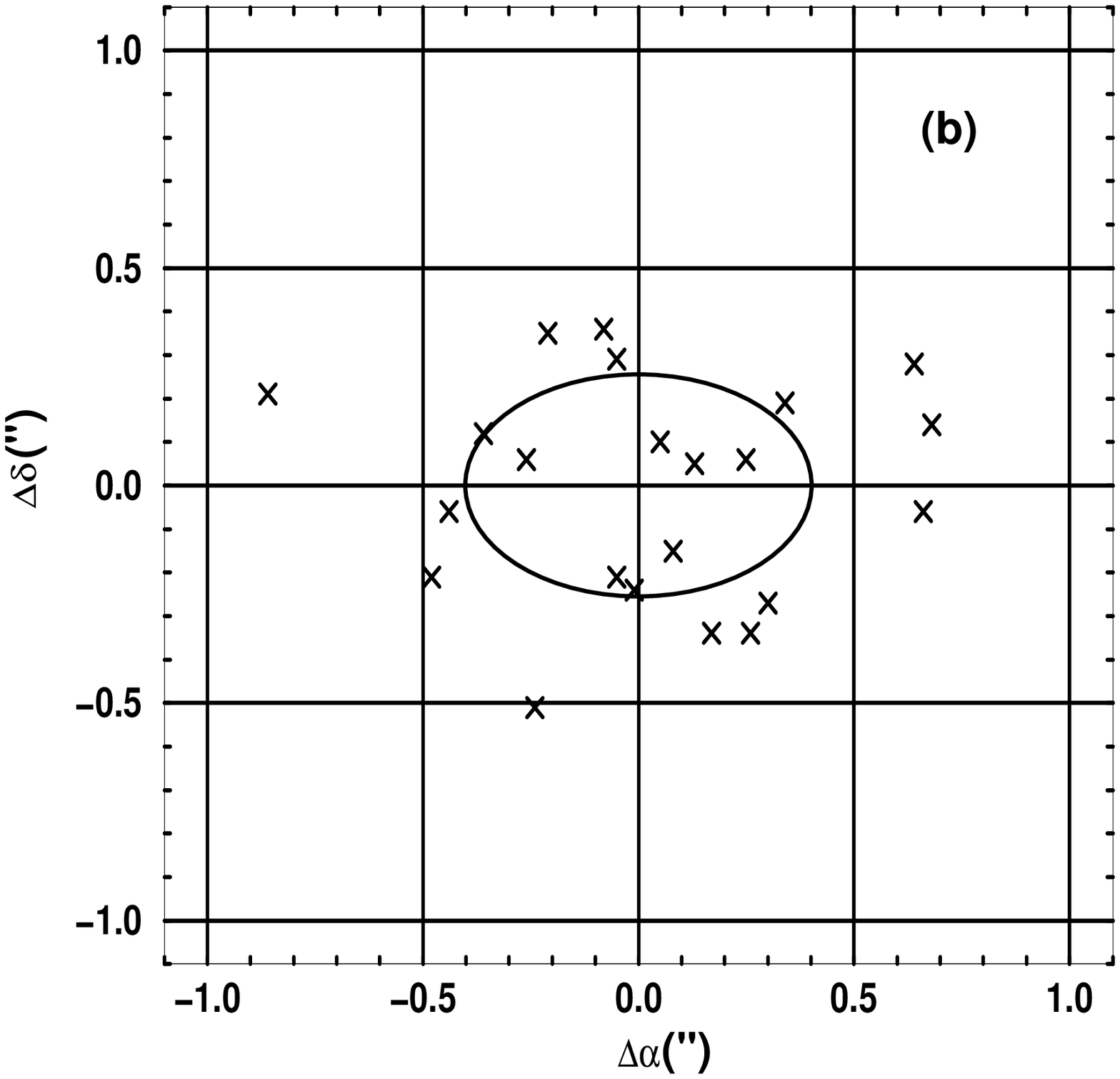}
\vspace{-0.5in}
\caption{(a) The sources used for registration and absolute alignment of the X-ray images
are shown in this 0.5--8~keV image.  The gaps between CCDs appear less sharp
in this image than in Figure~\ref{emap} as the overall background level is very low 
and the image scaling is logarithmic.  The diamonds indicate bright
sources present in all four \chandra\ observations used to register the
images.  The squares indicate X-ray sources with reliable optical
matches in Hogg00 or radio matches in R00 
 which were used for absolute alignment of the image.  A total of 
22 sources were used for absolute alignment of the \hbox{HDF-N} observation; 
14 are in the Caltech area.  (b) Positional dispersion of the 22 \chandra\ sources
used for absolute alignment with respect to Hogg00 and R00 coordinates; the ellipse
shows the RMS dispersion of the offsets.
\label{register}}
\end{figure}

The amount of ACIS exposure for a given patch of sky in the Caltech
area is influenced by two main factors: (1) telescope vignetting
reducing the effective area at larger off-axis angles,  and (2)
observatory dither changing the celestial location of bad pixels and 
gaps between the CCDs and thus reducing  
the amount of time a given area is visible (see Figure~\ref{emap}).  An
``effective exposure time," defined as the equivalent amount
of exposure time for a source located at the aim point, can be
calculated taking these two effects into account.   

For this data set, effective area maps were generated using a script
developed by T. S.  Koch, P. S. Broos, and D. P.
Huenemoerder.\footnote{The script is available at
http://www.astro.psu.edu/xray/acis/recipes/exposure.html}
 This script uses a combination of {\sc ftools} ({\sc ftools} Group
2000) and {\sc ciao datamodel} tools to generate QE and effective area
maps for each of the individual observations.  The {\sc ciao} tool {\sc
mkexpmap}, one of the tools used in the aforementioned script,
 makes a map of average effective area over an individual observation
(with units~cm$^{2}$).  We generated an effective area map assuming
a $\Gamma = 1.4$ power-law spectrum for the full 0.5--8 keV energy range.

To translate effective area into effective time, these effective area
maps are binned into 1$^{\prime \prime}$ square pixels, and a median
map is generated by replacing each effective area value with the median
of the effective area values in a circular region of diameter
5$^{\prime \prime}$, a dimension chosen to be representative of a
source detection cell size.  The motivation for making the median map
was to produce a two-dimensional ``look-up" table whose individual
entries would be appropriate for the range of \chandra\ PSF values over
the entire survey area ($\approx 1.0$--$5.0^{\prime \prime}$).  The
median statistic was chosen over the average since it is more robust.
The median map is divided by the maximum effective area and multiplied
by the total exposure time.  The four (see Table~\ref{chandralog}) individual
exposure maps were added to produce the full exposure map (see
Figure~\ref{emap}) for the entire 221.9~ks observation.
Figure~\ref{emap} also shows the survey area as a function of effective
exposure time and the distribution of exposure times for the 82 X-ray
sources described in \S~\ref{xrpointsources} (effective exposure times
for the individual sources are given in Table~\ref{xraydata}). The few
sources with exposure times much shorter than 221.9~ks are located on
or near the gaps between the CCDs. These gaps have effective widths of 
$\approx 32^{\prime \prime}$; 
 the observatory dither reduces their sharpness  
and produces scatter in the exposure times.
 The mean effective exposure time in the entire region is
 203$\pm14$~ks.  The median effective exposure time is 207~ks, and the
minimum effective exposure possible is 89~ks in the center of the gaps
between the CCDs.

\subsection{X-ray Emission from Point Sources \label{xrpointsources}}

We searched the images described in \S3.1 with the {\sc wavdetect}
software (Dobrzycki et~al. 1999; Freeman et~al. 2001) using the same
methods and safety checks as Brandt et~al. (2000) and H00.  We have
used a probability threshold of $1\times 10^{-7}$ for source
detection and searched using a ``$\sqrt{2}$~sequence" of wavelet
scales (scales of 1,
$\sqrt{2}$, 2, $2\sqrt{2}$, 4, $4\sqrt{2}$, 8, $8\sqrt{2}$, and 16 pixels).
The X-ray images shown in Figure~3 have
dimensions of $\approx 1032 \times 1044$ pixels; this means that we
expect $\approx 0.1$ false detections per image for the case of a uniform, static  
background.  In reality, the background is neither perfectly uniform or
static.  The background level decreases in the gaps between CCDs and
increases slightly near bright point sources due to the point spread
function (PSF) wings.  This latter effect should be small in the current data,
since there are few bright sources. The 90\% encircled energy radius 
 of the PSF, even at $4^\prime$ off-axis and for the hardest energies
under analysis here, is only
$5^{\prime \prime}$.

We mentioned in \S\ref{ACISobs} that there were no significant
background flares during any of the observations.  However, there are
other transient and more isolated phenomena, the most important of
which is the occasional flaring pixel (CXC, private communication).
Flaring pixels occur in front-side illuminated ACIS CCDs when 
charge from an incident charged  
particle is released slowly in a fashion that resembles real X-rays.  The
result is a series of spurious events occurring in the same CCD pixel
over 2--7 sequential frames. We have performed a flaring pixel analysis
using code developed by T.  Miyaji, which searches temporally for
events showing the characteristic energy decay of a flaring pixel
event.\footnote{This code is available at
ftp://maibock.phys.cmu.edu/pub/miyaji/Flagflare.tar.gz}
 Using this code, we were able to reject nine sources where flaring
pixel events combined with background events to exceed the source
detection threshold (see Table~\ref{flaring}).  All of the rejected sources had fewer
than nine counts, and only one had more than seven counts. All but one were
also detected  only in the soft band,  and all sources were only present
in one of the four observations. Any sources with fewer than seven counts,
especially if only detected in the soft band, are still considered
marginal but are included in the X-ray source catalog.

Even in a 221.9~ks observation the ACIS data are far from being
background limited (see \S\ref{ACISobs}).  Over the entire Caltech area
the detection limit for point sources with our selection criteria is
$\approx$~5--7 counts in the full, soft and hard bands.  For a
power-law model with photon index $\Gamma=1.4$ and the Galactic column
density, seven counts in 221.9~ks corresponds to a soft (hard) observed
flux detection limit of $1.3\times 10^{-16}$~erg~cm$^{-2}$~s$^{-1}$
($6.5\times 10^{-16}$~erg~cm$^{-2}$~s$^{-1}$).  For the cosmology of \S
1, at $z=1$ the corresponding rest-frame luminosity limits are
$L_{0.5-2}=3.4\times 10^{41}$~erg~s$^{-1}$ 
($L_{2-8}=1.8\times 10^{42}$~erg~s$^{-1}$); even fairly low-luminosity
Seyfert galaxies should be detected at this redshift.

 A summary of the numbers of detected X-ray sources and the background
levels in each of the three bands is given in
Table~\ref{xraydetections}.  We have also searched in the 5--8~keV band,
described as the ``ultra-hard'' band in Table~\ref{xraydetections}, finding
10 sources.  All but one of these 5--8~keV sources have strong 2--8~keV counterparts;
the one remaining source was detected with only 3.8~counts in the ultra-hard band.  We have not 
included this faint ultra-hard-band source in the list as its detection is
 marginal at this point.  

There are a total of 82 \chandra\ sources in the Caltech area, 73 of which
are detected in the full band.   Of the 45 hard-band sources,
13 are not detected in the soft band.  Of the 62 soft-band sources, 31 are
not detected in the hard band. The fraction of soft-band
 sources detected only in the soft band is somewhat lower than the
fraction of G01, but the two are roughly
consistent.  The fraction of hard-band sources detected only in the hard
band is somewhat higher than that of G01 (15/45 as
compared to G01's 15/91).  This difference could be due to the
hardening of sources at faint fluxes;  the data presented in this paper
are for a 221.9~ks observation, as compared to the 130~ks observation
presented in G01. 

The catalog of source detections and individual source properties is
given in Table~\ref{xraydata}.  The source positions listed are the
{\sc wavdetect} full-band X-ray source positions except when the source
was only detected in the soft band (there were no sources detected in
the hard band but not in the full band). Source photometry was
performed by {\sc wavdetect} whose accuracy was verified 
with our own aperture photometry. Good agreement was found for 
 sources with greater than 20 counts when using a circular
aperture of radius 3.0$^{\prime \prime}$ and background annulus of inner
and outer radius 4.0$^{\prime \prime}$ and 6.0$^{\prime \prime}$, 
respectively.  For sources with
fewer than 20 counts, the agreement is not quite as tight, but it is
still consistent within the errors quoted by {\sc wavdetect}.  Three
sources detected by {\sc wavdetect} with fewer than five counts were
found to have 5--8 counts when aperture photometry was performed.
This discrepancy is due to a failure in the {\sc wavdetect} source
characterization code which occasionally ``collapses" one or both of
the axes of the ``source ellipse" to zero, resulting in a gross
underestimate of the number of counts present even though a source was
detected with a probability threshold of $1 \times 10^{-7}$.  We have
manually inspected these three discrepant sources and indeed find
sources at the positions that {\sc wavdetect} reports.  For these
sources, we have used the counts determined by the aperture photometry;
all three are marked in Table~\ref{xraydata}.   For sources that were
not detected in a given band, we determine upper limits at the 99\%
confidence level by interpolating between the values in Table~3 of
Kraft, Burrows, \& Nousek (1991).  All X-ray upper limits in this paper
used this same methodology.  
The aperture sizes used to determine the
upper limits depended on how far off axis the source was.   Aperture
diameters of 2$^{\prime \prime}$, 3$^{\prime \prime}$, and 
4$^{\prime \prime}$  were used for sources with off-axis angle, $\theta$, 
satisfying $\theta < 3^{\prime}$ ,  
$3^{\prime} \le \theta < 5^{\prime}$ and $5^{\prime} \le \theta < 6^{\prime}$,
respectively.

The X-ray band ratio in Table~\ref{xraydata} is defined as the ratio of 
hard-band to soft-band counts, and it is plotted versus soft-band count rate for all 
the \chandra\ sources in Figure~\ref{bandratios}.  We find, with two
exceptions (see the caption of Figure~\ref{bandratios}), 
that the sources become harder at low soft-band count
rates, consistent with the results of G01.
The power-law photon indices in Table~\ref{xraydata} have been
determined from these X-ray band ratios using the 
AO2 version of the 
Portable, Interactive, Multi-Mission Simulator
({\sc pimms}; Mukai 2000).  
In the case of ``low" numbers of
counts we assume $\Gamma = 1.4$; this value is the
counts-weighted mean photon index for sources with fewer than 300 full-band 
counts (see Paper III).
A source with a ``low" number of counts is
defined as being (1) detected in only the hard band with less than 15
counts,  (2) detected in only the soft band with less than 30 counts,
or (3) detected in both the hard and soft bands, but with less than 15
counts in each.   In the current sample, 45 of the 82 sources are
considered to have ``low" numbers of counts by these criteria. A more
complete spectral analysis will be presented in Paper III.

In Table~\ref{xraydata} we also present observed fluxes for the full, hard
and soft bands. These have been determined from the given count rates
and photon indices using the AO2 version of {\sc pimms}, except in the 
case of the two
spectroscopically identified M~stars (see \S\ref{galactic}) where we
have assumed a $kT = 0.5$~keV thermal bremsstrahlung model.  The listed fluxes
 have not been corrected for Galactic absorption; this effect is small due to
the low Galactic column density along the line of sight to the
\hbox{HDF-N} (see \S\ref{intro}).   We estimate the fluxes to be
accurate to roughly~30\%, as residual uncertainties of CTI-induced QE
loss and other calibration issues still exist.  The errors introduced
in the fluxes by using a $\Gamma = 1.4$ power law for a source with a 
$\Gamma = 1.7$ power-law continuum are 14\%, 5\%, and 2\% for the 
full, hard, and 
soft bands, respectively.

Using the normalization of Ishisaki (1997), the 45 hard-band X-ray 
sources in the Caltech area are found to account for $\simgt 80$\% of 
the diffuse extragalactic X-ray background in the hard band (Paper III;
after adding the fraction of the background already resolved in
wide-field surveys). 

\subsection{Extended X-ray Emission \label{xrextended}}

To investigate possible extended X-ray emission from the \chandra\ sources, we 
have identified the sources whose extents as determined by {\sc wavdetect}  
(see \S9.2 of Dobrzycki et~al. 1999) significantly exceed the 50\% 
encircled-energy radius of the \chandra\ PSF at 1.5~keV (see Figure~4.8 of 
the Version 3 \chandra\ Proposers' Observatory Guide for the behavior of 
the \chandra\ PSF with source off-axis distance).  Recall that this is
restricted to features found within ``Mexican hat" regions having radii 
$\leq 8^{\prime \prime}$ (the maximum {\sc wavdetect} scale was 16 pixels; 
 see \S \ref{xrpointsources}).
 We have manually 
inspected the resulting nine sources, and we do not find any with highly 
significant evidence for extent. All of the nine sources are located close 
to the center of the field of view, where small errors in image cross 
registration and satellite aspect reconstruction can most easily lead to false 
apparent extent. In addition, because the aim point position varied 
between observations (see Table~\ref{chandralog}), there is 
not a unique PSF applicable to all observations of any given source; 
apparent source extent on the scale of about an arcsecond can be 
introduced by variations in the applicable PSF from observation to 
observation. Therefore, we conclude that there are no sources found here
by {\sc wavdetect} with highly significant spatial extent. We note, 
however, that our constraints on the spatial extents of nearly 
point-like sources are limited in many cases due to small numbers 
of counts ($\simlt 20$).

To investigate further any extended X-ray sources, we have searched the
soft-band and hard-band images described above using the CXC's Voronoi
Tessellation and Percolation detection algorithm ({\sc vtpdetect};
Ebeling \& Wiedenmann 1993; Dobrzycki et~al. 1999). We have used a
probability threshold (i.e., a {\tt limit} parameter) of $1\times
10^{-6}$, and we require a minimum number of events per source (i.e., a
{\tt coarse} parameter) of 10.  In our searching, we have included a  
``buffer" region around the Caltech area to minimize any edge effects
that could affect extended sources in the Caltech area. We do not find
any highly extended sources in either the soft or the hard bands. We
comment that much of the apparent diffuse emission seen in some of the
adaptively smoothed images (e.g., Figure~\ref{imagesmooth}) can be
explained via a combination of instrumental background and emission in
the wings of the PSFs of bright point sources.

While the methods we have used above to search for extended 
emission are effective and appropriate for the scope of
this paper, we note that optimal extended source searching 
requires precise modeling of the underlying background. 
As improved understanding of the \chandra\ background develops, 
we will improve our extended source searching appropriately. 
 
\section{Multiwavelength Counterparts of the Chandra Sources \label{multiwavmatch}}
\subsection{Source Matching Procedure \label{sourcematching}}

	The multiwavelength studies of the \hbox{HDF-N} and its vicinity that
were correlated with the X-ray detections are presented in
Table~\ref{catalogs}. Listed are the number of objects in each catalog
located within the  Caltech area, the positional accuracy assumed, the
number of objects with \chandra\ source matches, and the estimated false
match rate.  The positional accuracy we assumed for a given catalog was 
generally slightly larger than that quoted by the particular author.  
We assumed $\approx 0.5^{\prime \prime}$ positional
accuracy for the X-ray sources (see \S~\ref{xrimagecreate}) and used the positional error of each
catalog object to determine counterparts of X-ray detections; if a
catalog object's positional offset from the X-ray source was less than
the error on the object's position listed in Table~\ref{catalogs}, then a
 match was assumed.   In practice, the X-ray positional error was only
considered when an individual catalog's positional uncertainty was
less than 1.0$^{\prime \prime}$ (e.g., in the \hbox{HDF-N} itself and for the
very precise radio catalogs).    Additional criteria besides spatial
coincidence within a catalog's errors were used when possible (e.g.,  the
much deeper optical data in the \hbox{HDF-N} supersedes more shallow surveys
in the optical).  The estimated false match rates in Table~\ref{catalogs} 
were calculated using a ``random-step" technique.  All X-ray sources were 
shifted by 7$^{\prime \prime}$ to the northeast, northwest, southeast 
and southwest, and for each of these shifts the source matching algorithm 
was run.  The resulting numbers of falsely matched sources were then 
averaged to estimate the false match rate.    

The expected numbers of falsely matched sources are generally quite low, 
so false matches should not materially affect our results.  For a few
catalogs, however, the false match rate appears to be fairly high when
compared to the number of \chandra\ source matches.  For instance, the
number of false matches expected for the W96 \hbox{HDF-N} catalog is
2.5, but the median $R$ magnitude for the false matches is 28.5 whereas
the \chandra-W96 matches had a median $R$ magnitude of 23.4.  There
are zero \chandra-W96 false matches expected at this bright optical
flux level, where the optical source density is much lower. The same
effect is seen for the Hogg00 and Barger et~al. (1999) catalogs.  For
the Hogg00 catalog, the falsely matched sources had a median $R$ of
24.8 whereas the \chandra-Hogg00 matches had a median $R$ of 23.3.
For the Barger et~al. (1999) catalog, the falsely matched sources had a
median $R$ of 23.6 and the \chandra-Barger matches had a median $R$ of
22.6.  We thus are confident that the vast majority of the
\chandra\ source identifications are correct.

\subsection{Optical Photometric Data \label{optphotom}}

   Deep $I$ (Kron-Cousins; see Bessel 1990) and $V$ (Kron-Cousins)
 band images covering the field
were obtained in 1997 Apr with the UH8K CCD
mosaic camera on the Canada-France-Hawaii Telescope. The 8.3 hour $I$-band and
11 hour $V$-band images were cross calibrated to shallower
$I$ and $V$ images obtained with the University of Hawaii~2.2m telescope.
A more detailed description of these imaging data may be found in Barger et al.
(1999). Deep $R$ (Kron-Cousins) band imaging was obtained
using Keck LRIS during 1998 Feb and 1999 Feb. The combined 
7080~s exposure, which
was calibrated using Landholt standards, covers approximately
half of the field.  More discussion of the $R$-band imaging data
may be found in BCR00.

  For each object in the X-ray sample $I$, $R$ and $V$ magnitudes were
measured using 3$^{\prime \prime}$ apertures centered on the nominal
X-ray position. These aperture magnitudes were then
corrected to approximate total magnitudes using an
offset of  $-0.19$ mags for $I$, $-0.22$  mags for $R$, and  $-0.24$
mags for $V$. For objects brighter than $R=21$, isophotal
magnitudes computed to 1\% of the peak surface
brightness were also calculated and used in place of
the aperture magnitudes. For the fainter objects the
isophotal magnitudes generally agree well with the 
aperture magnitudes. The apertures are shown superimposed
on $19^{\prime \prime} \times 19^{\prime \prime}$ 
$I$-band thumbnail images of the surrounding field
in Figure~\ref{optical_cutouts1}. For a small number of objects there is an
apparent optical counterpart lying within the aperture
but sufficiently close to the edge that if this
is the counterpart we would overestimate the magnitude
of the object. In these cases we have remeasured the
magnitudes with apertures centered on the optical positions;
these objects have been marked in Table~\ref{multiwavtable}.
For each image we measured a  $1 \sigma$ noise level by
placing random apertures in the field and measuring
the distribution of corrected aperture magnitudes
at these positions. The $1 \sigma$ levels measured
in this way are 26.0 for $I$, 
27.3 for $R$, and 27.1 for $V$.

  In Table 6 we list the $R$ and $I$ magnitudes for the 82 \chandra\ 
sources. Where no direct $R$ measurement is available we have used 
the relation:

\begin{equation}
 R = I - 0.2 + 0.5(V-I)
\end{equation}

\noindent
This relation was computed as a fit to the $R$ magnitudes in the observed region
to determine the $R$ magnitudes from the $V$ and $I$ observations.
The $1 \sigma$ dispersion of the fitted objects is 0.18 mags, 
and 97\% lie within 0.4 mags of this relation. The
computed $R$ magnitudes should be accurate to this level
which is sufficient for the present work.   

When needed, we quote $2 \sigma$ lower limits
on $R$ and $I$ magnitudes in Table~\ref{multiwavtable}. 
There are a few
additional sources for which deeper data were available (e.g., in the
\hbox{HDF-N} itself).

Six of the 62 soft-band sources have no optical counterparts to
$R\approx 26.5$.    Four of these soft-band sources have neither
hard-band nor full-band counterparts and were detected with fewer
 than 6 counts; some of these may be false X-ray detections (see
 \S~\ref{xrpointsources}).  Only one of the 45 hard-band sources has no
optical counterpart to  $R\approx 26.5$ (\hbox{CXOHDFN}~J123648.3+621456);  
it has a full-band counterpart.  
There were an additional two sources lacking
optical counterparts which were detected in the full band (0.5--8 keV)
but in neither the soft band nor the hard band. 

There are a total of nine \chandra\ sources with no optical counterparts
to $R\approx 26.5$ and $I \approx 25.3$.  These nine sources also had no counterparts in
the Hogg00 $K_{\rm s}$ image as determined from our manual inspection. Our
results on optical counterparts are consistent with those of G01, who
find optical counterparts for 90\% of their X-ray sources.
Due to the potentially false detections of some of the faintest X-ray
sources in our sample, we also consider only X-ray sources detected with full-band fluxes $> 7
\times 10^{-16}$~erg~cm$^{-2}$~s$^{-1}$; we find that only 1 of 57 (1.8\%) lacks 
an optical counterpart to $R \approx 26.5$.

We have compared our sources with those of Barger et~al. (2001, hereafter B01)
 using their criteria for
``optically faint" (\hbox{$I > 23.5$}).  
Of the 45 hard-band sources, 18
(40\%) have \hbox{$I > 23.5$} compared with seven of the 20 (35\%) B01 hard-band
sources.  The two surveys are consistent
within the statistics; note that the current survey is more than twice
as deep as that of B01. More optically faint counterparts are expected
as extremely faint hard-band X-ray flux limits are reached (see
\S~\ref{optfluxcolors} for discussion of $R$ versus X-ray flux).

The other main source of optical photometric data used in this paper is 
the catalog of Hogg00.  We also used the deep ${\cal R}$ image of Hogg00
(kindly provided to us by C. Steidel and D. Hogg) for cross-calibration 
purposes.  Colors discussed in this paper are in the
$U_{\rm n}$, $G$, and ${\cal R}$ filters as used by Hogg00 and
described in Steidel \& Hamilton (1993), except where otherwise
indicated.  
The $U_{\rm n}$, $G$, and ${\cal R}$ filters have effective
wavelengths of 3570, 4830, and 6930~\AA, FWHM bandpasses of 700, 1200,
and 1500~\AA, and zero-magnitude flux densities of roughly 1550, 3890,
and 2750 Jy.  These magnitudes are normalized to Vega.  Also included
in our data analysis are ${\cal R} - K_{\rm s}$ colors as calculated by
Hogg00; the $K_{\rm s}$ filter has an effective wavelength of
2.15~$\mu$m, a FWHM bandpass of 0.3~$\mu$m, and a zero-magnitude flux
density of roughly 708 Jy.  The $\cal R$ band is related to the 
Kron-Cousins $R$ and $I$ bands by the following relation from Steidel \& Hamilton (1993):

\begin{equation}
{\cal R} = R + 0.24 - 0.28(R - I)
\end{equation}

\section{Optical Properties of the Chandra Sources\label{multiwavproperties}}

In this section we describe the bulk optical properties of the \chandra\
sources in the Caltech area.  Detailed descriptions of individual sources 
are given in \S A, and an update on X-ray source detections in
the \hbox{HDF-N} itself is given in \S\ref{xrayhdfn}.   

\subsection{Optical Fluxes and Colors \label{optfluxcolors}}

The median $R$ magnitude of the 73 \chandra\ sources in the Caltech area
 with optical detections is $\approx23.7$. The color distributions of our
X-ray-optical matches are shown in Figure~\ref{colors}; these colors were
calculated by Hogg00.  The two X-ray
detected M~stars stand out in several plots; they have colors that are
redder than most of the catalog objects.  X-ray detections with $20
< {\cal R} < 23$ tend to be redder than the average galaxy (most easily seen
in the ${\cal R}-K_{\rm s}$ versus ${\cal R}$ plot).  There also seems to be an excess
of objects with $4 < {\cal R}-K_{\rm s} \simlt 5$ as compared to the field 
galaxy population.  Three X-ray sources have colors that
are bluer than the general galaxy population; these bluer objects
are most evident in the $U_{\rm n}-G$ versus ${\cal R}$ and $G-{\cal R}$ 
versus ${\cal R}$ plots.
Optical spectra of all of these blue objects show broad lines; they are
all AGN and are fairly luminous ($L_{0.5-8} \approx
10^{43-44}$~erg~s$^{-1}$).

Figure~\ref{akiyama_fancy} shows the relationship between $R$ magnitude
and soft-band and hard-band X-ray flux.  We have plotted several values
of the X-ray to optical flux ratio, $\log{({{f_{\rm X}}\over{f_{\rm
R}}})}$, on this figure.  This ratio was derived using the Kron-Cousins $R$ filter
transmission function:

\begin{equation}
\log{\left({{f_{\rm X}}\over{f_{\rm R}}}\right)} = \log{f_{\rm X}} + 5.50  + {R\over{2.5}}
\end{equation}

\noindent
(compare with Figure~1 of Maccacaro et~al. 1988). The majority of the X-ray sources in the current survey 
 fall within $\log{({{f_{\rm X}}\over{f_{\rm R}}})} = 0.0\pm 1$. This range of values is 
 typical of sources found in the wider field X-ray surveys of Schmidt et~al. (1998) using \rosat\ 
and Akiyama et~al. (2000) using \asca;  it is that typical of AGN in the local  
Universe.  There are a few sources,
however, that deviate from
$\log{({{f_{\rm X}}\over{f_{\rm R}}})} = 0.0\pm 1$.  Two of these
sources are spectroscopically identified M~stars (see \S \ref{spectradetails} and
 Table~\ref{multiwavtable}), which have the expected $\log{({{f_{\rm
X}}\over{f_{\rm R}}})} \approx -2$.

At the faintest soft-band fluxes ($< 3 \times 10^{-16}$~erg~cm$^{-2}$~s$^{-1}$), 
there is an interesting group of 7
 sources which are X-ray weak for their optical fluxes (see
Figure~\ref{akiyama_fancy}). 
These are seen only below the flux limits
of the deepest pre-\chandra\ soft-band surveys (e.g., 
Schmidt et~al. 1998).   These
sources are all relatively nearby ($z < 0.5$) low X-ray luminosity
objects that are described in \S\ref{LLXR}.

There are also three objects which are notably faint in the optical 
with respect to
their X-ray flux values;  these sources
 are of interest because they may represent a new class of
astronomical object (e.g., extremely high redshift quasars).  
The three sources are found to be optically faint with 
respect to their hard-band fluxes, one of these is also found to 
be optically faint with respect to its soft-band flux. 
 This soft-band source, \hbox{CXOHDFN}~J123704.8+621601,
is relatively bright and 
has X-ray band ratio of $0.45^{+0.14}_{-0.11}$. Its
optical counterpart has $R=25.4$ and is discussed in \S A.3.
One of the other two sources is the Type 2
QSO candidate in the \hbox{HDF-N} itself (\hbox{CXOHDFN}~J123651.7+621221; see
H00). The third source is \hbox{CXOHDFN}~J123616.0+621107, which in addition
to being optically faint is anomalously hard for its soft-band count
rate (see Figure~\ref{bandratios}).  The X-ray and optical properties
point to a possible extreme AGN; this object is discussed in detail in
\S A.3.

\subsection{Optical Spectroscopic Observations with Keck and the Hobby-Eberly Telescope \label{optspectra}}

Extensive spectroscopic and photometric redshift studies have been
carried out in the \hbox{HDF-N} region (e.g., Fern\'andez-Soto et~al.
1999; C00).  We have supplemented these efforts by obtaining spectra of
X-ray sources in the Caltech area using the Keck and Hobby-Eberly
Telescopes.  This section describes our spectroscopic work.

Of the 82 X-ray sources, 25 have optical counterparts brighter than
$R=23$ (including two stars, which are excluded from most of the
analysis in this paper).  A total of 32 redshifts were found from both
the literature and our optical spectroscopy, including 24 of the 25
sources with $R < 23$ and eight sources with $R > 23$.  The majority of the
redshifts are from C00; we provide an additional six previously
unpublished redshifts from four Keck spectra and two HET spectra.  We also
present a few additional spectra which were taken in order to provide
more information about the nature of the individual objects.

Keck Low Resolution Imaging Spectrograph (LRIS; Oke et~al.  1995)
spectra were obtained on 2000 March 7 and earlier (L.L. Cowie 2000,
private communication).  These were taken using two
slit-masks oriented at PA$=90^{\circ}$.  Each mask was observed at
three positions (steps were $\approx 1.5^{\prime \prime}$ along the
slits) with 1200 s
 per exposure.  The 300 line mm$^{-1}$ grating was used with central
wavelength 6700~\AA.  The night was photometric with seeing of $\approx
0.8^{\prime \prime}$.  Keck Echellette Spectrograph and Imager (ESI;
Epps et~al. 1998) observations were taken on 2000 April 1--2 in
low-resolution mode using a 1$^{\prime \prime}$ slit.  Exposure times
were 300, 420, or 600 s at each position; most targets were observed at
three positions 2--3$^{\prime \prime}$ apart along the slit (brighter
targets were observed at two positions only), and the spectral
resolution was $\approx 17$~\AA.  Both nights were photometric with
seeing of 0.6$^{\prime \prime}$ the first night and 0.9$^{\prime
\prime}$ the second night.  The band pass for both the Keck LRIS and 
ESI is $\approx$ 5000--10000~ \AA.

Spectra were also obtained using the Marcario Low Resolution
Spectrograph (LRS; Hill et~al. 1998a,b; Schneider et~al. 2000) of
the Hobby-Eberly Telescope (HET; Ramsey et~al. 1998).  The HET operates
in queue mode, so the observations occurred on multiple evenings
between 2000 February 7 and 2000 June 2.
  A $2.0^{\prime\prime}$ slit and a 300~line~mm$^{-1}$ grism/GG385
blocking filter produced spectra from 4400--9000~\AA\ at
17~\AA\ resolution; points redward of  7700~\AA\ are suspect because of
possible second-order contamination.  
The exposure time per source ranged from 5--60~m.
 The image quality was typically $2.5^{\prime\prime}$ (FWHM).
Wavelength calibration was performed using HgCdZn and Ne lamps, and
relative flux calibration was performed using spectrophotometric
standards.

The spectra are shown in Figure~\ref{spectra1}, and
individual descriptions are given in \S\ref{spectradetails}.

\subsection{Redshifts, Luminosities, and Classifications of the Chandra Sources \label{zlumxray}}

\subsubsection{Redshifts and Luminosities}

The 32 sources in the X-ray sample for which we have redshifts
have their redshift information listed in Table~\ref{multiwavtable}.
Only two redshifts were derived using photometric data; all others are spectroscopic.
The sources with redshifts are fairly evenly distributed with respect to X-ray count rate
and band ratio (see Table~\ref{xraydata}); at least in terms of 
these properties, they appear to be representative of
the X-ray source population as a whole.

The median redshift of the sample is 0.76, but the spread is
quite large (see Figure~\ref{zmap}). About 90\% of the redshifts are
uniformly distributed in redshift space for $0<z<1.5$.  There exist two
groups of two objects each at $z \approx 2.6$ and $z\approx 3.4$. 
This possible redshift ``clustering'' is interesting, but until more 
redshifts are obtained above $z \approx 1.1$ we cannot determine whether
this clustering is real or merely an artifact of the small number of sources.

Full-band, soft-band and hard-band luminosities are given for the 
sources with redshifts in Table~\ref{multiwavtable}; these are 
calculated for the rest frame. The soft-band and hard-band luminosities 
are plotted versus redshift in Figure \ref{Lz}. We have found no strong 
correlations between either soft-band or hard-band luminosity and band ratio.

\subsubsection{Classification Scheme \label{idscheme}}

It is physically unreasonable to try to cast all of the X-ray sources
into a few sharply defined classes.  The classifications chosen below 
are designed to provide some insight into the general nature of the
population, but many of the sources could possibly belong to several of
the categories (e.g., a heavily obscured AGN that also shows 
circumnuclear starburst activity).   Objects that are
not confirmed AGN by our optical spectroscopic criteria are placed in
two ``galaxy" categories based on their X-ray luminosities.   
These ``galaxies" lack known AGN-defining spectral features 
in the optical band. Finally, we have a category for Galactic stars.

These various categories are briefly described as follows:

$\bullet$ {\bf BL AGN and NL AGN:} We follow the classification schemes of Schmidt et~al.
(1998) and B01 in that AGN must have
(1)~broadened emission lines (FWHM $>500$~km~s$^{-1}$) or (2)~emission
from either [\ion{Ne}{5}] or [\ion{Ne}{3}].  AGN are classified as
either broad-line (BL AGN, FWHM of emission lines $>1000$~km~s$^{-1}$) 
or narrow-line (NL AGN). 
We identify 15 of the 30 extragalactic sources as AGN, 
including eight spectroscopically identified BL AGN and QSOs.   

$\bullet$ {\bf Low X-ray Luminosity Galaxies (LL):} 
Low-luminosity galaxies (${L}_{0.5-8} < 
10^{41.5}$~erg~s$^{-1}$; see \S\ref{LLXR}) are
objects in which the X-ray source
can plausibly be explained by moderate-strength starbursts, hot gas in
elliptical galaxies, or other sources besides accretion onto a nuclear 
supermassive black hole (although accretion onto a supermassive black
hole is still a viable possibility for some of these objects).
There are eight sources in this category; only one is detected 
in the hard band.

$\bullet$ {\bf High X-ray Luminosity Galaxies (HL): } 
The seven objects classified as HL galaxies are all 
detected in the hard band and have hard-band X-ray luminosities
$\simgt 10^{42}$~erg~s$^{-1}$ (three are detected in {\it only} the hard
band), strongly suggesting that these objects are actually
AGN.  They stand in contrast to the low-luminosity X-ray galaxies, which
are generally detected only in the soft band (see \S~\ref{LLXR}). 
In some cases, we
were unable to assess the AGN nature of the object due to lack of
access to the optical spectrum. 

$\bullet$ {\bf Stars: }
Two M4 stars are included in the sample.  A curious 
difference between our sample and that of G01 is that none of the optically
bright, X-ray faint objects in the G01 sample are identified as stars, whereas 
two of our optically bright, X-ray faint sources are spectroscopically 
identified as stars (one would
expect more in the larger field of G01).  Figure~\ref{akiyama_fancy} shows
the location of the stars in a plot of $R$ magnitude versus X-ray flux.

A summary of the optical spectroscopic classifications of the
\chandra\ sources in the Caltech area is given in
Table~\ref{spectralclasses}.  This table separates the objects by the
X-ray band (soft or hard) in which they were detected, in addition to
giving the total number of objects in each class (regardless of X-ray
nature).  

  We plot ${\cal R}$ magnitude versus redshift in Figure~\ref{Rz}, breaking
the source redshifts shown in Figure~\ref{zmap} into separate 
spectroscopic classes.  The four higher redshift ($z > 2$) objects are
all luminous QSO-type AGN. 
At $z \approx 1$, there is a mixture of BL AGN, NL AGN,
and HL galaxies.
At $z \approx 0.5$, there is a mixture of NL AGN, HL galaxies, and LL galaxies 
(see \S \ref{LLXR}). Finally, at the brightest
${\cal R}$ magnitudes and lowest redshifts, the sources are LL galaxies and
stars.  We estimate the fraction of optically selected sources that are 
X-ray sources in a manner similar to that of B01 (see their \S~5). We find
that the fraction of galaxies with $-22.5 < M_{\rm {\cal R}} < -24$ having 
counterparts in the X-ray band is $0.06\pm{0.03}$, fully consistent 
with the $0.07^{+0.05}_{-0.03}$ fraction found by B01. 

\subsubsection{Comparison of Classifications with Other Samples}

We can compare our 20 hard-band classifications with the 13 sources in
the complete 2--10 keV sample of B01.  The survey of B01 reached
$3.8\times 10^{-15}$~erg~cm$^{-2}$~s$^{-1}$ in the 2--10 keV band (our
limit in the 2--10~keV band is $\approx 8.1\times
10^{-16}$~erg~cm$^{-2}$~s$^{-1}$).  We
find a comparable fraction of AGN in the hard band; 11 of 20 (55\%)
hard-band objects with identifications in our sample are
spectroscopically confirmed AGN, while 7 of 13 (54\%) hard-band objects in  
B01 are spectroscopically confirmed AGN.

Our soft-band classifications are most suitably compared with the  
\rosat\ Ultra Deep Survey (UDS) work of Lehmann et~al. (2001), who have
 obtained optical spectra for 85 of the 91 soft-band X-ray
sources in the Lockman Hole. The limiting 0.5--2 keV flux for the
\rosat\ UDS is $1.2 \times 10^{-15}$~erg~cm$^{-2}$~s$^{-1}$ (compared 
to $1.3\times 10^{-16}$~erg~cm$^{-2}$~s$^{-1}$ for this paper).  To
account for the increased number of low-luminosity galaxies seen at
very faint soft-band fluxes (see \S~\ref{LLXR}), we compare only our 13
sources with soft-band fluxes $> 1.2 \times
10^{-15}$~erg~cm$^{-2}$~s$^{-1}$ to the Lehmann et~al. (2001) sample.
We have spectroscopic identifications for 10 of those 13 sources; 6 are
broad-line AGN, one is a narrow-line AGN, and three objects fall into
the ``galaxies" categories.  Of 85 objects with
spectroscopic identification in the \rosat\ UDS, 56 are broad-line AGN,
and 13 are Type 2 AGN;  our survey is thus consistent with the
\rosat\ UDS after accounting for the small number of sources in our survey at
the relevant flux level.  In our entire identified soft-band sample
(see Table~\ref{spectralclasses}), however, the percentage of AGN is
not quite as high as that of the Lehmann et~al. (2001) sample, mainly
due to the increased prevalence of low-luminosity X-ray emitting
galaxies at very faint soft-band fluxes ($< 3 \times
10^{-16}$~erg~cm$^{-2}$~s$^{-1}$; see \S~\ref{LLXR}).

\subsection{Low-Luminosity X-ray Emitting Galaxies \label{LLXR}}

	As mentioned in \S\ref{zlumxray}, a group of
low-luminosity (${L}_{0.5-2} < 10^{41.5}$~erg~s$^{-1}$) soft X-ray 
sources appears at faint soft-band
flux levels.  Seven of these sources are visible as a group in
Figure~\ref{akiyama_fancy}a, appearing at faint X-ray fluxes ($\simlt 3
\times 10^{-16}$~erg~cm$^{-2}$~s$^{-1}$) and bright optical fluxes ($R
\simlt 21$).  There is one additional faint X-ray source that was not
detected independently in either the soft or the hard X-ray bands; this
full-band source, \hbox{CXOHDFN}~J123702.0+621123, has properties 
similar to the other low-luminosity sources.   We also place \hbox{CXOHDFN}~J123656.9+621301,
an elliptical galaxy in the \hbox{HDF-N}, in this category based on its
lower X-ray luminosity, although it
is more optically faint ($R=23.4$).

	All of these objects are galaxies at moderately low redshift,
$z=0.089$--0.56, and most have absorption-dominated or
``intermediate'' type optical spectra, where the spectrum is not
dominated by emission (C00).   The constraints on their X-ray spectral
properties are not strong due to low numbers of counts; the typical
X-ray band ratio upper limit is $<1.20$.    All but one
source (\hbox{CXOHDFN}~J123656.9+621301) is undetected in the hard band; 
the hard-band luminosity upper limits are $\leq 10^{41.7}$~erg~s$^{-1}$.
The one hard-band detected source has $L_{2-8} \approx 10^{41.6}$~erg~s$^{-1}$.
Furthermore, their optical luminosities are typical of
galaxies rather than AGN; most have $L_{\rm R} \approx 0.3 L_{\rm *}$.
While the newly discovered
X-ray source in the \hbox{HDF-N} itself
(\hbox{CXOHDFN}~J123644.0+621249; see \S\ref{xrayhdfn}) does show signs
of containing a weak AGN, several of these sources do not 
show optical evidence of harboring AGN in fairly high quality optical spectra.
  In the rest of the cases, however,
we lack the high-quality optical spectroscopy to determine if there is a weak
AGN present.  
  This group of objects is important regardless of the presence
of AGN, as they are either examples of very low-luminosity AGN or the bright
end of the ``normal" galaxy X-ray luminosity function.   

	G01 also mention this population (see their Figure~7).  Of the
7--8 optically bright, X-ray faint sources in the G01 sample, five have
soft-band X-ray fluxes $\simgt 3 \times
10^{-16}$~erg~cm$^{-2}$~s$^{-1}$.  These sources are identified by G01
as galaxies.  The nature of this optically bright, X-ray faint
population can only be studied with \chandra\ exposure times longer
than $100$~ks, necessary to reach the required soft-band X-ray flux
level ($\approx 3 \times 10^{-16}$~erg~cm$^{-2}$~s$^{-1}$).  Even
deeper observations may reveal that AGN cease to dominate the X-ray
number counts at extremely faint soft-band fluxes in a manner analogous
to the dominance of star-forming galaxies rather than AGN at very faint
radio flux densities (see \S\ref{microJy} for a discussion of $\mu$Jy
radio sources).  This is perhaps not surprising, as one would expect
less luminous, nearby X-ray sources (similar to M82, for example;
 e.g., Griffiths et~al. 2000)  to 
appear as X-ray surveys achieve greater sensitivities.  It had previously
been proposed
that starburst galaxies may contribute to the
X-ray background; some of these observed objects could very well be the 
early evidence of this  population
(e.g., Moran, Lehnert, \& Helfand 1999 and references therein).


\section{Chandra Sources in the \hbox{HDF-N} \label{xrayhdfn}}

\subsection{Newly Discovered X-ray Sources \label{newxrayhdfn}}	

With the addition of 58.2~ks of \chandra\ exposure to the 164.4~ks discussed
by H00, we have detected two new sources in the \hbox{HDF-N}: 
\hbox{CXOHDFN}~J123639.6+621230 and \hbox{CXOHDFN}~J123644.0+621249 
(see Figure~\ref{HDFzoom}). A total of eight X-ray sources have 
now been discovered in the \hbox{HDF-N}; all six \hbox{HDF-N} sources of H00 are 
still detected in the 221.9~ks observation.  At present, all eight of the 
X-ray sources found in the \hbox{HDF-N} have likely optical counterparts.

CXOHDFN~J123639.6+621230 is identified with an $R=24.3$
broad-line AGN classified by C00; the X-ray detection supports
the AGN nature of this source. The W96 match, 
4-852.12, is $0.41^{\prime \prime}$ from the X-ray source position 
(see Figure~\ref{HDFzoom}). C00 report $z=0.943$ from the detection
of a single broad emission line which they take to be Mg~II at 
2800~\AA. Fern\'andez-Soto et~al. (2001) note that the expected colors of 
an AGN at $z=0.943$ do not match the photometric properties of this 
source in the near-infrared and that a much better match is found at
$z=3.475$, in which case the broad emission line is Ly$\alpha$. 
The C00 redshift has recently been revised to $z=3.479$ (J.G. Cohen 2000,
private communication).  The corresponding rest-frame 
0.5--8~keV luminosity is $1.4\times 10^{44}$~erg~s$^{-1}$.

CXOHDFN~J123644.0+621249 was detected with only five X-ray counts and 
only in the soft band. It is located $0.34^{\prime \prime}$ 
from an $R=21.3$ object with characteristics of an AGN, however, so we have 
fairly high confidence that the detection is real. This source is 
a 1.4~GHz and 8.4~GHz emitter with $z=0.557$ (R98; R00; C00). 
A Keck spectrum of the object  
is presented in Figure~8. The spectrum shows clear [\ion{O}{2}] 
emission and has weak H$\beta$ and perhaps [\ion{O}{3}] lines.  
The H$\beta$ line is resolved with a FWHM of $\approx 600$~km~s$^{-1}$, 
suggesting the presence of an AGN. This source is also discussed by R98, 
who report [\ion{O}{2}], [\ion{O}{3}] and H$\beta$ lines with 
P~Cygni profiles; we do not see such profiles in the current data (see
Figure~8). 
The 8.5~GHz radio luminosity is $10^{22.5}$~W~Hz$^{-1}$, and R98 
estimate a star-formation rate of 30~$M_{\odot}$~yr$^{-1}$.   
The rest-frame soft-band luminosity of this source is 
$1.3\times 10^{42}$~erg~s$^{-1}$, which would be high for
a pure starburst galaxy. The broad H$\beta$ line,
combined with the large X-ray luminosity, suggest the presence 
of an AGN.

\subsection{New Hard X-ray Detections of Elliptical Galaxies\label{HDFNellipticals}}

All three of the \hbox{HDF-N} elliptical galaxies reported by H00 
were undetected in the hard band with the 164.4~ks of exposure time
used by H00. With the addition of 57.5~ks of exposure, two of these sources,  
\hbox{CXOHDFN}~J123655.4+621311 and \hbox{CXOHDFN}~J123656.9+621301, are now detected 
in the hard band with $5.5\pm 2.5$ and $7.4\pm 2.8$ hard-band 
counts, respectively.  Their hard-band luminosities are 
$1.9\times 10^{42}$~erg~s$^{-1}$ and $4.0\times 10^{41}$~erg~s$^{-1}$, 
respectively. The luminous hard components revealed from these sources 
strengthen the case that they harbor AGN; an AGN was already suspected
in \hbox{CXOHDFN}~J123655.4+621311 given its flat radio spectrum (R98; H00). 
The soft-band luminosities of \hbox{CXOHDFN}~J123655.4+621311 and 
\hbox{CXOHDFN}~J123656.9+621301 are somewhat higher than expected for normal 
ellipticals but are not outside the range seen for these objects. 

Fern\'andez-Soto et~al. (2001) have argued that \hbox{CXOHDFN}~J123656.9+621301
is at a higher redshift than the value of $z=0.474$ stated by C00, 
giving $z=1.270$. They suggest that the spectroscopic redshift is 
in error due to confusion with a nearby source, even though C00
list it as being secure and based on multiple spectral features. 
We cannot resolve this issue with the X-ray data, although we note
that the X-ray detection may make an unusual intrinsic spectral
energy distribution for this object more plausible. If 
\hbox{CXOHDFN}~J123656.9+621301 is indeed at $z=1.270$ rather than $z=0.474$, 
it would have a full-band X-ray luminosity of 
$5.2\times 10^{42}$~erg~s$^{-1}$ rather than 
$6.5\times 10^{41}$~erg~s$^{-1}$ 
($3.2\times 10^{41}$~erg~s$^{-1}$ was quoted in H00 where 
$q_{0}=0.5$ and $\Gamma=2$ were used).
This would make the case for the presence of an AGN even stronger.

\section{Chandra Constraints on Source Populations at Other Wavelengths\label{otherlambda}}

\subsection{The \boldmath$\mu$Jy Radio Sources\label{microJy}}

While surveys of radio sources are dominated by powerful AGN at flux 
densities above a few mJy, a large population of lower luminosity galaxies 
becomes visible at fainter limits (e.g., Windhorst et~al. 1985). About
60\% of these $\mu$Jy radio sources are thought to be starburst-type
systems associated with disk galaxies at moderate redshift
(e.g., Muxlow et~al. 1999; R00; and references therein); and 
approximately 30\% of $\mu$Jy radio sources are thought to be low-luminosity 
AGN often identified with field ellipticals. The remaining 10\% are
associated with red, optically faint ($I>25$; $I-K>3$) systems 
which may be luminous dust-shrouded starbursts at $z\approx$~1--3, 
luminous obscured AGN, or extreme redshift ($z>6$) AGN
(e.g., Richards et~al. 1999).

Extremely deep radio coverage of the \hbox{HDF-N} and its environs has 
been obtained using the VLA (R98; R00), 
MERLIN (Muxlow et~al. 1999; Richards 1999), and the WSRT 
(Garrett et~al. 2000). We have used the \chandra\ data to set 
constraints on the X-ray emission properties of the sources detected 
in these surveys. Most of the radio surveys above have been 
performed at 1.4~GHz, so we shall first focus on the X-ray properties
of sources selected at this frequency. In \S7.1.4  we shall
return to discuss radio sources detected at 8.5~GHz that are not 
detected at 1.4~GHz (R98).

The $\mu$Jy radio sources in the \hbox{HDF-N} and its environs may be 
classified based upon their radio spectral indices and radio 
morphologies. Starburst-type objects have steep radio spectra 
($S_{\nu} \propto \nu^{-\alpha}$, $\alpha>0.5$) and appear extended on sub-galactic scales. 
AGN candidates generally have flat or inverted radio spectra ($\alpha<0.5$) 
together with a compact core and one or two-sided extended radio structure. 
In some cases optical morphology and infrared properties can also be used 
in the classification process (see Richards 1999).
Muxlow et~al. (1999) and Richards (1999) have reported 31 
starburst-type 1.4~GHz sources and six AGN-candidate 1.4~GHz sources 
in the Caltech area (see Chapter~4 of Richards 1999). They also have 
found 30 1.4~GHz sources in the Caltech area that they cannot easily 
classify as either of these types (hereafter these are referred to 
as objects of unknown type). Their classifications were intended to identify 
the origin of the bulk of the radio emission from a particular galaxy; 
they did not preclude starburst-type objects from containing embedded 
AGN or AGN candidates from also having starburst activity. 
With \chandra\ we detect 11 of the starburst-type sources, two of the 
AGN candidates, and three of the unknown-type objects 
(see Table~\ref{radiotable} and Figure~\ref{sbagn}).  
It is worth noting that the X-ray detection fraction for the $\mu$Jy
sources of unknown type is significantly lower than for the 
starburst-type and AGN-candidate $\mu$Jy sources, but it is difficult
to examine this issue in detail due to the limited information on the
unknown-type objects. Below we will discuss the starburst-type and
AGN-candidate $\mu$Jy sources in further detail. 

\subsubsection{Radio Starburst-Type Systems\label{radiostarbursts}}

In Figure~\ref{sbagn} we plot the 1.4--8.5~GHz spectral index versus 
the 1.4~GHz flux density for the 1.4~GHz sources in the Caltech 
area; we have marked the sources detected by \chandra. Notably, 
within our limited statistics, the $\approx 35$\% \chandra\ 
detection fraction for the starburst-type radio sources is 
comparable to that for the radio AGN candidates (see \S\ref{radioagn}). 
The X-ray detection fractions are comparable in both the soft and the hard 
bands. 

The 11 \chandra-detected, starburst-type radio sources have typical 
redshifts of 0.3--1.0 and a range of X-ray band ratios. They 
are uniformly distributed among the population in terms of 1.4~GHz 
flux density and 1.4--8.5~GHz spectral index (see Figure~\ref{sbagn}). The high X-ray luminosities 
($\simgt 5\times 10^{42}$~erg~s$^{-1}$) and large band ratios of 
several of these systems strongly suggest that they contain 
previously unrecognized AGN. Others have X-ray 
luminosities that are comparable to, or only slightly higher than,
those of the most powerful starbursts in the local Universe 
(e.g., Moran, Lehnert, \& Helfand 1999).

Considering the starburst-type radio population as a whole, including 
the $\approx 65$\% of sources not detected in X-rays, the present \chandra\ 
data are still consistent with the idea that the majority of these 
objects are predominantly powered by star formation. 

\subsubsection{Radio AGN Candidates\label{radioagn}}

Two of the six radio AGN candidates seen at 1.4~GHz in the Caltech area are detected 
by \chandra: J123646.3+621404 and J123652.8+621444 (see Figure~\ref{sbagn}). The 
first is a well-known AGN in a spiral galaxy at $z=0.960$ that is located in 
the \hbox{HDF-N}; its X-ray detection was reported by H00. 
J123652.8+621444 is at $z=0.322$ and has an $R=19.5$ elliptical host. 
Its radio spectrum is inverted with $\alpha=-0.12$, and it shows
radio variability on a timescale of months. 

The non-detections of several radio AGN candidates by \chandra\ are notable 
given the fact that a few of these systems 
are at low redshift and are optically bright. For example, 
J123716.3+621512 ($z=0.232$; $I=19.8$; ${\cal R}=20.3$; $\alpha=0.41$)
is undetected; assuming a $\Gamma = 1.4$ power law (see \S \ref{xrpointsources}),
 the limit on its 0.5--8~keV emission is $<3.9\times 10^{40}$~erg~s$^{-1}$.
If this source is indeed an AGN, it must be either intrinsically
X-ray weak, perhaps due to low radiative efficiency accretion, or 
it must have heavy obscuration up to $\approx 10$~keV with little 
X-ray scattering. 

Another remarkable AGN in the Caltech area that is not detected in the \chandra\ 
data is the Fanaroff-Riley~I (FR~I)
Wide Angle Tail (WAT) radio source J123725.7+621128 (Muxlow et~al. 1999; 
Richards 1999). With a 1.4~GHz flux density of 6.0~mJy, this object is one 
of the brightest radio sources in our field (it is not shown in Figure~\ref{sbagn}
because it lies beyond the right-hand edge of the plot). J123725.7+621128 is 
located in the \hst\ flanking fields, and its host galaxy is a $K=18.8$, 
$I=22.9$, ${\cal R}=24.5$ elliptical of unknown redshift. Application of the $K$-$z$ 
relation (e.g., Eales et~al. 1997) to this galaxy suggests a redshift of 
$z=$~1--2. FR~I sources generally have low X-ray luminosities 
($10^{41}$--$10^{42}$~erg~s$^{-1}$ from 2--10~keV; e.g., 
Sambruna, Eracleous \& Mushotzky 1999), and an X-ray luminosity
somewhat below our detection limit would still be consistent with the
X-ray properties of some local FR~Is (e.g., the FR~I in Hydra~A; McNamara et~al. 2000).
WAT sources are often associated with 
rich groups or clusters; we discuss the nondetection of extended X-ray 
emission from a group or cluster in \S\ref{clusters}. 

Similarly, the 0.42 mJy FR~I in the \hbox{HDF-N}, J123644.3+621133 (R98), 
is not detected
even with 221.9~ks of exposure time, although there is a hint of a 
soft-band photon excess ($\approx 5$ counts) at its position 
(see \S3.2 of H00). 

\subsubsection{Optically Faint $\mu$Jy Sources\label{optfaintradio}}

The \hbox{HDF-N} itself contains three optically faint $\mu$Jy sources, as
defined above, at 1.4~GHz. As described by H00, one 
of these three (J123651.7+621221) is detected by \chandra\ and is a good 
candidate for a Type~2 QSO at $z\approx 2.7$. Encouraged by this result, we 
have searched for other optically faint $\mu$Jy radio sources with X-ray 
matches, but we find only one further X-ray match among the 17 sources 
listed in Richards et~al. (1999). The low matching fraction with X-ray 
sources is consistent with these objects being predominantly powered by 
starburst activity, although our constraints are difficult to quantify
as most of these objects do not have measured redshifts. 

The one further X-ray match is with J123711.9+621325 (see Figure~3 of 
Richards et~al. 1999) which has a 1.4~GHz flux density of 54~$\mu$Jy 
and a steep radio spectrum with $\alpha>1.16$. J123711.9+621325 
does not have a measured redshift at present. It presumably contains an AGN in 
addition to likely starburst activity, but the limited amount of 
information available for this source combined with its low X-ray 
flux (see Table~\ref{radiotable}) prohibit detailed interpretation.  

\subsubsection{Sources Detected at 8.5~GHz but not 1.4~GHz\label{radio8p5}}

In addition to the radio sources in the 
Caltech area that are detected at 1.4~GHz, there are 25 faint radio sources
that R98 detect at 8.5~GHz that are not also detected at 1.4~GHz. 
Eight of these sources are in their main catalog (i.e., their Table~3) and 17
are in their supplemental catalog (i.e., their Table~5). These sources,
of course, tend to have flatter radio spectra than those selected
at 1.4~GHz and thus might contain a higher fraction of AGN
(e.g., \S6 of R00). 

With \chandra\ we only detect two of these sources: 
J123655.4+621311 and J123658.8+621435.\footnote{We note that the
8.5~GHz source J123644.0+621249
is not detected by Richards et~al. (1999) at 1.4~GHz but is
detected by Garrett et~al. (2000) at this frequency.  For this
reason, we do not include it in the sample of X-ray sources detected
at 8.5~GHz but not 1.4~GHz.} 
J123655.4+621311 is in the \hbox{HDF-N} itself and is discussed above 
in \S\ref{HDFNellipticals} and in H00; it appears
to contain a fairly low-luminosity AGN. J123658.8+621435 is identified 
with a face-on spiral at $z=0.678$. Its radio spectrum is flat
($\alpha<0.44$), but we are not aware of any claimed AGN
in this object; C00 state that the spectrum
is intermediate between an emission-line and an absorption-line
spectrum.   The HET spectrum of this object is given in Figure~8, 
and more discussion is included in \S \ref{spectradetails}.

If we alternatively look at the entire 8.5~GHz source population
in the Caltech area, irrespective of 1.4~GHz detections, ten of 
the 48 sources at 8.5~GHz are detected by \chandra. 

\subsubsection{Far-Infrared Luminosities Estimated from 1.4~GHz Flux Densities}

We have used 1.4~GHz flux densities to estimate the bolometric 
far-infrared luminosities of the \chandra\ sources with redshifts (Figure~\ref{FIR}), 
following \S6.3 of B01. Two of the radio-detected sources in the
current sample appear to be ultraluminous far-infrared galaxies (ULIGs) 
while the rest have lower luminosities. These two ULIGs, 
\hbox{CXOHDFN}~J123646.3+621404 and \hbox{CXOHDFN}~J123651.7+621221, 
are both in the
\hbox{HDF-N} itself and are detailed in H00; the ULIG 
status of \hbox{CXOHDFN}~J123651.7+621221 should be treated with caution 
since its redshift is photometric. Figure~\ref{FIR}b does not reveal any 
strong difference between the far-infrared luminosities of \chandra\ 
sources selected in the soft and hard bands.

\subsection{The Submillimeter Sources \label{submmxray}}

In H00 we presented 164.4~ks \chandra\ constraints on the five submillimeter 
sources in the \hbox{HDF-N} itself (Hughes et~al. 1998) as well as on the five most 
significant ($>3\sigma$) submillimeter sources reported in the vicinity of the 
\hbox{HDF-N}
(BCR00). None of these sources was detected at high significance with 
164.4~ks of data, and this also holds with the 221.9~ks of data now available. 
With the new data the constraints on $\alpha_{\rm sx}$, the 850~$\mu$m 
to 2~keV spectral index, given in Table~1 of H00 are tightened by 
$+0.022$ on average. This tightening of constraints does not qualitatively 
change the basic conclusion reported by H00: these submillimeter sources appear to
be dominated by star formation or have Compton-thick nuclear obscuration and little
circumnuclear X-ray scattering.

We do note, however, that manual inspection of the full-band \chandra\
image suggests an X-ray photon excess coincident with the BCR00 
submillimeter source J123646.1+621448. We have run {\sc wavdetect} over
this area of the full-band image using a less conservative probability 
threshold of $5\times 10^{-6}$, and we detect a 6.9-count source coincident 
with J123646.1+621448. J123646.1+621448 is optically faint with $R=25.5$, 
and its 850~$\mu$m flux density is 10.7~mJy. A spectroscopic redshift
is not available for this source, but BCR00 estimate $z=$~1.6--3.1
with a millimetric redshift technique. If this redshift is correct, 
the implied X-ray luminosity of $\sim 10^{43}$~erg s$^{-1}$ 
suggests that a moderate-strength AGN is present in this source.

We have stacked \chandra\ images of nine of the submillimeter sources mentioned
above to derive the tightest possible constraints upon $\alpha_{\rm sx}$ 
for the average X-ray weak submillimeter source; we excluded
J123646.1+621448 (see the previous paragraph). Our effective 
exposure time is 1.86~Ms, derived from the stacked exposure map. 
No photon excesses
are apparent in the stacked soft-band or hard-band submillimeter
source images. Unfortunately,
the limited accuracy of the submillimeter source positions limits
the efficacy of the stacking technique since one must allow for
source-to-source offsets that are considerably larger than the \chandra\ PSF. 
We have used a $3^{\prime\prime}$-radius source cell for count 
extraction from the stacked images, and we are background limited in this source cell. Our
$3\sigma$ upper limits on the numbers of soft-band and hard-band
counts in the stacked images are 13.4 counts ($3.2 \times 10^{-17}$~erg~cm$^{-2}$~s$^{-1}$)
 and 17.2 counts ($1.9 \times 10^{-16}$~erg~cm$^{-2}$~s$^{-1}$), respectively. 
The corresponding soft-band $\alpha_{\rm sx}$ limit is 
$\alpha_{\rm sx}>1.45$, and the hard-band limit is 
$\alpha_{\rm sx}>1.33$ (compare with Figure~2 of H00). 
These $\alpha_{\rm sx}$ limits have been computed with 
the same X-ray spectral shape assumptions used by H00, and we 
have used the weighted mean 850~$\mu$m flux density of 3.92~mJy. 

H00 also noted \chandra\ detections of two of the $2.8\sigma$ submillimeter
sources reported by BCR00: J123616.1+621513 and J123629.1+621045. The X-ray
properties of these sources, derived from 221.9~ks of \chandra\ data, are
reported in Table~\ref{xraydata}. 

More detailed analysis and interpretation of the submillimeter sources 
in the \hbox{HDF-N} and its vicinity will be presented by Barger et~al., in 
preparation.

\subsection{The \iso-CAM Sources\label{isoxray}}

Infrared surveys, like X-ray surveys, are a powerful probe of the obscured 
Universe. The {\it Infrared Space Observatory\/} (\iso) has performed a deep 
survey of the \hbox{HDF-N} and its environs with the infrared imaging instrument
\iso-CAM; at present this is the deepest survey ever made at 15$\mu$m and is 
also very deep at 6.75$\mu$m (see A99 and 
references therein). The \iso-CAM survey covers all of the \hbox{HDF-N} as well as 
some of its flanking fields, although the total area covered is still 
only about a quarter of the Caltech area. The sensitivity of the \iso-CAM 
survey was generally highest for the \hbox{HDF-N} itself and dropped off outside 
it (see \S6 of A99 for details). A99 have presented the most 
recent analysis of the \iso-CAM survey data, finding 49 ``main" 
sources and 51 ``supplementary'' ones; Aussel (1999) provides an update 
on the supplementary sources and states that the vast majority with
flux densities above 100~$\mu$Jy are real. We have matched the \chandra\
sources with the \iso-CAM sources of A99, adopting
$3.5^{\prime\prime}$-radius error circles for the \iso-CAM sources
following Aussel (1999). In total, we find seven matches with the main 
\iso-CAM sources and four with the supplementary \iso-CAM sources. 

The \hbox{HDF-N} itself contains eight \chandra\ sources, 18 main \iso-CAM sources, 
and 15 supplementary \iso-CAM sources. A remarkable finding is that six
of the eight \chandra\ sources in the \hbox{HDF-N} itself have matches with
\iso-CAM sources (see Table~\ref{ISOtable} and Figure~\ref{ISOCAM_figure}). 
Three of these matches are with main 
\iso-CAM sources, and three are with supplementary \iso-CAM sources. This high 
matching fraction bodes well for future sensitive infrared observations 
(e.g., with the {\it Space Infrared Telescope Facility\/}) 
of sources found in deep X-ray surveys,
as well as for large-area infrared/X-ray surveys such as the
European Large-Area \iso\ Survey (ELAIS). Furthermore, we note that five of 
the six \chandra/\iso-CAM matches in the \hbox{HDF-N} are also detected at 8.5~GHz 
by R98. Most \iso-CAM sources, however, do not have 
\chandra\ source matches; this is as expected given that the surface 
density of \iso-CAM sources on the sky is much higher than that 
of \chandra\ sources. 
The nature of the \hbox{HDF-N} \chandra\ sources with \iso-CAM matches is 
discussed in detail in \S\ref{xrayhdfn} and in H00; at 
least four of them show fairly secure signs of hosting AGN. These AGN 
presumably heat dust in their environments to produce the 
emission seen by \iso-CAM (see Wilman, Fabian, \& Gandhi 2000
and Alexander et~al. 2001).  

Outside the \hbox{HDF-N}, where the sensitivity of the \iso-CAM survey is 
lower, the fraction of \chandra\ sources with \iso-CAM matches is lower but is
still $\approx 25$\% (see Table~\ref{ISOtable} and Figure~\ref{ISOCAM_figure}). 
Here we have four \chandra\ source matches 
with main \iso-CAM sources and one \chandra\ source match with a supplementary
\iso-CAM source. Again we note that a high fraction (3/5) of the  
\chandra/\iso-CAM matches are also detected at 8.5~GHz. A few of the 
\chandra/\iso-CAM matches outside the \hbox{HDF-N} are remarkable. 
For example, HDF\_PM3\_2 (J123634.4+621212; see A99) is associated with a 
$z=0.458$ (C00), $R=19.5$ disk galaxy showing evidence for a double nucleus 
which may indicate a recent merger. R98 suggest 
that this galaxy has a star formation rate of 
$\approx 95$~$M_\odot$~yr$^{-1}$, about an order of magnitude larger 
than that of M82. \chandra\ has only detected J123634.4+621212 in the
soft-band, consistent with a starburst nature, and its X-ray luminosity 
in this band of $\approx 8\times 10^{40}$~erg~s$^{-1}$ is also about
an order of magnitude larger than that of M82.  
HDF\_PM3\_6 (J123636.5+621348) is associated with a $z=0.960$, $R=21.5$ 
AGN discussed by Phillips et~al. (1997); intervening absorption lines 
are seen towards this AGN at $z=0.846$. 

Overall, the \chandra/\iso-CAM matches span a wide range of full-band X-ray 
luminosity from $\approx 10^{40}$--$10^{43}$~erg~s$^{-1}$. AGN comprise
the more luminous X-ray sources and starbursts the less luminous
ones, as expected. Our \chandra/\iso-CAM matches also have a 
range of X-ray band ratios, but as a population they do not 
obviously stand out from the other X-ray sources. 

\subsection{Very Red Objects\label{VROxray}}

C00 and Hogg00 have identified and discussed 33
very red objects (VROs), defined as having ${\cal R}-K_{\rm s}>5.0$
(down to ${\cal R} =25.5$ and $K_{\rm s}=20$). These were found in a region slightly
smaller than the Caltech area (see Figure~2 of Hogg00), due
to the fact that their $K_{\rm s}$ coverage did not encompass the entire Caltech
area. The nature of most of these VROs is uncertain due to the difficulty
of spectroscopic follow up as most have ${\cal R} >23.5$. The majority
of VROs have been proposed to be old galaxies at $z\sim 1$,
in which case they can sometimes point to high-redshift clusters
(e.g., Stanford et~al. 1997;  Hasinger et~al. 1998; Cimatti et~al. 1999; and
references therein).
Some, however, may be distant AGN and starbursts that are shrouded in dust
(e.g., Newsam et~al. 1997; Richards et~al. 1999; H00; Lehmann et~al. 2000;
and references therein). Sensitive X-ray observations of VROs, particularly
in the hard-band, are a powerful way to assess their AGN content.

\subsubsection{Individual \chandra\ Detections of VROs}

We detect four of the 33 VROs in X-rays (see Table~\ref{VROtable}). The small \chandra\ detection fraction 
suggests a relatively small AGN content in the optically selected VRO population, 
although AGN suffering from heavy obscuration 
($N_{\rm H}\simgt 5\times 10^{23}$~cm$^{-2}$) might escape detection in even 
these data. The four VRO detections are among the hardest \chandra\ sources (see Table~\ref{VROtable} and 
Figure~\ref{bandratios}); none is formally detected in the soft-band, although J123629.1+621045 
has a hint of a soft-band photon excess at its position. The band ratios of 
the X-ray detected VROs are larger than expected for typical elliptical galaxies 
at moderate redshift; for $z\approx 1$, the hard-band counts arise from above 
$\approx 4$~keV in the rest frame, whereas typical ellipticals produce 
the bulk of their X-ray emission below 4~keV.   

Only two of the X-ray detected VROs, the Hogg00 sources 
J123629.1+621045 and J123715.9+621213, are bright enough for optical
spectroscopy (see Table~\ref{multiwavtable}). Both are classified as having 
``emission-line" spectra by C00, but these authors did 
not note any evidence for AGN. Analyses of the Keck/LRIS spectra
in Figure~8 show the presence of one significant narrow emission
line in each spectrum that we interpret as [\ion{O}{2}]~3727~\AA;
our redshifts are identical to those of C00. 
There is no evidence for the presence of a [\ion{Ne}{5}] line in either
spectrum (see \S4.2 of Schmidt et~al. 1998), but this is not a 
stringent constraint given the relatively low signal-to-noise of the data. 
 The spectrum of J123715.9+621213 contains a weak feature that may correspond to 
[\ion{Ne}{3}]~3869~\AA. J123629.1+621045 is also notable as an 
$81.4\pm 8.7$~$\mu$Jy source at 1.4~GHz and a probable 
submillimeter source (R00; see \S\ref{submmxray}). The large 
band ratios of these two VROs, combined with their observed X-ray 
luminosities, suggest that they are moderately luminous obscured AGN. 
The same conclusion is also likely to hold for the two X-ray detected 
VROs without redshift information, although clearly follow-up work is 
required for these. 

While the four X-ray detected VROs are hard, we note that objects are detected 
in the X-ray with ${\cal R}-K_{\rm s}=$~4--5 and a range of X-ray hardness; 
there are ten such objects in the soft band and eight 
in the hard band (compare with Newsam et al. 1997).  
Hogg00 report only one extremely red object in the Caltech area with ${\cal R}-K_{\rm s}>6$
(J123711.1+620933); this object is not detected in our source searching,
and manual inspection does not show any hint of a photon excess at its
position. 

\subsubsection{Average X-ray Properties of the X-ray Weak VROs}

We have stacked \chandra\ images of the 29 X-ray undetected VROs in
the soft and hard bands to place further constraints on their
X-ray emission (see Figure~\ref{VROpostagestamp}). Before 
stacking we have manually verified that none of these 29 sources 
shows apparent emission just below the detection threshold,
and we have also manually verified that flaring pixels 
(see \S\ref{xrpointsources}) are not corrupting the stacking procedure.
The stacked images have effective exposure times of 6.02~Ms, and we
have searched these with {\sc wavdetect} using a probability
threshold of $1\times 10^{-7}$. A 31.3-count source is detected
in the soft-band stacked image; the resulting
average soft-band flux of an X-ray weak VRO is
$\approx 1.9\times 10^{-17}$~erg~cm$^{-2}$~s$^{-1}$.
At $z\approx 1$ and $z\approx 2$, the corresponding
rest-frame soft-band X-ray luminosities are
$\approx 6\times 10^{40}$~erg~s$^{-1}$ and
$\approx 3\times 10^{41}$~erg~s$^{-1}$, respectively.
No sources are detected in the hard-band stacked image;
the $3\sigma$ upper limit for the hard band is 16.4 counts, corresponding
to an average upper limit on the flux of $5.7 \times 10^{-17}$~erg~cm$^{-2}$~s$^{-1}$.
The band ratio for the average X-ray weak VRO is $<0.52$; for a 
power-law model with the Galactic column density, this corresponds
to $\Gamma>1.4$. 

Our detection of the average X-ray weak VRO in the soft band but 
not the hard band stands in contrast to the results for the 4
individual VRO detections described above; the latter were some
of the hardest X-ray sources and were not detected in the soft
band. The band-ratio upper limit for the average X-ray weak VRO
is highly disjoint from the lower limits of the individually detected 
VROs (see Table~\ref{VROtable}). 
There thus appear to be at least two separate classes of VROs in terms of 
their observed X-ray emission properties. The softer X-ray spectral shapes 
and likely X-ray luminosities of the X-ray weak VROs are consistent
with the emission properties of moderate redshift ($z\sim 1$) ellipticals 
where the X-rays arise from a hot interstellar medium. The X-ray 
data therefore provide supportive evidence for an elliptical galaxy 
interpretation of many of the VROs, as has been argued based on observations 
at other wavelengths (see \S\ref{VROxray}). 
However, given that the constraints on the X-ray spectral shape of the
average X-ray weak VRO are not tight ($\Gamma>1.4$; see above), it is
still possible that at least some of them are obscured moderate-luminosity 
AGN at fairly high redshift ($z\simgt$~2--3); their apparently softer X-ray 
spectra would then simply arise as a result of the photoelectric absorption 
cutoff being moved to lower energies by the redshift. At $z\simgt 3$, for 
example, X-ray emission from AGN having $N_{\rm H}\simlt 10^{23}$~cm$^{-2}$ 
would be redshifted into the observed soft band. Such AGN would not 
appear to be particularly hard sources, especially if there were also 
a contribution from scattered, unabsorbed flux. If there is a significant 
population of high-redshift, obscured AGN among the VROs, it is somewhat 
surprising that the band ratios of the average X-ray weak VRO and the 
individually detected VROs are so disjoint; we would have perhaps 
expected to detect a few VROs with intermediate band ratios, but this 
could simply be due to limited source statistics. 

The properties of the three \chandra-detected elliptical galaxies in the
\hbox{HDF-N} itself can be compared with the X-ray constraints we place on
the rest of the VRO population.   The three elliptical galaxies in the \hbox{HDF-N}
are \hbox{CXOHDFN}~J123648.0+621309 (${\cal R}-K_{\rm s}=3.23$; $z=0.475$),
\hbox{CXOHDFN}~J123655.4+621311 (${\cal R}-K_{\rm s}=4.66$; $z=0.968$) and
\hbox{CXOHDFN}~J123656.9+621301 (${\cal R}-K_{\rm s}=4.41$; $z=0.474$);
they would be very red at $z > 1$.
Their X-ray band ratios
of $< 1.03$, $0.26^{+0.22}_{-0.14}$ and $0.59^{+0.22}_{-0.14}$ are
much softer than those of the X-ray detected VROs (see Table~\ref{VROtable})
but consistent with that of the average X-ray weak VRO.

\subsection{Optical Active Galaxy Candidates \label{AGN}}

	The \chandra\ sources were cross-correlated with five lists of
 AGN candidates identified at optical wavelengths (see
Table~\ref{catalogs}), and only three objects were found with matches.
These sources are 
a $z=2.580$
QSO (\hbox{CXOHDFN}~J123622.9+621527; see \S A.2 and Liu et~al. 1999),
the $z=0.96$ AGN in the \hbox{HDF-N} itself
(\hbox{CXOHDFN}~J123646.3+621404; see H00) and
a $z=1.020$ QSO (\hbox{CXOHDFN}~J123706.8+621702; see \S A.3 and Vanden
Berk et~al. 2000).

	The AGN candidates identified at optical wavelengths are generally 
 optically faint ($R\approx27$).  We have plotted them along with X-ray sources
from this paper and others in Figure~\ref{akiyama_fancy}.  Most of the
candidates with $R\simgt 25$ are still found to be consistent with the
range of $\log{({{f_{\rm {\rm X}}}\over{f_{\rm R}}})}$  typical of X-ray detected
AGN (Schmidt et~al. 1998; Akiyama et~al. 2000), so we cannot rule out
the presence of AGN in these optically faint candidates.

We can place somewhat stronger constraints statistically by stacking
the images of the X-ray undetected AGN candidates in the same manner as
for the submillimeter sources and the VROs (see \S\ref{submmxray} and
\S\ref{VROxray}).  The results of stacking the X-ray images of the 
X-ray undetected AGN candidates are
given in Table~\ref{AGNcandidatetable}. There were no sources detected
in any of the relevant X-ray bands when {\sc wavdetect} was run on
the stacked images, so we place upper limits on the average X-ray emission from
these sources. The typical flux upper limit is 
$\approx 2\times 10^{-17}$~erg~cm$^{-2}$~s$^{-1}$ in the soft band and
$\approx 1\times 10^{-16}$~erg~cm$^{-2}$~s$^{-1}$ in the hard band.
We did not use the stacking technique for the X-ray undetected Vanden Berk
et~al. (2000) objects as there are only two such sources in the Caltech
area, and the stacking technique provides large improvements over the raw
limits only when the effective integration time can be increased
significantly.
 For the Jarvis \& MacAlpine (1998) sample, all objects
are claimed to be at $z > 3.5$ so the resulting average rest-frame X-ray
luminosity in the 0.5--8~keV band is $\simlt 3\times
10^{44}$~erg~s$^{-1}$.   For the Sarajedini et~al. (2000) sample, the
most distant object has a photometric redshift of $z\approx 1.9$, and
the mean redshift is $z\approx 1.1$.   The average X-ray luminosity limits in
the full-band for these sources are $\simlt 1\times
10^{44}$~erg~s$^{-1}$ for $z\approx 1.9$ and 
$\simlt 3\times 10^{43}$~erg~s$^{-1}$ for $z\approx 1.1$.
 The Conti et~al. (1999) objects are split into
high-redshift ($z > 3$) and low-redshift ($z < 3$) classes, so the
constraints on these objects can be compared to the Jarvis \& MacAlpine
(1998) limits in the high-redshift case and to the Sarajedini et~al.
(2000) limits in the low-redshift case.   All of these authors have pointed
out that the typical range of absolute magnitudes for these AGN candidates 
($M_{\rm V} \approx -18$) are consistent with low-luminosity 
Seyfert 1 galaxies or Seyfert 2 galaxies rather than with luminous QSOs, so the X-ray
limits are still consistent with their findings.   Extremely deep ($\simgt 1$~Ms) X-ray
observations will be necessary to test these AGN search techniques effectively.

The two X-ray undetected Vanden Berk et~al. (2000) objects are  
notable since they are bright in the optical; ultraviolet excess QSO
candidate J123628.1+621433 has $R=21.3$ and high-redshift QSO
candidate J123723.7+621544 has $R=19.3$ (see Vanden Berk et~al. 2000
for a discussion of these object classifications).   We have 
placed upper limits on their X-ray emission;
 the limit is $\approx
2.4\times 10^{-16}$~erg~cm$^{-2}$~s$^{-1}$ in the full band.
J123628.1+621433 has a measured redshift of $z=0.243$ (C00); the
corresponding X-ray luminosity upper limit is $3.4\times 10^{40}$
erg~s$^{-1}$.   According to Vanden Berk et~al. (2000),
J123723.7+621544 is at $z > 3$; the corresponding X-ray luminosity
upper limit at $z=3$ is $9\times 10^{42}$~erg~s$^{-1}$.  We find that
J123628.1+621433 most likely does not contain a luminous AGN. While
J123723.7+621544 may harbor a moderately luminous AGN, it most
likely does not contain a luminous quasar even if it is at very high
redshift.

Two other notable AGN that have not been detected in the X-ray, the
WAT radio source J123725.7+621128 (Muxlow et~al. 1999;
R00) and the FR~I in the \hbox{HDF-N}, J123644.3+621133, are 
discussed in \S\ref{radioagn}.

\subsection{Clusters and Groups \label{clusters}}
As described in \S\ref{xrextended}, we have not found any highly significant extended 
X-ray sources in the Caltech area. We are not aware of any firmly 
established clusters or groups in this area found from observations 
at other wavelengths. 

WAT radio sources, such as J123725.7+621128  
(see \S\ref{radioagn}), are often found in clusters. We have manually 
inspected the area around J123725.7+621128, and we find no hint
of any extended X-ray emission centered near this source. 
Given the depth of the current observation, we can rule out moderately 
luminous cluster X-ray emission near J123725.7+621128 to 
$z\approx 1.5$. 

We have also manually searched for any diffuse X-ray emission near 
the possible $z\approx 0.85$ cluster discussed by 
Dawson et~al. (2001). Again, we see no hint of diffuse
X-ray emission from this object in the current data. 

Extrapolating the known number counts for clusters 
(e.g., Rosati et~al. 1998), it is not surprising that none has 
been detected by \chandra\ in the Caltech area; only 0--2 
are expected.  
Similarly, comparing the solid angle and depth of the current survey to 
those of deep \rosat\ surveys, it is plausible (although mildly
surprising) that no extended X-ray sources have been detected. 
Scaling from the results of Schmidt et~al. (1998), for example, 
we would have expected $\sim 3$ extended X-ray sources in the Caltech area. 

\subsection{Galactic Objects \label{galactic}}

There are two X-ray detections of M~stars in the Caltech area (see Figure~8
for their optical spectra): \hbox{CXOHDFN}~J123625.4+621404 and \hbox{CXOHDFN}~J123725.7+621648. 
The TiO absorption features characteristic of an M4 star are apparent in both
spectra, and the M star classification is clearly consistent with their 
X-ray to optical flux ratios, 
$\log{({{f_{\rm X}}\over{f_{\rm R}}})}\approx -2$ in both cases.   
Curiously, H$\alpha$ emission is
not detected from either; this emission is generally associated with plages and/or 
flares on magnetically active stars.   We estimate their distances, based on their
optical magnitudes, to be $\approx 240$ and $\approx 520$ pc.  The corresponding 
 soft-band X-ray luminosities are $\approx$ 10$^{28}$~erg s$^{-1}$ in both cases. 
The second star is one of the most distant X-ray detected M~stars ever reported.   
More information on the properties of the \chandra-detected
stars in our X-ray survey area will be presented in Feigelson et~al., in preparation.

Ibata et~al. (1999) have put forward five objects that appear to show proper 
motion in \hst\ images of the \hbox{HDF-N}. These objects are very faint ($V\sim 28$) 
and blue, but their nature is uncertain; they may be old white dwarfs that
represent a population comprising much of the missing Galactic dynamical
mass. None of these five objects is 
detected in the source searching described in \S~\ref{xrpointsources}, but, given their potential 
importance, we have manually inspected their positions to search for any 
hints of photon excesses. We find none. 

\section{Conclusions}

\subsection{Overview}

This work represents the most sensitive study to date of the 
extragalactic X-ray background source population. We reach 
limiting soft (hard) flux levels of 
$\approx 1.3\times 10^{-16}$~erg~cm$^{-2}$~s$^{-1}$  
($\approx~6.5\times 10^{-16}$~erg~cm$^{-2}$~s$^{-1}$) 
for individual point sources, and 
$\approx 1.1\times 10^{-17}$~erg~cm$^{-2}$~s$^{-1}$ 
($\approx 5.7\times 10^{-17}$~erg~cm$^{-2}$~s$^{-1}$) 
for summed classes of objects (e.g., VROs).  
The sources in this study comprise $>90$\% of the 0.5--2~keV
X-ray background and $>80$\% of the 2--8~keV background
(Paper III; after adding the fraction of the background already
resolved in wide-field surveys).
Further observations of the \hbox{HDF-N} area, which will
bring the total exposure time to $\approx1$~Ms, have been approved 
for the coming year.

For the 39\% of sources with redshift measurements, we find 
redshifts ranging from 0.1--3.5 and 0.5--8~keV luminosities 
ranging from $10^{40}$--$10^{45}$~erg~s$^{-1}$. 
These spectroscopically identified sources 
generally have luminosities well below those of 
quasars; they resemble weak Seyferts (with both Type~1 
and Type~2 spectra), starburst galaxies, and elliptical galaxies. 
As with the $\mu$Jy radio sources and ultraluminous far-infrared 
galaxies studied over the past two decades, it is often unclear 
whether the faintest X-ray sources are powered by accretion 
onto a massive black hole or by powerful bursts of star 
formation. We detect some definite members of both classes, 
but many sources could be either AGN, starbursts or a combination 
of both. At 0.5--2~keV fluxes below 
$\approx 3\times 10^{-16}$~erg~cm$^{-2}$~s$^{-1}$, we find
that a significant fraction of the new sources are associated 
with low-redshift, optically bright ($R\simlt 21$) galaxies.
Optically faint ($R>26.5$) X-ray 
sources that might represent a new class of astronomical object 
(e.g., extremely high redshift quasars) comprise $<11$\% of the  
current sample and $<2$\% of the sources with 0.5--8~keV fluxes 
$>7\times 10^{-16}$~erg~cm$^{-2}$~s$^{-1}$. 

\subsection{Specific Findings}

\begin{itemize}

	\item{Eighty-two X-ray sources have been discovered in the Caltech area down to  
limiting soft (hard) flux levels of $\approx 1.3\times 10^{-16}$~erg~cm$^{-2}$~s$^{-1}$ 
($\approx 6.5\times 10^{-16}$~erg~cm$^{-2}$~s$^{-1}$).  The sources become
X-ray harder at low soft-band count rates, consistent with the hardening seen in
the \chandra\ Deep Field South survey of G01. The 62 soft-band sources
account for $>90$\% of the soft X-ray background, and the 45 hard-band sources 
account for $>80$\% of the hard X-ray background (after adding the fraction of 
the background already resolved in wide-field surveys). See \S3.3.}

	\item{Redshifts and identification information for 32 of the 82 
X-ray sources in the Caltech area are provided, including 
six redshifts previously unpublished for
this region.  Included in the sample with redshifts are 96\% 
of the sources with $R<23$. The median optical magnitude of our sample is $R\approx 23.5$, and
73 of the 82 sources have optical counterparts brighter than $R\approx 26.5$.
See \S \ref{multiwavmatch} and \S \ref{multiwavproperties}. }

	\item{At faint soft-band fluxes ($\simlt 3\times 10^{-16}$~erg~cm$^{-2}$~s$^{-1}$), 
we observe the emergence of a population of apparently ``normal"
galaxies, including moderately star-forming galaxies and elliptical galaxies.   This population
will only be apparent in \chandra\ surveys with $>100$~ks observation lengths.
All of these objects are optically bright
($R \simlt 21$) compared to AGN with the same soft-band X-ray fluxes, and they 
are found to have low
soft-band X-ray luminosities of $\simlt 3 \times 10^{41}$~erg~s$^{-1}$.
See \S \ref{LLXR}. }

	\item{We have not found any highly convincing Type 2 QSOs thus far.  Our best candidate,
\hbox{CXOHDFN}~J123651.7+621221, has only a moderate observed X-ray luminosity and an uncertain redshift. See \S \ref{optfaintradio}.}

	\item{Two new X-ray sources in the \hbox{HDF-N} are presented in this paper:   
\hbox{CXOHDFN}~J123639.6+621230 is identified with an $R=24.3$
broad-line AGN (C00), and \hbox{CXOHDFN}~J123644.0+621249 is identified with a star-forming galaxy (R98)
which may also contain an AGN.  The total number of X-ray sources found in the \hbox{HDF-N}
is now eight. See \S \ref{newxrayhdfn}.}

	\item{This study provides the tightest X-ray constraints to date on the $\mu$Jy radio source population.  With \chandra\ we detect 11 of the starburst-type sources, two of the 
AGN candidates, and three of the unknown-type objects.  The high X-ray
 luminosities 
($\simgt 5\times 10^{42}$~erg~s$^{-1}$) and large band ratios of 
several of the starburst-type systems strongly suggest that they contain 
previously unrecognized AGN.  In general, however, the \chandra\ 
data are consistent with the idea that the majority of these 
objects are predominantly powered by star formation.  We detect only two of the 17
optically faint $\mu$Jy sources, consistent with these objects being predominantly powered by 
starburst activity, although our constraints are difficult to quantify
as most of these objects do not have redshifts. See \S \ref{microJy}. } 

	\item{None of the ten reported submillimeter sources in the Caltech area (Hughes et~al. 1998;
BCR00) 
is detected with high significance in our 221.9~ks observation.  The tightening of the X-ray
constraints relative to H00 does not qualitatively 
change the basic conclusion reported by H00: these submillimeter sources appear to
be dominated by star formation or have Compton-thick nuclear obscuration and little
circumnuclear X-ray scattering. See \S \ref{submmxray}.  }

	\item{A remarkable finding is that six  
of the eight \chandra\ sources in the \hbox{HDF-N} itself have matches with
\iso-CAM sources.  This high 
matching fraction bodes well for future sensitive infrared observations 
(e.g., with the {\it Space Infrared Telescope Facility \/}) of sources found in deep X-ray surveys.  
See \S \ref{isoxray}. }

	\item{In our data, four of the 33 VROs with ${\cal R}-K_{\rm s}>5.0$ are 
detected in X-rays; they are X-ray hard and
moderately luminous ($L_{0.5-8} \simgt 10^{42}$~erg~s$^{-1}$).  
Stacking the \chandra\ images of the 29 X-ray undetected VROs in
the soft and hard bands allowed us to place further constraints on their
X-ray emission.  The stacked images have effective exposure times of 6.02~Ms,
and an ``average" soft-band source was detected with a soft-band flux  of 
$\approx 1.9\times 10^{-17}$~erg~cm$^{-2}$~s$^{-1}$.  Our detection of the 
average X-ray weak VRO in the soft band but not
in the hard band stands in contrast to our results for the 4
individual VRO detections. There thus appear 
to be at least two separate classes of VROs in terms of their observed 
X-ray emission properties.  See \S \ref{VROxray}.}

	\item{Only three of the $\simgt 30$ AGN candidates identified at optical wavelengths are detected.
In most cases, the X-ray constraints are not tight due to the extreme optical faintness ($R\approx 27$)
of these candidates.  See \S \ref{AGN}.}

	\item{Thus far we have not found any X-ray sources 
with highly signficant evidence for X-ray spatial extent.
We do not detect two objects which might be expected to be associated with  
clusters of galaxies. See \S \ref{xrextended} and \S \ref{clusters}. }

	\item{Two M~stars are found in the ACIS image, but none of the 
possible old white dwarfs discussed by Ibata et~al. (1999) are detected. See 
\S \ref{galactic}.}

\end{itemize}


\acknowledgments

We gratefully acknowledge the financial support of
NASA grant NAS~8-38252 (GPG, PI),
NASA GSRP grant NGT5-50247 (AEH),
NSF CAREER award AST-9983783 and the Alfred P. Sloan Foundation (WNB),  
NSF grant AST99-00703~(DPS), and NSF grants AST-0084816 and AST-0084847
(AJB, LLC). 
AJB acknowledges support through Hubble Fellowship grant 
HF-01117.01-A awarded by the Space Telescope Science Institute, which 
is operated by the Association of Universities for Research in Astronomy, 
Inc., for NASA under contract NAS 5-26555.  

We thank
D.~Alexander,
M.~Eracleous,
G.~Hasinger,
D.~Hogg, 
C.~Liu, 
T.~Miyaji,
E.~Richards and  
C.~Steidel for 
helpful discussions and kindly providing data.
This work would not have been possible without the enormous efforts of the entire
\chandra\ team. 
We gratefully acknowledge all the creators and operators of the
W. M. Keck Observatory.
The Hobby-Eberly Telescope (HET) is a joint project of the University of Texas
at Austin,
the Pennsylvania State University,  Stanford University,
Ludwig-Maximillians-Universit\"at M\"unchen, and Georg-August-Universit\"at
G\"ottingen.  The HET is named in honor of its principal benefactors,
William P. Hobby and Robert E. Eberly.  
The LRS is named for Mike Marcario of High Lonesome Optics who fabricated 
several optics for the instrument but died before its completion.

\newpage
\appendix

\section{Notes on Individual Sources \label{appendix}}

	All X-ray luminosities in this Appendix are quoted for the 0.5--8~keV
rest frame of the object unless otherwise indicated.   Band ratios and inferred
photon indices for all sources can be found in Table~\ref{xraydata}.

	Sources located in the \hbox{HDF-N} itself are discussed in H00 and in
 \S\ref{xrayhdfn} of this paper.

	The classification system used is described in \S \ref{zlumxray}.  

\subsection{Identifications from Optical Spectra in Figure~8 \label{spectradetails}}

       {\bf {CXOHDFN}~J123617.0+621010} ($z=0.845$, HL Galaxy):  Only 
one clear emission line is present in this optical spectrum;
this narrow line is found at 6879.3~\AA\  and is assumed to be
[\ion{O}{2}].  The \ion{Ca}{2} H and K absorption bands are also
present, and these match the $z=0.845$ required for the emission line
to be [\ion{O}{2}].  We do not detect a [Ne V] emission line, but this
object's appreciable X-ray luminosity of  $1.2 \times 10^{43}$~erg
s$^{-1}$ strongly suggests that it is an AGN.  The X-ray spectrum is
not particularly hard.  R00 identifies this
object in the radio as a radio starburst.  Due to the lack of a
[\ion{Ne}{5}] or [\ion{Ne}{3}] feature, we must identify this object as
a galaxy, but we do note that the signal-to-noise in the spectrum is
not of sufficient quality to rule out Ne emission.

	{\bf {CXOHDFN}~J123618.0+621635} ($z=0.679$, NL AGN): The two 
[\ion{O}{3}] emission lines are seen clearly, placing this object at
$z=0.679$.  There is a weak feature at 5755.7~\AA\ which we identify as
[\ion{Ne}{5}] in emission.  The presence of this higher ionization optical
emission line and the X-ray luminosity of $2.5 \times 10^{43}$~erg
s$^{-1}$ indicate that this object is a narrow-line AGN.   
This source is also detected at 1.4~GHz by R00.  

	{\bf {CXOHDFN}~J123618.5+621115} ($z=1.022$, BL AGN):  The
spectrum of this object has a broad line located at $\approx$
5668~\AA.   The line is assumed to be \ion{Mg}{2}; the corresponding
FWHM velocity is $\approx$ 4200~km~s$^{-1}$.
 A faint feature at 7539.2~\AA\ is identified as [\ion{O}{2}], placing
this object at $z=1.022$.  It is identified as a broad-line AGN.

	{\bf {CXOHDFN}~123625.4+621404} (Star):  This object is an 
M~star of type M4--M5.  Strong TiO absorption bands dominate the
optical spectrum. This source is also one of only
two X-ray/2MASS detections in the Caltech area.

	{\bf {CXOHDFN}~J123629.0+621046} ($z=1.103$, NL AGN):  This
object was identified by Hogg00 as a VRO (see \S\ref{VROxray}).
This spectrum shows an [\ion{O}{2}] emission line  and the \ion{Ca}{2}
H and K absorption features.  A weak feature at 7782.2~\AA\ is
interpreted as [Ne ~{\sc iii}] emission.  The redshift of this source
is $z=1.103$ (in agreement with the value published by C00); the X-ray
luminosity in the 2--8~keV band is $5.0 \times 10^{42}$~erg s$^{-1}$.
This X-ray source is quite hard, with a band ratio of $1.26\pm 0.31$.
The corresponding absorption column at $z=0$ for a $\Gamma = 2$ power
law would be $>~10^{22}$~cm$^{-2}$.  The absence of broad lines, the
[Ne ~{\sc iii}] emission feature, the hard X-ray spectrum, and the
moderate X-ray luminosity lead to the classification of this object as
an absorbed, narrow-line AGN.

	{\bf {CXOHDFN}~J123629.2+621613} ($z=0.848$, NL AGN):  This
spectrum shows an [\ion{O}{2}] emission line but few other
features.  The Ca~H absorption band appears near 7340~\AA, but Ca~K
overlaps with the atmospheric B band at this object's redshift of
0.848.   There may be a [\ion{Ne}{5}] feature at 6181.5~\AA, but since the
stronger [\ion{Ne}{5}] line at 3425~\AA\ is absent, this feature may
not be real.  This X-ray source is quite hard, with a band ratio
$>2.08\pm 0.47$; it has a hard-band luminosity of $4.4 \times
10^{42}$~erg s$^{-1}$ and is undetected in the soft-band.  The upper
limit on the soft-band luminosity is $4.1 \times 10^{41}$~erg s$^{-1}$,
and the lack of strong AGN features in the optical spectrum prevents us
from ruling out star formation as a strong contributor to this object's
emission.  We conclude, however, that the bulk of the emission in the
X-ray band is from an absorbed AGN due to the large amount
of hard X-ray emission from this source.

	{\bf {CXOHDFN}~J123633.4+621418} ($z = 3.408$, BL AGN): This
spectrum shows broad C~{\sc iv} and Ly$\alpha$ emission lines
and the characteristic continuum dip on the blue side of the Ly$\alpha$
line produced by the Ly$\alpha$ forest.  The measured redshift matches
that of C00 ($z=3.408$).  The identification is a broad-line AGN.

	{\bf {CXOHDFN}~J123636.7+621156} ($z=0.556$, NL AGN):  This
spectrum has an [\ion{O}{2}] emission line at 5798.9~\AA\ and an
H$\beta$ line at 7564.6~\AA.  The redshift of 0.556 agrees with the
value found by Barger et~al. (1999) and C00.  The H$\beta$ line is
moderately broadened (FWHM $\approx 1100$~km~s$^{-1}$).  Absorption
features of \ion{Ca}{2} H and K and CH (rest frame 4280--4300~\AA) are
present.  This object's X-ray spectrum is fairly
soft, and it has an X-ray luminosity of $2.1\times 10^{42}$~erg
s$^{-1}$.  This object is not detected at either 8.5~GHz or 1.4~GHz.
The lack of evidence for X-ray absorption, despite the moderate X-ray
luminosity, indicates that star formation may be a significant
contributor to this object's emission, but the width of the H$\beta$
line indicates a contribution from an AGN.  We classify this object as
a narrow-line AGN.

	{\bf {CXOHDFN}~J123644.0+621249} ($z=0.555$, NL AGN):  This
object is in the \hbox{HDF-N} and is discussed in
\S\ref{newxrayhdfn}.   It has a somewhat broadened H$\beta$ line and
weak [\ion{Ne}{3}] emission, so we classify it as a narrow-line AGN.

       {\bf {CXOHDFN}~J123655.4+621311} ($z=0.968$, HL Galaxy): This source is
an elliptical galaxy in the \hbox{HDF-N} (see H00 and
\S\ref{HDFNellipticals} of this paper)  the only clearly visible feature
in its optical spectrum is an  [\ion{O}{2}] 3727~\AA\ emission line.  
Its flat radio spectrum indicates the possible presence of an AGN.  The redshift
we find ($z=0.955$) is similar to the $z=0.968$ value of C00.

	{\bf {CXOHDFN}~J123658.8+621435} ($z=0.678$, BL AGN):  R98
detect this source at 8.5~GHz (but not at 1.4~GHz; see
\S\ref{radio8p5}) and identify it with a face-on spiral galaxy at a
redshift of 0.678 (C00).  This source has an X-ray luminosity of
$1.0\times 10^{43}$~erg s$^{-1}$.  The HET spectrum has a fairly low
signal-to-noise ratio, but we do detect the \ion{Ca}{2} break,
[\ion{O}{2}] emission, and a broad \ion{Mg}{2} emission line.  We
identify this object as a broad-line  AGN.

	{\bf {CXOHDFN}~J123702.7+621543} ($z=0.514$, HL Galaxy):  This spectrum shows
emission lines from [\ion{O}{3}] 5007, 4959, and 4363~\AA, [\ion{O}{2}]
3727~\AA\, and H$\beta$. It also shows the Balmer series (H$\delta$
through H12) and \ion{Ca}{2} H and K in absorption. The redshift we
measure, $z=0.514$, matches that of C00. The X-ray spectrum of this
object is fairly hard ($\Gamma \approx 1.2$), and the X-ray luminosity
in the 2--8~keV rest frame band is $1.2 \times 10^{43}$~erg s$^{-1}$.  
 The Balmer jump is indicative of a ``post-starburst'' galaxy (for a
 discussion of post-starburst galaxies see Fisher et~al. 1998 and
references therein), dominated
 by an A~star population, while the X-ray properties and the strong
 emission lines both indicate an AGN is also present.   This object
thus appears to be a ``post-starburst AGN" (e.g., Boroson \& Oke 1984;
Brotherton et~al. 1999; Dewangan et~al. 2000).  Despite the appreciable
X-ray luminosity, no broad lines are seen (the only likely strong broad
line in our observational window is H$\beta$, which is unresolved).

	{\bf {CXOHDFN}~J123704.6+621652} ($z=0.377$, NL AGN):  The main
features in this spectrum are [\ion{O}{2}] 3727~\AA, [\ion{Ne}{2}]
3867.8~\AA, and the Ca~{\sc ii} K absorption band.  There is a weak
H$\beta$ line, but the [\ion{O}{3}] 5007~\AA\
 and [Ne {\sc v}] lines are not detected.  The X-ray luminosity of this
object is $1.1 \times 10^{42}$~erg s$^{-1}$, and it has a moderately
soft X-ray spectrum.  The presence of H$\beta$ and lack of [\ion{O}{3}]
emission lines point toward a soft ionizing continuum but may be
inconsistent with the detection of [\ion{Ne}{3}].  We deduce that this
object is undergoing moderate star formation and possibly harbors a
low-luminosity AGN. 

	{\bf {CXOHDFN}~J123706.8+621702} ($z=1.020$, BL AGN):  The spectrum contains a strong
broad line at 5641~\AA; we identify the feature as \ion{Mg}{2} at $z=1.02$.
This broad-line AGN was also identified as a QSO
candidate by Vanden Berk et~al. (2000).

	{\bf {CXOHDFN}~J123715.9+621213} ($z=1.020$, {\rm HL Galaxy}):  There
 is a single narrow line at 7529~\AA\ that is probably
[\ion{O}{2}] at $z=1.020$. This source
 is not detected in the soft X-ray band.  Its hard-band X-ray
luminosity is $4.0 \times 10^{42}$~erg s$^{-1}$.  We suspect this
object harbors an AGN and note that the signal-to-noise of the
spectrum is too low to preclude the existence of [\ion{Ne}{5}] or
[\ion{Ne}{3}].   We classify it as a galaxy to follow the optical
classification system.

	{\bf {CXOHDFN}~123725.7+621648} (Star): This object is an 
M~star of type M4--M5.  Strong TiO absorption bands dominate the
optical spectrum.

\subsection{Identifications from the Literature\label{IDsdetails}}

       {\bf {CXOHDFN}~J123621.3+621109} ($z=1.014$, HL Galaxy):  This object
is identified as having a spectrum that is of ``intermediate'' type
(intermediate between being emission and absorption dominated) by C00.
It is fairly red, with ${\cal R}-K_{\rm s}=4.72$ (Hogg00). This source is
only detected in the hard band, with a hard X-ray luminosity of $4.0
\times 10^{42}$~erg~s$^{-1}$.  In the radio, it is matched to an
extended (3.2$^{\prime \prime}$) source at 1.4~GHz but lacks a detection
at 8.5~GHz (R98; R00), giving a lower limit on the radio
spectral index of $\alpha > 0.86$.  Since the hard X-ray luminosity is
rather large for a starburst, we suspect this object is an obscured
AGN.  The upper limit on this object's X-ray luminosity in the soft
band is $5.6 \times 10^{41}$~erg s$^{-1}$, so it is quite plausible
that star formation is making a significant contribution to the X-ray
emission.  We must classify it as a galaxy since we are do not have
access to its optical spectrum and are thus unable to ascertain if it
exhibits [\ion{Ne}{5}] or [\ion{Ne}{3}] emission.

	{\bf {CXOHDFN}~J123622.9+621527} ($z=2.580$, BL AGN): This is
a spectroscopically confirmed QSO found by Liu et~al. (1999).

        {\bf {CXOHDFN}~J123634.5+621213} ($z=0.458$, LL Galaxy):  This object
is detected by \iso-CAM and is identified as a starburst galaxy.   We 
designate it as a low-luminosity X-ray galaxy. See
\S\ref{isoxray} for further discussion of its properties.

        {\bf {CXOHDFN}~J123635.3+621110} ($z=0.410$, LL Galaxy): This object's
 spectrum is dominated by absorption according to C00 and appears to be of
elliptical/S0 morphology on the Hogg00 ${\cal R}$ image.  It is detected in the
 soft band with an X-ray luminosity of $6.3 \times 10^{40}$~erg
 s$^{-1}$.  Since the spectrum apparently lacks any strong emission
 features, this object would not be classified as either a starburst
galaxy or an AGN.   Its large optical luminosity, $\log(L_{\rm B}) =
44.8$, predicts a much higher X-ray luminosity if this object were
dominated by X-ray emission from the hot gas in an elliptical galaxy
(Eskridge, Fabbiano, \& Kim 1995).

	{\bf {CXOHDFN}~J123636.6+621347} ($z=0.960$, BL AGN): This
 object's spectrum is described in detail by Phillips et~al.
(1997).  The principal emission lines in their spectrum are broad
\ion{Mg}{2}, narrow [\ion{O}{2}], [\ion{Ne}{5}], [\ion{Ne}{3}],
H~$\delta$, and H~$\gamma$.  It also shows absorption from a nearby
disk galaxy at $z=0.846$ in \ion{Mg}{2} and Fe~{\sc ii}.   This source
is classified as a broad-line AGN by C00 and is discussed in
\S\ref{isoxray}.

	{\bf {CXOHDFN}~J123639.6+621230} ($z=3.479$, BL AGN):  This
broad-line AGN (C00) is in the \hbox{HDF-N} 
(see \S~\ref{newxrayhdfn} for discussion).

	{\bf {CXOHDFN}~J123641.8+621131} ($z=0.089$, LL Galaxy):  This
 object is in the \hbox{HDF-N}.  The X-ray emission is identified with a ``knot''
most easily visible in the $U$ band (see H00).  

	{\bf {CXOHDFN}~J123642.2+621545} ($z=0.857$, HL Galaxy):  This
 source is one of the radio starbursts (see \S\ref{microJy}).
The optical spectrum is classified by C00 as being of ``intermediate''
type.  This source is also detected by \iso-CAM (see \S\ref{isoxray}).

        {\bf {CXOHDFN}~J123646.3+621404} ($z=0.962$, BL AGN):  This source is 
 in the \hbox{HDF-N} (see H00).

	{\bf {CXOHDFN}~J123648.0+621309} ($z=0.475$, LL Galaxy):  This source is an elliptical galaxy in the 
\hbox{HDF-N} (see H00).  

	{\bf {CXOHDFN}~J123651.7+621221} ($z\approx 2.7$, NL AGN):  This source is a 
Type~2 QSO candidate in the \hbox{HDF-N} (see H00 and references therein). 

        {\bf {CXOHDFN}~J123652.9+621445} ($z=0.322$, LL Galaxy):  This source
is identified as an AGN by R98 due to the variability of its radio
emission on timescales of months.  C00 identify it as having an
absorption-dominated spectrum.  This object is underluminous in the
X-ray for an AGN with  $\log{({{f_{\rm X}}\over{f_{\rm R}}})} = -2.4$,
and it is undetected in the hard X-ray band.  This object is identified as
an LL galaxy.

        {\bf {CXOHDFN}~J123656.9+621301} ($z=0.474$, LL Galaxy):  This source
is an elliptical galaxy in the \hbox{HDF-N} (see H00 and
\S\ref{HDFNellipticals} of this paper).  It most likely contains an
 AGN. 

	{\bf  {CXOHDFN}~J123658.3+620958} ($z=0.137$, LL Galaxy):  This
 object is detected only in the soft band and is associated with
a bright S0 or elliptical galaxy ($R = 18.0$).  C00 identify it as an
emission-line dominated galaxy. The X-ray luminosity is only $6.3
\times 10^{39}$~erg~s$^{-1}$ in the soft band,  extremely faint for an
AGN. 
 This source is also one of only
two X-ray/2MASS detections in the Caltech area.

        {\bf {CXOHDFN}~J123702.0+621123} ($z=0.136$, LL Galaxy):  This object
is identified by C00 as having a spectrum of ``intermediate" type. It
has a low X-ray luminosity of $8.6\times 10^{39}$~erg~s$^{-1}$ in the
full band.  It is identified as an LL galaxy.

	{\bf {CXOHDFN}~J123713.7+621424} ($z=0.475$, HL Galaxy): This
 object has an intermediate-type spectrum (C00).  This source is
only detected in the hard band, and it has a hard-band luminosity of $6.8
\times 10^{41}$~erg~s$^{-1}$.  

\subsection{Other Interesting Objects}

	{\bf {CXOHDFN}~J123616.0+621107} ($z$ unknown): This
 object is one of the brightest 2--8~keV sources in this sample
with a full-band observed X-ray flux of $1.3 \times
10^{-14}$~erg~cm$^{-2}$~s$^{-1}$, but its optical counterpart is fairly
faint ($R=25.5$).  It is also very hard, anomalously so for its bright
X-ray flux (see Figure~\ref{bandratios}) with a band ratio of 
$2.14^{+0.63}_{-0.47}$; the corresponding $N_{\rm H}$ at $z=1$ is
$\approx 10^{23}$~cm$^{-2}$ for $\Gamma = 1.7$.  Note that
 the X-ray spectral nature of this source is not well understood; a
power law may not be the appropriate continuum shape.  We have analyzed
the light curve and find no strong evidence for variability.  This
object is not detected in the radio.  We suspect that this object does
harbor some type of extreme AGN, but further investigation is
required. 

	{\bf {CXOHDFN}~J123635.6+621424} ($z$ unknown): This
	object is quite red (${\cal R}-K_{\rm s}=4.74$; Hogg00) and also
rather hard (its band ratio is $0.75^{+0.71}_{-0.43}$ corresponding to 
$\Gamma = 0.75$).  We observed this 
object with the Keck LRIS and detected no prominent
emission features.  It has a rather steep radio spectrum ($\alpha>0.87$),
and the radio emission extends across 2.8$^{\prime \prime}$ (R98;
R00).  R98 identify it with a face-on spiral galaxy, and R00 identifies
it as a star-forming galaxy.  It is also detected by \iso-CAM (A99).
Given the object's spatial extent, it should be at moderate-to-low redshift;
at $z \approx 1$ it would have a hard-band
X-ray luminosity of $\approx 4 \times 10^{42}$~erg~s$^{-1}$ and  
soft-band luminosity $\approx 7 \times 10^{41}$~erg~s$^{-1}$.   The
radio and infrared data point to significant star formation
 in this object, and the hard X-ray emission indicates the possibility
of an obscured AGN.

        {\bf {CXOHDFN}~J123704.8+620941} ($z$ unknown): This object is
the only X-ray source with two possible optical counterparts with $R \approx 24$
(see Figure~\ref{optical_cutouts1} for an the $I$-band image).  
The northern counterpart is not clearly visible on the deep $V$ image but is 
clearly visible on the $I$ band image, indicating that this object is rather red.
Since both of these objects are of comparable brightness and within  
$\approx 1.0^{\prime \prime}$ of the X-ray source position, we cannot ascertain 
with the current data which object is the actual counterpart.  

	{\bf CXOHDFN~J123704.8+621601} ($z$ unknown): This object 
is optically faint for its
soft-band X-ray flux of $2.9 \times 10^{-15}$~erg~cm$^{-2}$~s$^{-1}$ and 
hard-band X-ray flux of $5.8 \times 10^{-15}$~erg~cm$^{-2}$~s$^{-1}$ (it
is marked with a box in Figure~\ref{akiyama_fancy}).  It is
 undetected at 1.4~GHz (R00).  
This object may be a highly absorbed AGN at fairly high
redshift ($z>3$);  the unabsorbed emission is redshifted into the
0.5--8~keV band, resulting in the only moderately hard nature of the X-ray
emission.



\begin{deluxetable}{lccclllc}
\tabletypesize{\footnotesize}
\tablewidth{0pt}
\tablecaption {Journal of \chandra\ observations \label{chandralog}}
%
\scriptsize
\tablehead{
\colhead{Obs.}                                 &
\colhead{Obs.}                                 &
\colhead{Exposure}                             &
\multicolumn{2}{c}{Aim Point}                  &
\multicolumn{2}{c}{ACIS-I Corners}             &
\colhead{Roll}                                 \\
\colhead{ID}                                 &
\colhead{Date}                               &
\colhead{Time (s)$^{\rm a}$ }                &
\colhead{$\alpha_{\rm 2000}$}                &
\colhead{$\delta_{\rm 2000}$}                &
\colhead{$\alpha_{\rm 2000}$}                &
\colhead{$\delta_{\rm 2000}$}                &
\colhead{Angle ($^\circ$)$^{\rm b}$}
}
\startdata
580 & 1999 Nov 13 & 49543 & 12 37 13.1 & +62 12 42.3 & 12 37 35.0 & +62 24 16 &  36.4 \\
    &             &       &            &             & 12 35 39.8 & +62 14 10 & \\
    &             &       &            &             & 12 37 05.9 & +62 00 37 & \\
    &             &       &            &             & 12 39 02.6 & +62 10 32 & \\
967 & 1999 Nov 14 & 57444 & 12 37 01.0 & +62 12 56.8 & 12 37 20.6 & +62 24 36 &  38.0 \\
    &             &       &            &             & 12 35 27.9 & +62 14 16 & \\
    &             &       &            &             & 12 36 57.6 & +62 00 53 & \\
    &             &       &            &             & 12 38 51.4 & +62 11 07 & \\
\enddata
\tablenotetext{a}{All observations were continuous. These times have been corrected for 
lost exposure due to the 0.041~s read-out time per CCD frame and the drifting of the 
back-end processor (BEP) clock. Note that times in H00 have not had these corrections made.}
\tablenotetext{b}{Roll angle describes the orientation of the \chandra\ instruments
on the sky.  The angle is between 0--360$^{\circ}$, and it increases to the West
of North---opposite to the sense of traditional position angle.  }

\end{deluxetable}


\begin{deluxetable}{rrrrcc}
\tabletypesize{\footnotesize}
\tablewidth{0pt}
\tablecaption {False detections due to flaring pixels \label{flaring}}
\scriptsize
\tablehead{
\multicolumn{2}{c}{$\alpha_{2000}$} & 
\multicolumn{2}{c}{$\delta_{2000}$} &
\colhead{{\sc wavdetect}}&
\colhead{Flaring Pixel } \\ 
\multicolumn{2}{c}{($- 12^{\rm h}$)}                    &
\multicolumn{2}{c}{($- 62^{\circ}$)}                &
\colhead{SB Counts$^{\rm a}$}                       &
\colhead{Counts}                                   
}
\startdata
36& 54.5\phn\phn  & 9 & 51.3    &    6.8       &   6      \\
36& 41.6\phn\phn  & 9 & 54.8    &    6.6       &   5      \\
36& 39.5\phn\phn  & 9 & 58.0    &    5.8       &   5      \\
36& 34.6\phn\phn  &13 & 32.6    &    8.7       &   8      \\
36& 58.2\phn\phn  & 9 &  4.0    &    5.7       &   5      \\
37& 20.6\phn\phn  &11 & 41.6    &    3.8       &   4      \\
36& 45.1\phn\phn  &13 & 12.2    &    5.3$^{\rm b}$ &   3$^{\rm b}$\\
36& 45.1\phn\phn  &13 & 50.6    &    4.8       &   5      \\
36& 49.8\phn\phn  &13 & 41.1    &    5.8       &   4      \\
\enddata
\tablenotetext{a}{The number of events detected in the soft-band (SB) X-ray image
by CXC's {\sc wavdetect}.  Note that this number is background subtracted and
has been convolved with a function representing the \chandra\ PSF 
(see Freeman et~al. 2001 for more information).}
\tablenotetext{b}{This source was only detected in the full band.
The quoted counts are for the full band. }
\end{deluxetable}


\begin{deluxetable}{lcccllllll}
\tabletypesize{\footnotesize}
\tablewidth{0pt}
\tablecaption {\chandra\ source detections in the Caltech area \label{xraydetections}}
\scriptsize
\tablehead{
\colhead{Energy}                                 &
\colhead{Number of}				&
\multicolumn{4}{c}{Detected Counts Per Source}      &
\colhead{Background}		 \\
\colhead{Band}                                              &
\colhead{Sources$^{\rm a}$ }                                      &
\colhead{Maximum}                                   &
\colhead{Minimum}                                   &
\colhead{Median}                              &
\colhead{Mean}					&
\colhead{Counts pixel$^{-1}$}
}
\startdata
0.5--8~keV (Full)       &       73&     866.4 & 4.4 &   27.9 &  78.1  & 0.053$\pm$0.004 \\
0.5--2~keV (Soft)       &       62&     673.6 & 4.5 &   17.2 &  58.4  & 0.019$\pm$0.002\\
2--8~keV (Hard)         &       45&     187.5 & 5.5 &   18.2 &  39.0  & 0.033$\pm$0.002 \\
5--8~keV (Ultra-Hard)  &	10&     23.3  & 3.8 &   13.2 &  12.4  & 0.019$\pm$0.002\\
\enddata
\tablenotetext{a}{There were 82 independent X-ray sources detected in total.}
\end{deluxetable}

\renewcommand{\baselinestretch}{0.95}


\begin{deluxetable}{clrrrrrlrrlrrrrr}
\rotate
\tablecolumns{16}
\tabletypesize{\scriptsize}
\tablewidth{0pt}
\tablecaption {Properties of \chandra\ sources in the Caltech area \label{xraydata}}

\tablehead{
\colhead{\#}		&
\colhead{CXOHDFN Name}               &
\colhead{FB }              &
\colhead{FB}                 &
\colhead{$\Delta$ HB$^{\rm b}$ }        &
\colhead{HB }              &
\colhead{HB }               &
\colhead{$\Delta$ SB$^{\rm b}$ }        &
\colhead{SB }              &
\colhead{SB }               &
\colhead{Exposure}          &
\colhead{Band}			&
\colhead{$\Gamma^{\rm d}$}              &
\multicolumn{3}{c}{Flux (10$^{-15}$ erg cm$^{-2}$ s$^{-1}$)} 

\\
\colhead{}                &
\colhead{(J2000)}                              &
\colhead{Counts$^{\rm a}$}                     &
\colhead{Error}                      &
\colhead{($^{\prime \prime}$)}   &
\colhead{Counts$^{\rm a}$}             &
\colhead{Error}                                 &
\colhead{($^{\prime\prime}$)} &
\colhead{Counts$^{\rm a}$}                   &
\colhead{Error}                                 &
\colhead{Time (ks)$^{\rm a}$}                             &
\colhead{Ratio$^{\rm c}$}                                 &
\colhead{}                                      &
\colhead{FB}              &
\colhead{HB}               &
\colhead{SB}
}
\startdata
1 & 123614.4+621046 &         19.9 &     5.4 & \nodata      & $<$13.82 & \nodata & \nodata      & $<$11.1      & \nodata &          200.6 & \nodata                 & 1.40 &    1.02 & $<$1.44 & $<$0.25 \\
2 & 123615.9+621515 &         33.6 &     6.8 &          0.3 &    20.05 &     5.2 & 2.1$^{\rm e}$ &          8.9 &     3.3 &          202.3 & 2.30$^{+2.28}_{-1.04}$  & 0.06 &    3.45 &    2.81 &    0.19 \\
3 & 123616.0+621107 &        162.5 &    13.6 &          0.2 &   101.94 &    10.6 &          0.7 &         48.6 &     7.4 &          201.8 & 2.14$^{+0.63}_{-0.47}$  & 0.09 &   16.48 &   14.23 &    1.05 \\
4 & 123617.0+621011 &        123.9 &    11.9 &          0.2 &    38.79 &     7.0 &          0.2 &         83.8 &     9.4 &          200.7 & 0.47$^{+0.15}_{-0.12}$  & 1.58 &    5.69 &    3.89 &    1.87 \\
5 & 123618.0+621635 &        400.1 &    20.9 &          0.3 &   123.00 &    11.8 &          0.4 &        282.0 &    17.2 &          200.5 & 0.45$^{+0.07}_{-0.06}$  & 1.63 &   17.94 &   12.24 &    6.32 \\
\\
6 & 123618.5+621115 &        436.2 &    21.4 &          0.1 &    92.15 &     9.9 &          0.1 &        340.9 &    18.8 &          203.0 & 0.27$^{+0.30}_{-0.04}$  & 1.93 &   16.27 &    8.53 &    7.66 \\
7 & 123619.1+621441 &         76.0 &     9.1 &          0.2 &    18.00 &     4.6 &          0.2 &         60.1 &     7.9 &          204.5 & 0.30$^{+0.13}_{-0.10}$  & 1.88 &    2.90 &    1.67 &    1.34 \\
\enddata
\end{deluxetable}

\clearpage


\renewcommand{\baselinestretch}{1.053}


\begin{deluxetable}{llclrr}
\tablewidth{0pt}
\tablecaption {Catalogs correlated against the \chandra\ sources \label{catalogs}}
%
\tabletypesize{\scriptsize}
\tablehead{
\multicolumn{1}{l}{Catalog}     &
\multicolumn{1}{l}{Catalog}     &
\colhead{}                      &     
\colhead{}                      &     
\colhead{}                      &     
\colhead{}                      \\	
\multicolumn{1}{l}{Description} &
\multicolumn{1}{l}{Reference}   &
\colhead{$N_{\rm S}^{\rm a}$}       &	
\colhead{$R_{\rm M}^{\rm b}$}       &      
\colhead{$N_{\rm M}^{\rm c}$}       &     
\colhead{$N_{\rm F}^{\rm d}$}       
}
\startdata
{\it Radio sources}                                       &                                 &     &           &     &       \\
\hspace*{0.15 cm} VLA (8.5~GHz)                           &  R98         		    & 48  & 0.4$^{\rm e}$ & 10  & 0.25  \\
\hspace*{0.15 cm} VLA (1.4~GHz)                           &  Richards et~al. (1999)         & 17  & 1.0$^{\rm e}$ &  2  & $<$0.25\\
                                                          &  R00                & 65  & 0.7$^{\rm e}$ & 16  & 0.25  \\
\hspace*{0.15 cm} VLA + MERLIN				  &  Richards (1999)                & 67  & 1.3       & 16  & 0.25  \\
							  &  Muxlow et al. (1999)	    & 	  & 	      &     &       \\
\hspace*{0.15 cm} WSRT (1.4 GHz)                          &  Garrett et~al. (2000)          & 65  & 3.0       & 11  & 0.75  \\
                                                          &                                 &     &           &     &       \\
\enddata
\tablenotetext{a}{Number of catalog sources in the Caltech area.}
\tablenotetext{b}{Positional error (in arcseconds) of catalog sources adopted for matching with the \chandra\ sources.}
\tablenotetext{c}{Number of catalog sources with \chandra\ source matches.}
\tablenotetext{d}{Statistically expected number of catalog sources with 
false \chandra\ source matches (see \S\ref{sourcematching}). Note that this technique is
of limited precision in some cases due to small number statistics. }
\tablenotetext{e}{Typical value. Errors differed from source to source as 
described in the respective reference.}
\tablenotetext{f}{In the \hbox{HDF-N} there are actually zero false matches expected at the optical flux levels
of the \chandra-W96 matches (see \S~\ref{sourcematching}).}
\tablenotetext{g}{For Jarvis \& MacAlpine (1998), 12 of the 32 objects are AGN candidates.  For Vanden Berk et~al. (2000), four of the five objects are QSO candidates.}
\end{deluxetable}

\clearpage


\begin{deluxetable}{clrrrlclrrr}
\tablecolumns{11}
\tabletypesize{\footnotesize}
\tablewidth{0pt}
\tablecaption {Multiwavelength properties of the Caltech area \chandra\ sources \label{multiwavtable}}
\tablehead{
\colhead{\#}		&
\colhead{CXOHDFN Name } 		&
\colhead{}                &
\colhead{}                &
\colhead{}                    &
\colhead{$z$}                &
\colhead{}                     &
\colhead{ID}                 &
\multicolumn{3}{c}{${\log{L}_{\rm X}}$(erg s$^{-1}$)$^{\rm e}$}
\\
\colhead{}				&
\colhead{(J2000) }			&
\colhead{$R^{\rm a}$}                    	&
\colhead{$I^{\rm a}$}				&
\colhead{$z$}				&
\colhead{Ref$^{\rm b}$}          	&
\colhead{ID$^{\rm c}$}                  &
\colhead{Ref$^{\rm d}$} 		&
\colhead{Full}				&
\colhead{Hard}				&
\colhead{Soft}			
}
\startdata
1 & 123614.4+621046$^{\rm f}$          &    24.8 & 24.2 &             \nodata & \nodata   & \nodata       & \nodata & \nodata & \nodata       & \nodata       \\
2 & 123615.9+621515$^{\rm f}$          &    25.2 & 24.4 &             \nodata & \nodata   & \nodata       & \nodata & \nodata & \nodata       & \nodata       \\
3 & 123616.0+621107          &    25.5 & 24.4  & \nodata & \nodata   & \nodata     & \nodata & \nodata       & \nodata       \\
4 & 123617.0+621011$^{\rm g}$ &   22.7 & 21.7 &   0.845 & H01    & HL        & H01     &    43.1 &          42.9 &          42.6 \\
5 & 123618.0+621635$^{\rm g}$ &   20.5 & 19.9 &   0.679 & H01    & NL AGN    & H01     &    43.4 &          43.2 &          42.9 \\
\\
6 & 123618.5+621115$^{\rm g}$ &   21.4 & 21.0 &   1.022 & H01    & BL AGN    & H01     &    43.8 &          43.6 &          43.5 \\
7 & 123619.1+621441          &    24.6 & 24.3 &      \nodata & \nodata & \nodata   & \nodata & \nodata & \nodata       & \nodata       \\
\enddata
\end{deluxetable}

\clearpage


\begin{deluxetable}{lcccc}
\tablecolumns{4}
\tabletypesize{\scriptsize}
\tablewidth{0pt}
\tablecaption {Summary of spectroscopic classifications of \chandra\ sources \label{spectralclasses}}
\tablehead{
\multicolumn{1}{c}{Object}              &
\colhead{Soft-band}                 &   
\colhead{Hard-band}		    &
\colhead{Total}	
\\
\multicolumn{1}{c}{Class} 			&
\colhead{Detections} 		&
\colhead{Detections}		&
\colhead{Detections}
}

\startdata
{\it AGN}			&		&		&		\\
\hspace*{0.15 cm} Broad-line (BL)&	8	&	6	&	8	\\
\hspace*{0.15 cm} Narrow-line (NL)&	5	&	6	&	7	\\
\hspace*{0.15 cm} Total		&	13	&	12	&	15	\\
\\
{\it Galaxies}&			&		&		&		\\
\hspace*{0.15 cm} Low-luminosity (LL)&	7	&	1	&	8	\\
\hspace*{0.15 cm} High-luminosity (HL)&	4	&	7	&	7	\\	
\hspace*{0.15 cm} Total		&	11	&	8	&	15	\\
\\
{\it Stars}	 		&	2	&	0	&	2	\\
\enddata
\end{deluxetable}


\begin{deluxetable}{lclrllrcl}
\tablecolumns{12}
\tabletypesize{\scriptsize}
\tablewidth{0pt}
\tablecaption {Chandra detections of $\mu$Jy radio sources \label{radiotable}}
%
%
\tablehead{
\colhead{}               			&
\colhead{$\Delta$ R-XR$^{\rm a}$ }        	&
\colhead{$S_{1.4}$}              		&
\colhead{}        				&
\colhead{$\theta^{\rm b}$ }              	&
\colhead{$z$}					&
\colhead{Band}					&
\colhead{}               	&
\colhead{$\log{L_{0.5-8}}$}			\\
\colhead{VLA object}        			&
\colhead{($^{\prime \prime}$) }        			&
\colhead{($\mu$Jy)}              			&
\colhead{${\alpha}_{\rm R}^{\rm b}$}  	             	&
\colhead{($^{\prime \prime}$) }              		&
\colhead{}                                   		&
\colhead{Ratio}                                  	&
\colhead{Class$^{\rm b}$ }               		&
\colhead{(erg s$^{-1}$)}		

}

\startdata
J123646.36+621404.9 &     0.2&  179.0 &      $-0.04$      &  0.2&     0.962    &   1.27      &AGN  & 43.6   \\
J123652.96+621444.3 &     0.4&  168.0 &      $-0.12$      &  2.5&     0.322    & $<$1.03     &AGN  & 41.2  \\
\\
J123616.14+621513.9$^{\rm c}$&      1.8&   53.9 &      \nodata      &  1.0&     \nodata  &   2.30      &S    & \nodata  \\
J123617.10+621011.5&      0.5&   62.0 &      \nodata      &  0.7&     0.845    &   0.47      &S    & 43.1 \\
J123621.27+621109.5&      0.2&   52.6 &      $>$0.86      &  3.2&     1.014    &   $>$1.32   &S    & 42.3 \\
J123631.27+620957.3$^{\rm d}$&   0.7&  152.0 &      $>$0.99        &2.5&     \nodata  &   $<$1.14   &S    & \nodata \\ 
J123634.46+621213.3&      0.4&  233.0 &      0.74         &  1.0&     0.458    &   $<$1.19   &S    & 41.4 \\
\enddata

\tablenotetext{a}{Offset between the radio and \chandra\ source positions.}
\tablenotetext{b}{${\alpha}_{\rm R}$ is the radio spectral index and $\theta$ 
is the radio spatial extent; 
both were determined by R00.   R00 classifications are `S' for `Starburst-type'
and `U' for `Unknown'.}
\tablenotetext{c}{Please see footnote ``e" in Table~\ref{xraydata} and
footnote ``f" in Table~\ref{multiwavtable} 
about this object (CXOHDFN~J\hbox{123615.9+621515}).  }
\tablenotetext{d}{This source was detected in neither the full nor the hard band.   }
\tablenotetext{e}{This source was not detected independently in either the soft or hard bands.}

\end{deluxetable}

\clearpage


\begin{deluxetable}{lcccccc}
\tablecolumns{7}
\tabletypesize{\scriptsize}
\tablewidth{0pt}
\tablecaption {\chandra\ detections of faint \iso-CAM sources \label{ISOtable}}
%
%
\tablehead{
\colhead{}              &
\colhead{$\Delta$ ISO-XR }        &
\colhead{}              &
\colhead{7 $\mu$m }   &
\colhead{15 $\mu$m }        &
\colhead{Band}	&
\colhead{$\log{L_{0.5-8}}$}           
\\
\colhead{ISOCAM object} 	&
\colhead{($^{\prime \prime}$)}      &
\colhead{$z^{\rm a}$}      &
\colhead{($\mu$Jy)}      &
\colhead{($\mu$Jy)}      &
\colhead{Ratio}      &
\colhead{(erg s$^{-1}$)}      
}

\startdata
J123634.4+621212 & 1.8  &   0.456  & \nodata            &  $448^{+68}_{-59}$  & $<1.19$   &  41.4 \\
J123635.6+621424 & 0.5  &  \nodata & \nodata            &  $441^{+43}_{-82}$  & $1.09$    &  \nodata \\
J123636.5+621348 & 1.4  &   0.960  & \nodata            &  $353^{+40}_{-66}$  & $0.24$    &  43.3 \\
J123642.1+621545 & 1.4  &   0.857  & \nodata            &  $459^{+46}_{-86}$  & $0.9$     &  42.6 \\
J123644.2+621251$^{\rm b}$ & 1.7  &   0.557  & $<50$    &  $282^{+60}_{-64}$  & $<1.15$   &  41.0 \\
J123646.4+621406 & 1.3  &   0.960  & $191^{+41}_{-86}$  &  $107^{+95}_{-20}$  & $1.27$    &  43.6 \\
J123647.7+621312 & 3.4  &   0.475  & $36^{+65}_{-30}$   &  $<74$              & $<1.03$   &  41.6 \\
J123651.9+621221 & 1.2  &   2.7$^{\rm c}$ & $<36$       &  $48^{+32}_{-9}$    & $2.62$    &  43.2 \\
\enddata
\tablenotetext{a}{See Table~\ref{multiwavtable} for individual redshift references.}
\tablenotetext{b}{This source was only detected in the soft band.  This luminosity is in the soft band.}
\tablenotetext{c}{This redshift is photometric (M. Dickinson 2000, private communication).}

\end{deluxetable}


\begin{deluxetable}{llllll}
\tablecolumns{6}
\tabletypesize{\scriptsize}
\tablewidth{0pt}
\tablecaption {\chandra\ detections of VROs \label{VROtable}}
%
%
\tablehead{

\colhead{Hogg00}              &
\colhead{}               &
\colhead{}               &
\colhead{Band}               &
\colhead{}               &
\colhead{$\log{L_{2-8}}$}              \\ 
\colhead{Source}              &
\colhead{${\cal R}-K_{\rm s}$}               &
\colhead{$K_{\rm s}$}               &
\colhead{Ratio}               &
\colhead{$z^{\rm a}$}               &
\colhead{erg s$^{-1}$}              
}

\startdata

J123629.1+621045 &    5.28 & 17.96 &  $>$1.26 &   1.013    &      42.7 \\
J123651.2+621051 &    5.11 & 19.56 &  $>$2.98 &   \nodata  &      \nodata \\
J123715.9+621213 &    5.05& 18.49 &  $>$2.27 &   1.020     &      42.6 \\
J123724.3+621359 &    5.24 & 19.96 &  $>$1.78 &   \nodata  &      \nodata \\

\enddata
\tablenotetext{a}{From C00; these were determined on the basis of a single 
emission line assumed to be [O{\sc ii}]~3727~\AA.}

\end{deluxetable}
\clearpage


\begin{deluxetable}{lcclc}
\tablecolumns{5}
\tabletypesize{\scriptsize}
\tablewidth{0pt}
\tablecaption {\chandra\ stacking constraints on AGN candidates \label{AGNcandidatetable}}
%
%
\tablehead{

\colhead{AGN}              &
\colhead{Number of }               &
\colhead{Stacked Exposure}               &
\multicolumn{2}{c}{3$\sigma$ Flux Limits (erg cm$^{-2}$ s$^{-1}$)} \\
\colhead{Search}              &
\colhead{Undetected Objects}               &
\colhead{Time (Ms)}               &
\colhead{Soft}               &
\colhead{Hard}             
}

\startdata
Conti et~al. (1999)		& 8  & 1.7 &	1.5$\times 10^{-17}$	& 8.7$\times 10^{-17}$\\
Jarvis \& MacAlpine (1998)	& 12 & 2.6 &	1.0$\times 10^{-17}$	& 8.5$\times 10^{-17}$\\
Sarajedini et~al. (2000) 	& 8  & 1.7 &	1.7$\times 10^{-17}$	& 1.2$\times 10^{-16}$
\enddata
\end{deluxetable}
\clearpage


\begin{figure}
\epsscale{1.0}
\vspace{-1.0in}
\figurenum{3}
\plotone{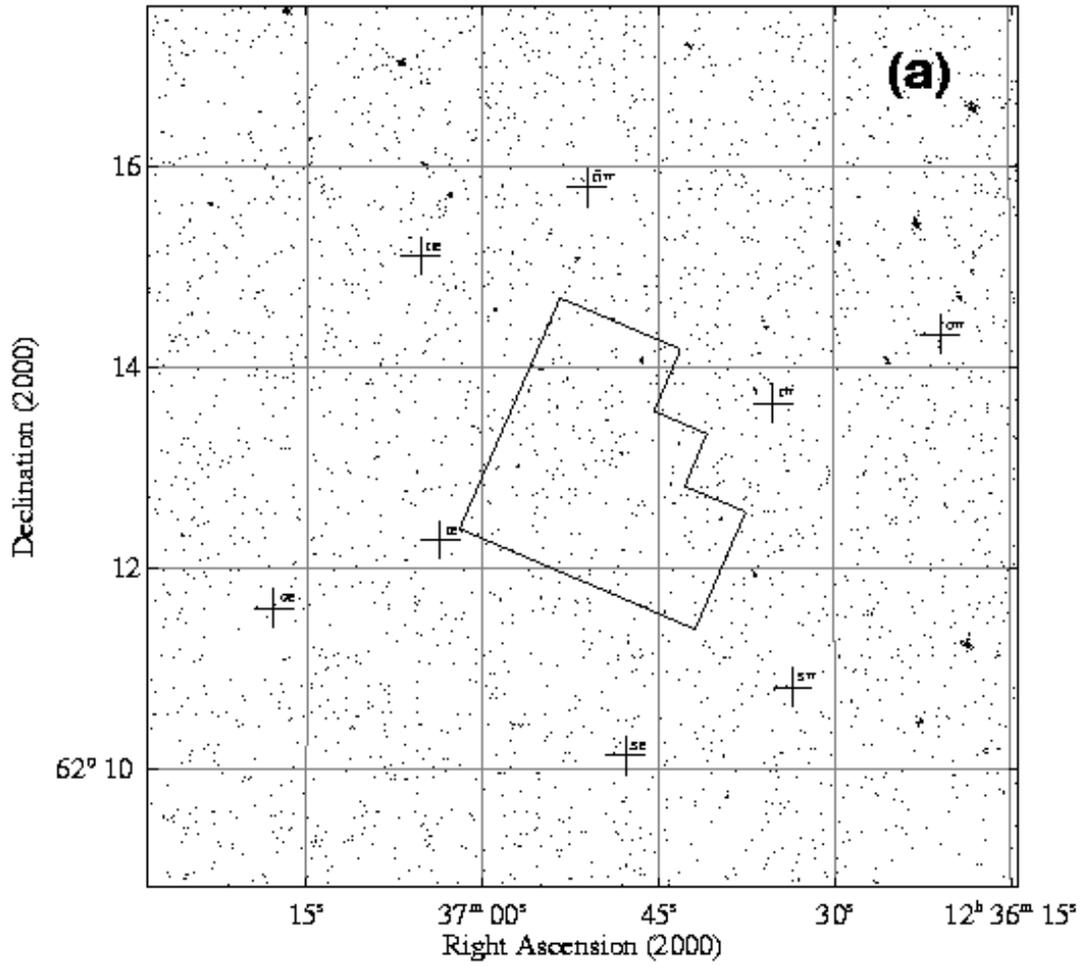}

\caption{The raw X-ray images in the (a) 0.5--2~keV (soft) band and (b) 2--8~keV (hard) band.
The area shown matches the Caltech area indicated on the
exposure map in Figure~\ref{emap}. The outlined region is the \hbox{HDF-N},
and the crosses mark the centers of each of the Hubble Flanking
Fields (see Table~2 of W96).
\label{imageraw1}}
\end{figure}


\begin{figure}
\epsscale{1.0}
\figurenum{3b}
\plotone{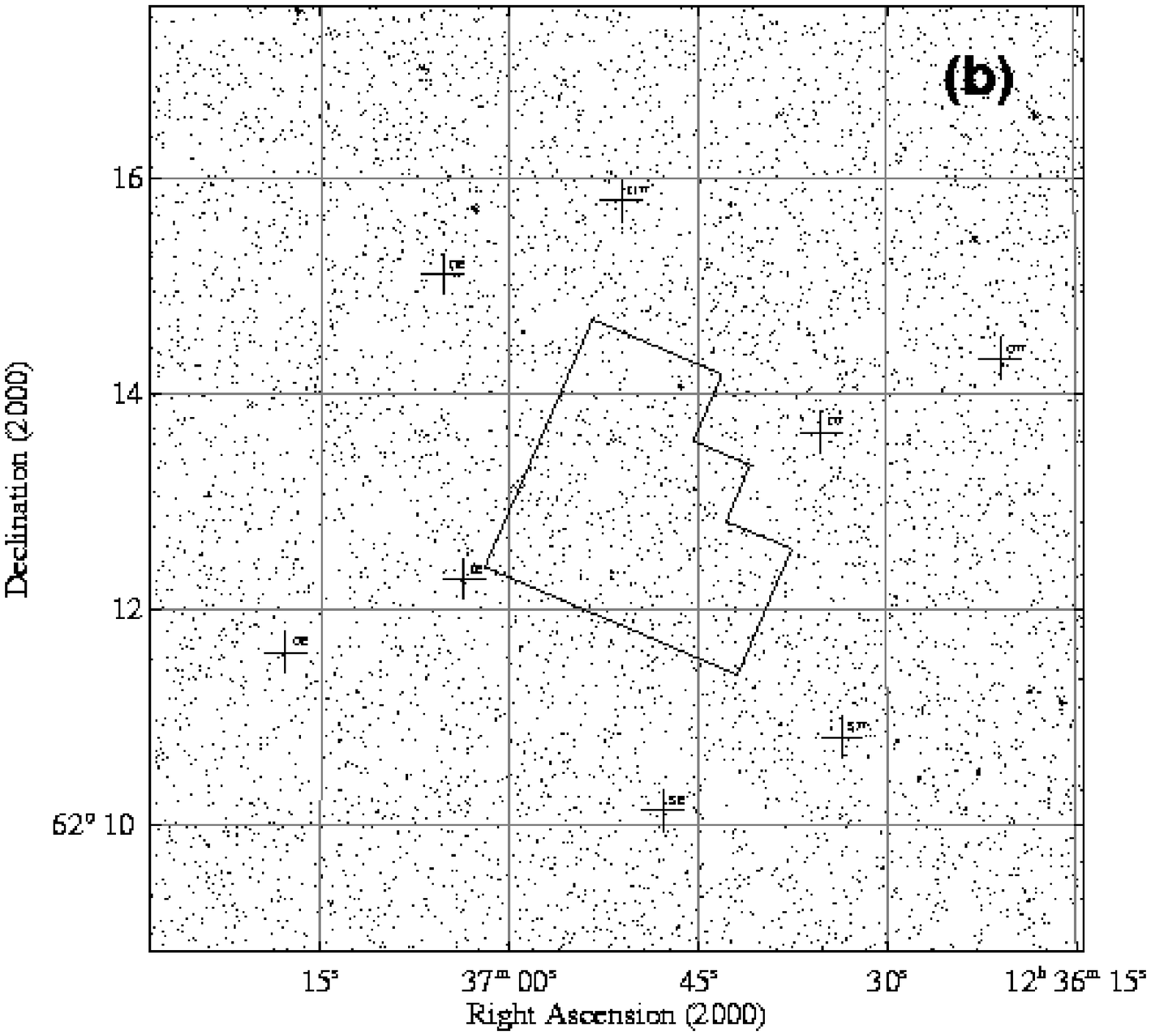}
\label{imageraw2}
\end{figure}


\begin{figure}
\vspace{-1.0in}
\epsscale{1.0}
\figurenum{4}
\plotone{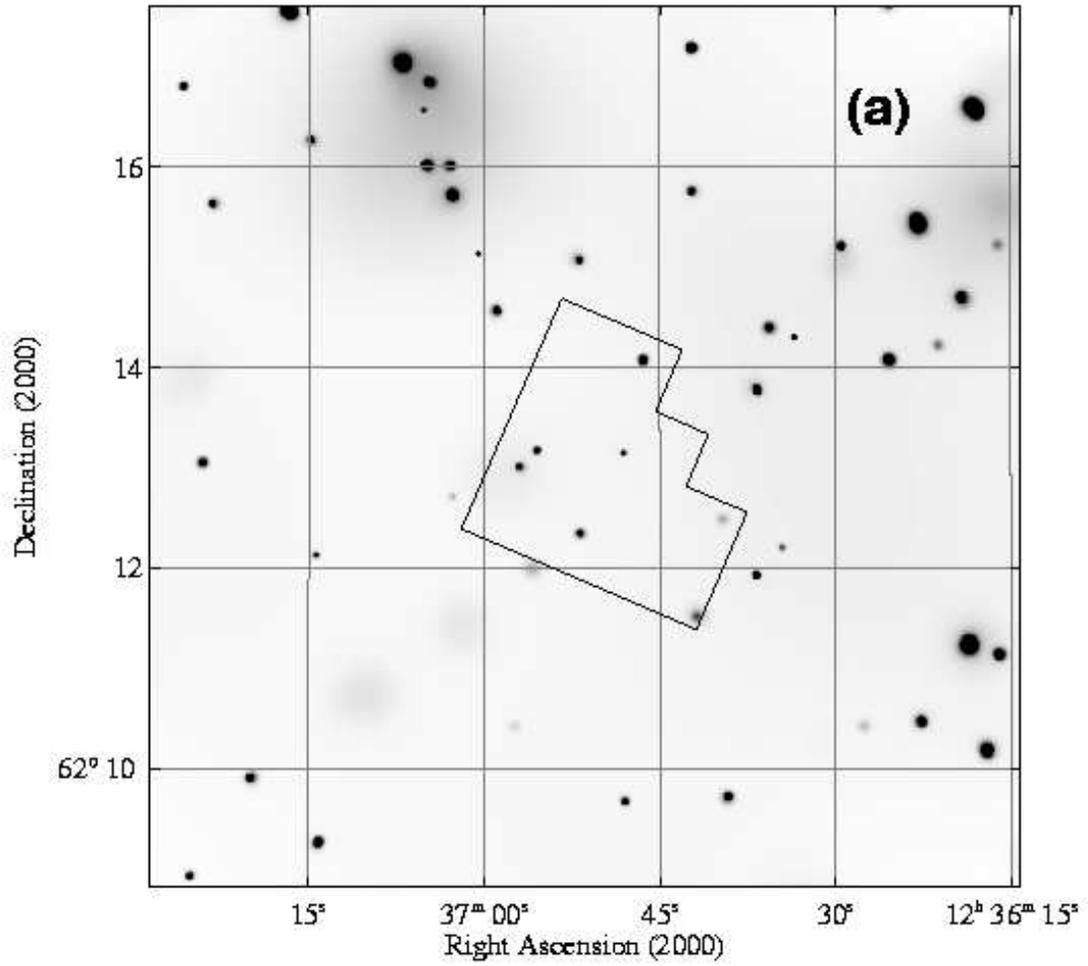}

\caption{Adaptively smoothed versions of (a) the soft-band image shown in
Figure~\ref{imageraw1}a and (b) the hard-band image shown in Figure~\ref{imageraw2}.
The adaptive smoothing is at the $3\sigma$ level using the code of Ebeling, White, \&
Rangarajan (2001).  
\label{imagesmooth}}
\end{figure}


\begin{figure}
\epsscale{1.0}
\figurenum{4b}
\plotone{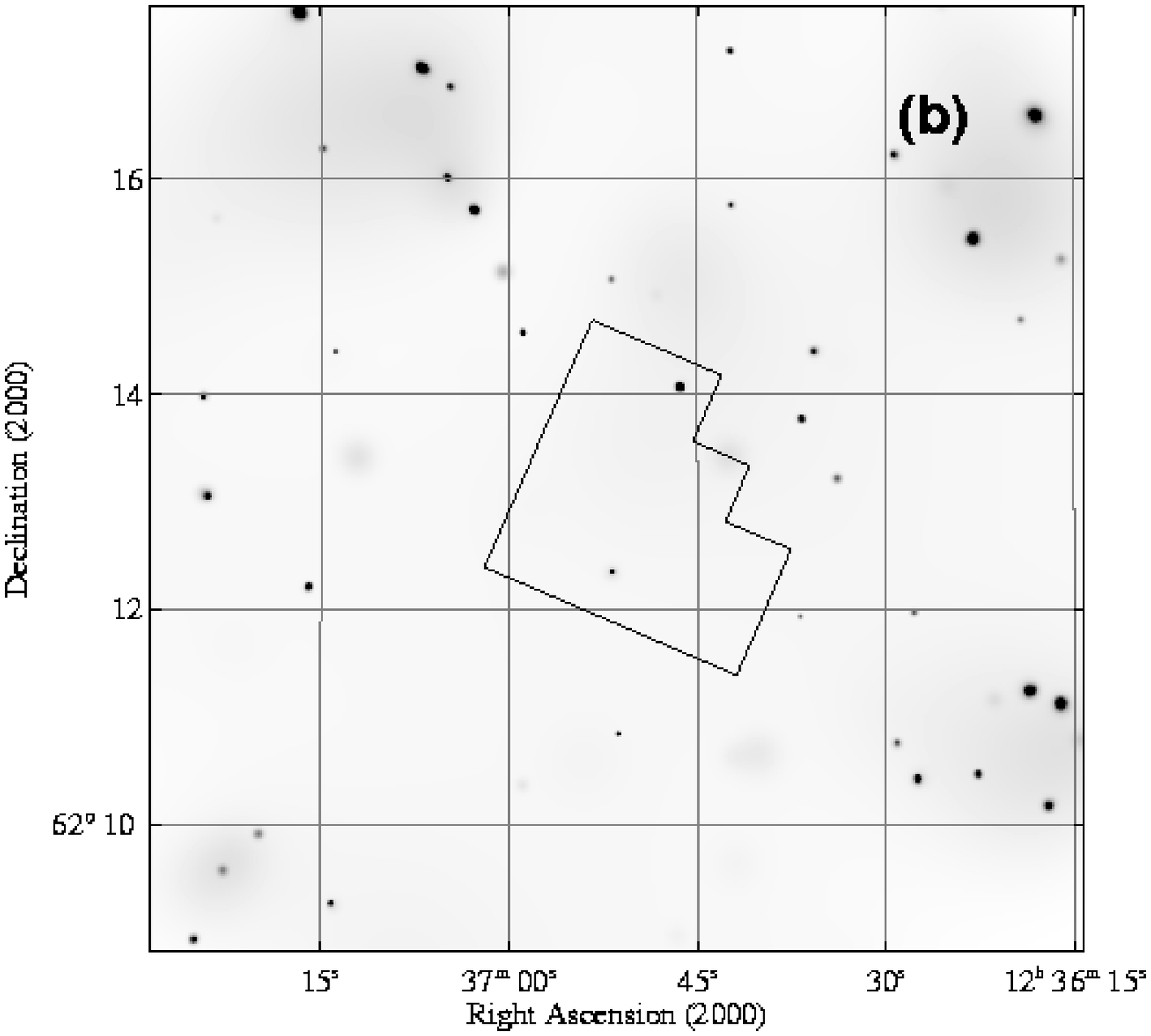}
\label{imagesmooth2}
\end{figure}

\begin{figure}
\epsscale{0.90}
\figurenum{5}
\plotone{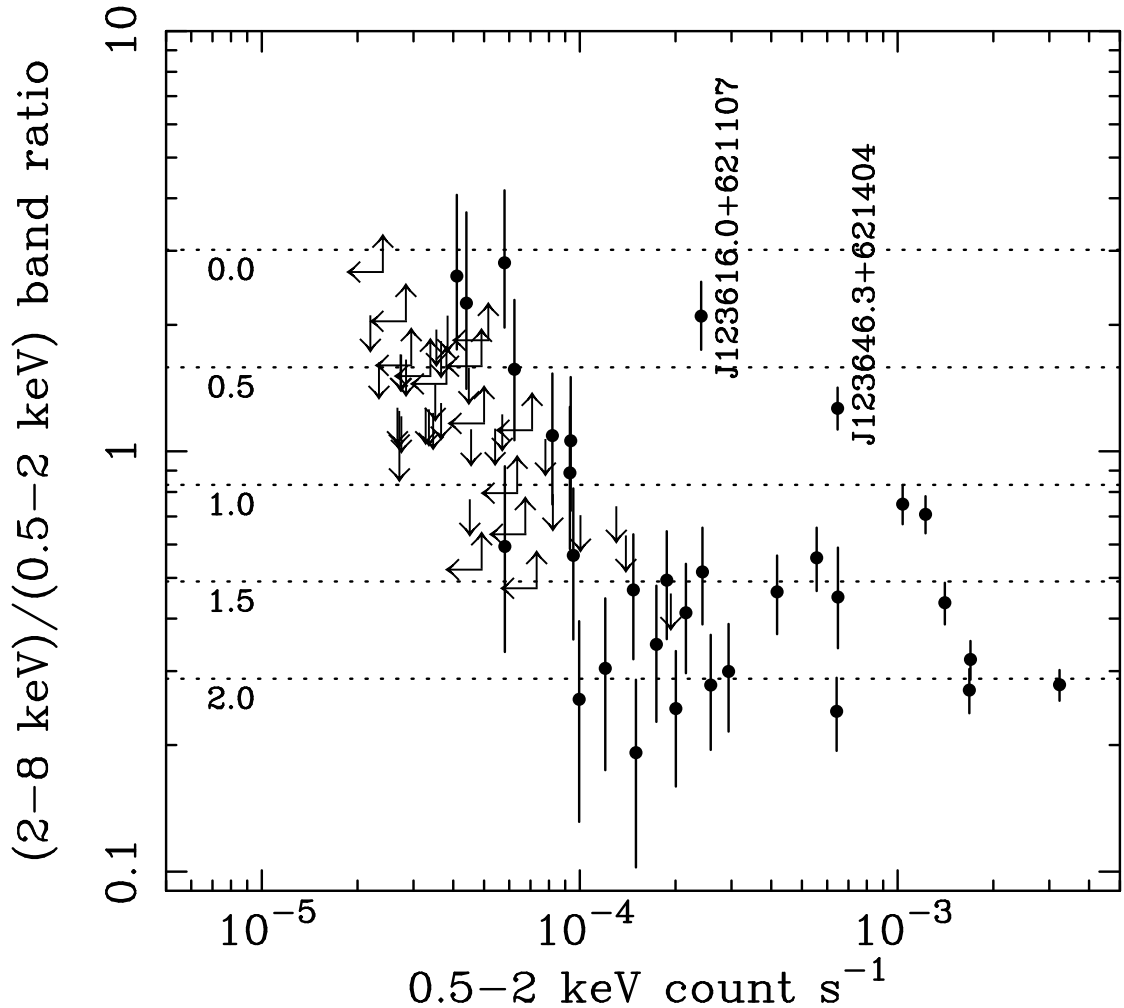}

\caption{ 
Band ratio as a function of soft-band (0.5--2~keV) count rate for 
the \chandra\ sources in the Caltech area.  Dotted lines are labeled
with the photon indices which correspond to a given band ratio. Sources only 
detected in one of the two bands are shown as limits 
(see Table~\ref{xraydata}). Note the general 
hardening of sources with low soft-band count rates (compare 
with Figure~2 of G00). We have marked two 
sources that are unusually hard for their soft-band count 
rates: \hbox{CXOHDFN}~J123616.0+621107 and \hbox{CXOHDFN}~J123646.3+621404. 
The first is also remarkable for its large 
$f_{\rm X}/f_{\rm R}$ ratio; its optical counterpart has $R=25.9$. 
The second is the $z=0.960$ AGN in the \hbox{HDF-N} 
itself (see H00).  
\label{bandratios}}
\end{figure}
\clearpage


\begin{figure}
\epsscale{1.0}
\figurenum{6}

\plotone{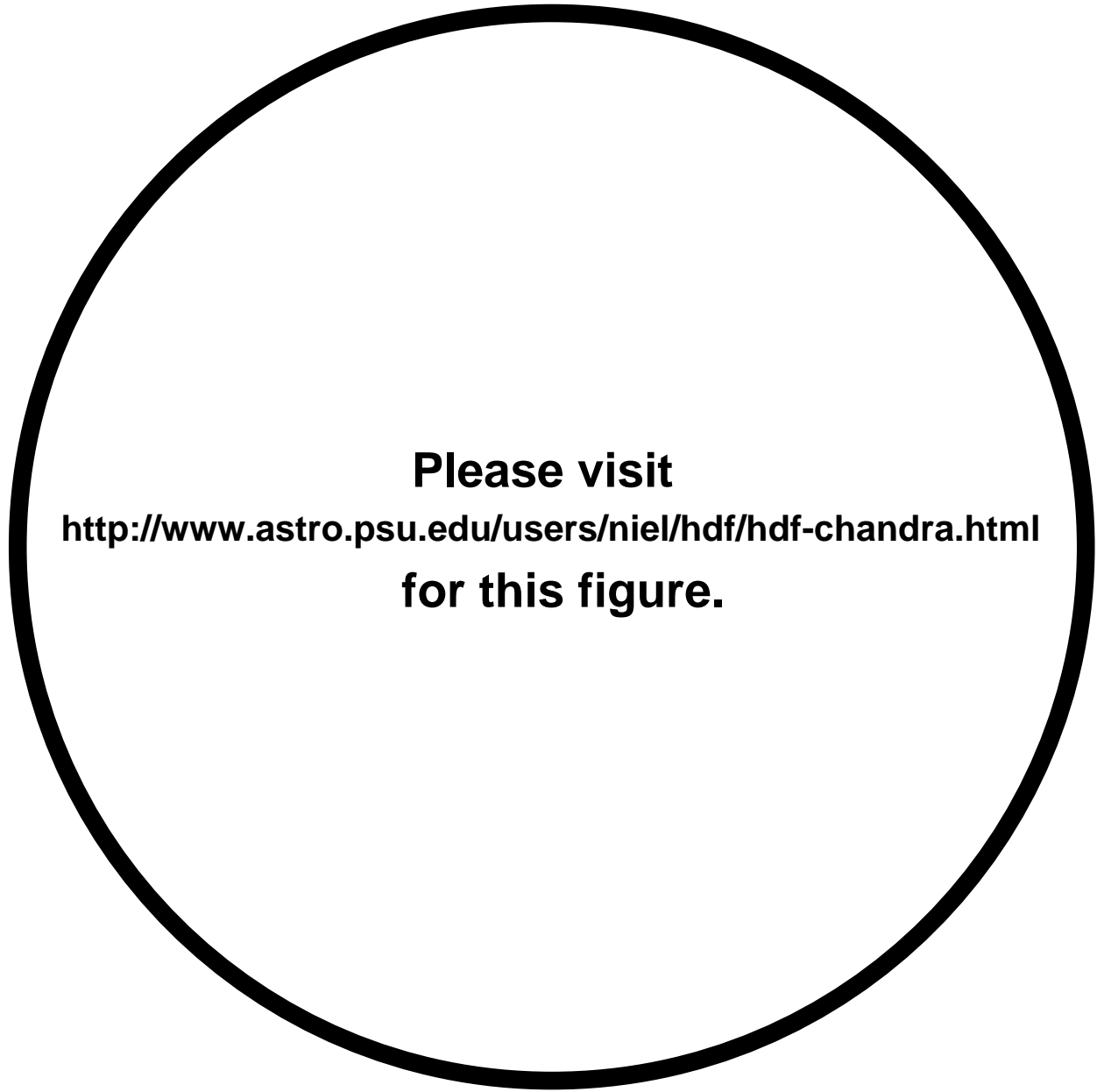}

\caption{Deep $I$-band ``thumbnail" images for the 82 \chandra\ sources.  
Each image is $19^{\prime \prime} \times 19^{\prime \prime}$, and the
circles have $3.0^{\prime \prime}$ diameters.  The images
are numbered as in Tables~\ref{xraydata} and \ref{multiwavtable}. 
Objects have been labeled with their spectroscopic redshifts where applicable.}
\label{optical_cutouts1}
\end{figure}


\begin{figure}
\epsscale{1.0}
\figurenum{6b}

\plotone{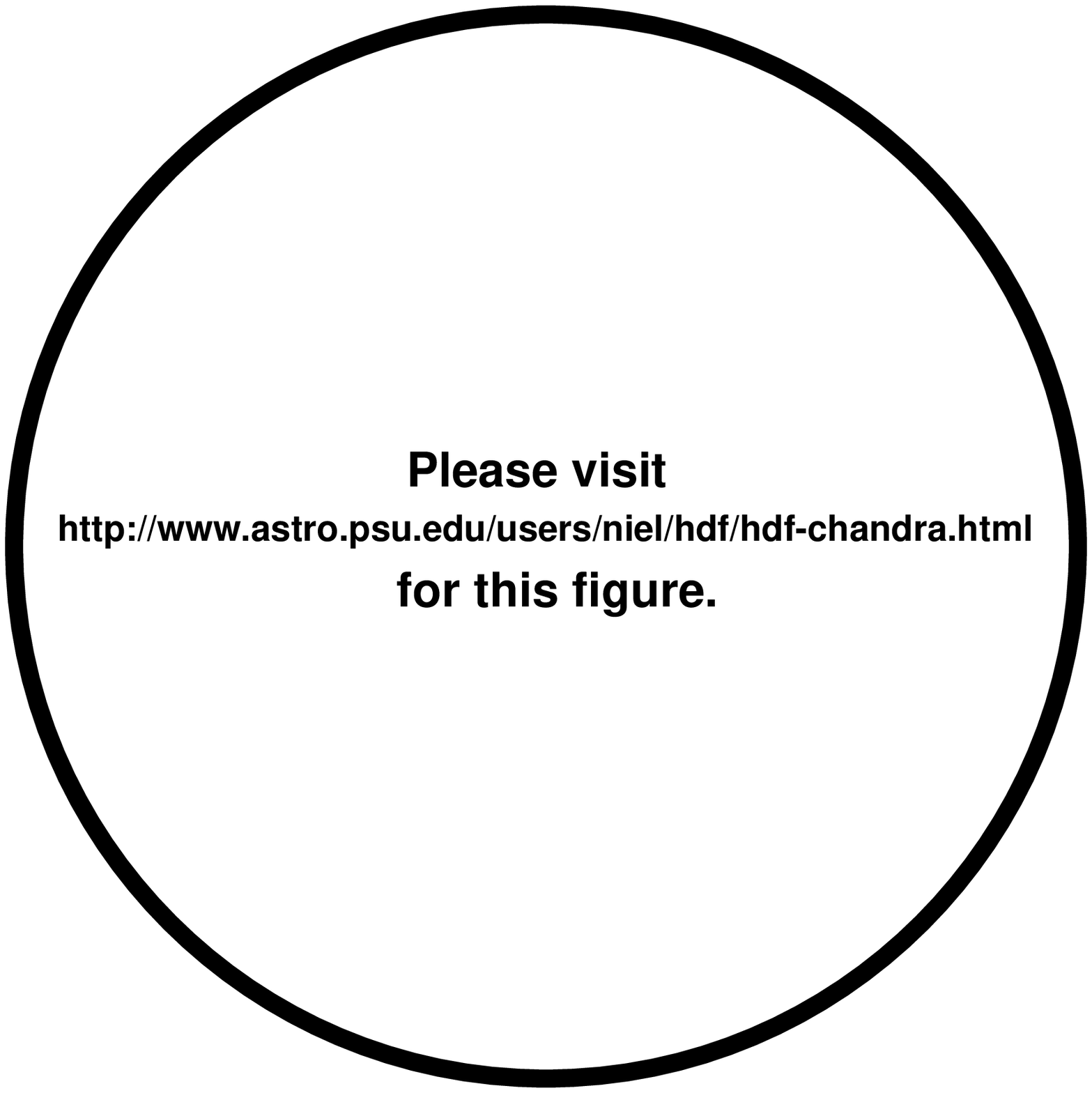}

\label{optical_cutouts2}
\end{figure}


\begin{figure}
\vspace{-1.0truein}
\epsscale{1.00}
\figurenum{7}

\plotone{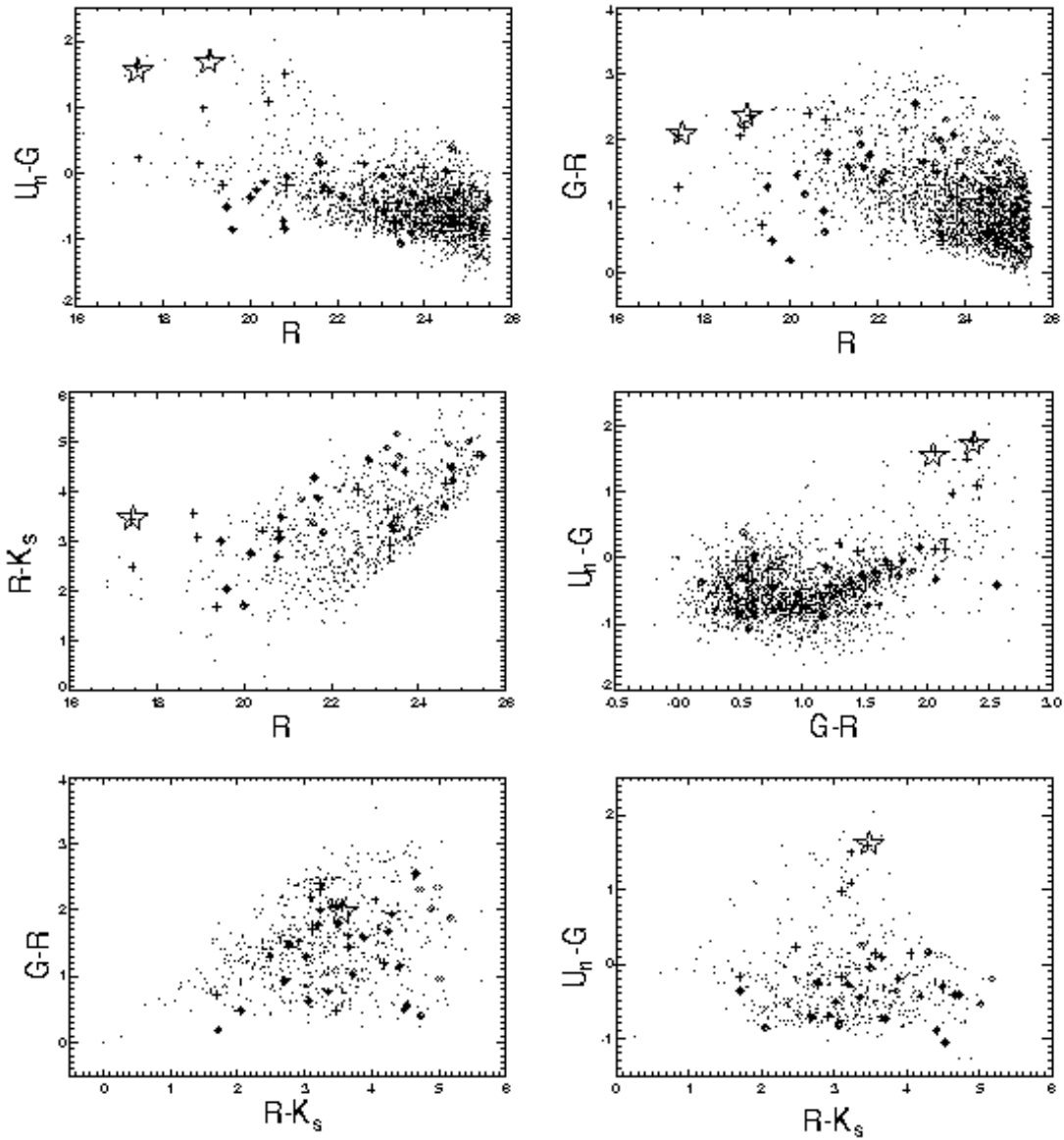}

\caption{Optical colors of the \chandra\ sources as reported 
by Hogg00.  All magnitudes in these plots are in the $U_{\rm n}$, $G$, ${\cal R}$
and $K_{\rm s}$ filters as described in Hogg00 and \S \ref{optphotom}.  
Dots show the entire Hogg00 catalog,
diamonds are hard-band detections, and plus symbols are soft-band detections.
The star symbols mark the two X-ray detected M~stars.
All colors available in the Hogg00 catalog are plotted in this figure to show all the 
possible permutations.  Note that not all X-ray sources in the current sample have
measured colors and/or magnitudes in the Hogg00 catalog.
\label{colors}}
\end{figure}


\begin{figure}
\epsscale{0.65}
\figurenum{8}

\plotone{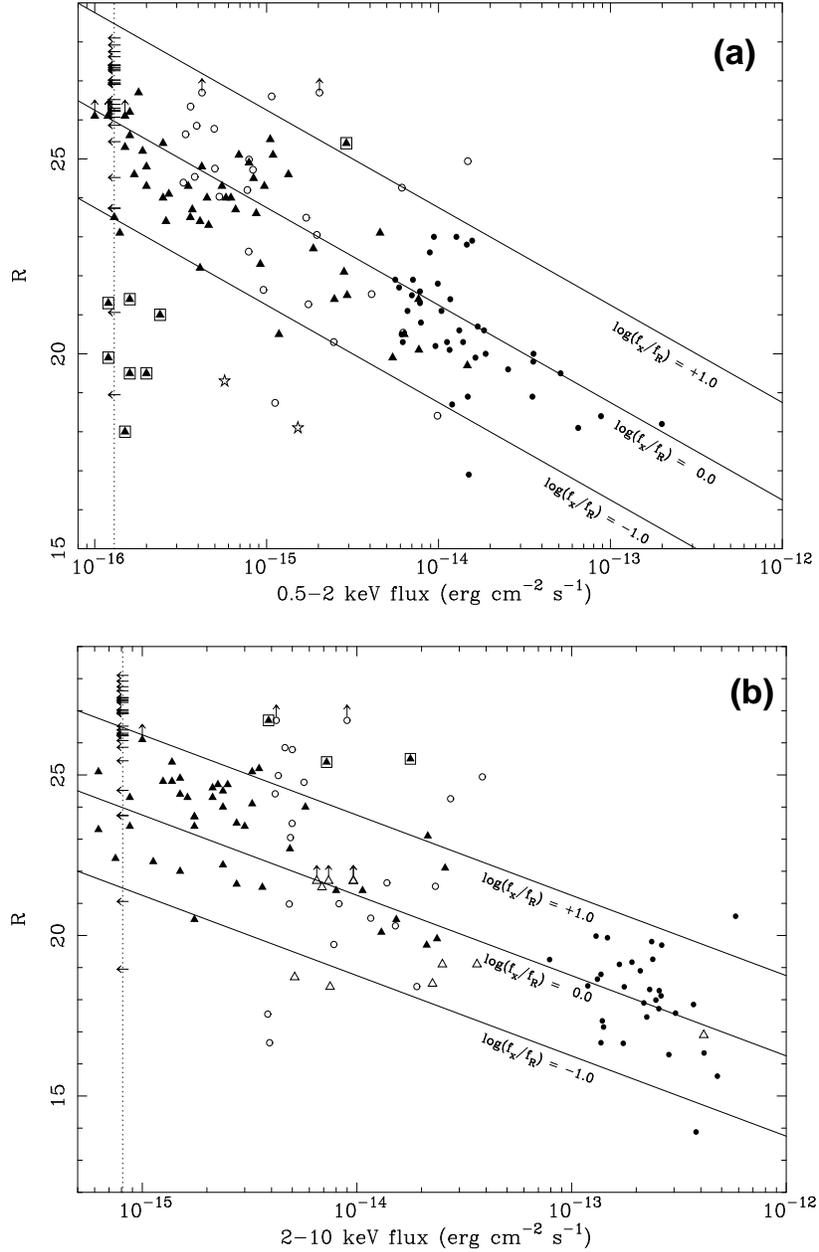}

\caption{Plots of $R$ magnitude versus (a) 0.5--2~keV flux and (b) 2--10~keV flux 
for X-ray detected sources. Solid triangles are sources from this 
paper, open triangles are those from Brandt et~al. (2000), and 
open circles are those from Mushotzky et~al. (2000).
 Solid dots in the upper and lower panels are sources 
from Schmidt et~al. (1998) and Akiyama et~al. (2000), respectively. 
We plot only the AGN from Schmidt et~al. (1998), excluding galaxies, stars,
and groups. 
The two stars in the upper panel are spectroscopically identified stars 
(see Table~\ref{multiwavtable} and \S\ref{spectradetails}), and the left-pointing arrows near the left-hand edges
of the plots are AGN candidates discussed in \S\ref{AGN}; F606W magnitudes
have been converted to $R$ assuming $F606W-I$ = 0.22. The slanted
lines are calculated for 0.5--2~keV (2--10~keV) X-ray fluxes ($f_{\rm X}$) in the top (bottom) 
panel. The vertical dotted lines show
our detection limits for an assumed $\Gamma=1.4$ power-law
spectrum; sources slightly beyond these lines have spectral
shapes differing from a $\Gamma=1.4$ power law.  The boxes mark our sources with 
extreme $\log{({{f_{\rm X}}\over{f_{\rm R}}})}$ ratios as discussed in \S \ref{optfluxcolors} and \S\ref{LLXR}. 
\label{akiyama_fancy}}
\end{figure}


\begin{figure}
\epsscale{1.2}
\figurenum{9}

\plotone{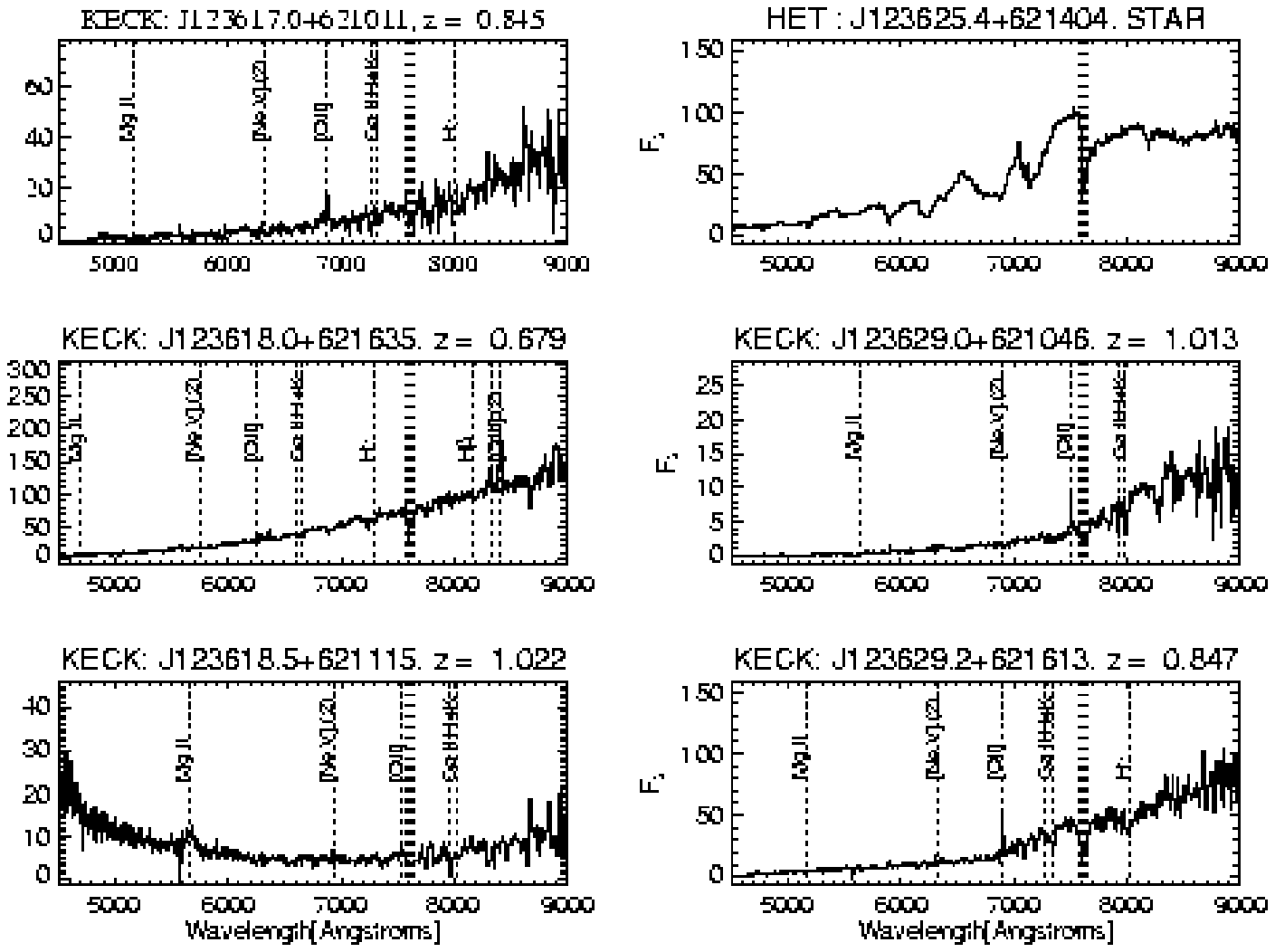}

\vspace{-4.0in}
\caption{Optical spectra of \chandra\ sources.  For all spectra, the ordinate is 
normalized to F$_{\lambda}$ (erg~cm$^{-2}$ s$^{-1}$ \AA$^{-1}$).
No attempt has been made to place the spectra on an absolute 
spectrophotometric scale.  Several key optical transitions have been labeled at their
observed-frame wavelengths; a ``(2)" indicates that the next line redward 
is from the same element and ionization state.  The hash marked regions indicate
atmospheric absorption.
The spectral resolution is 
$\approx 17$~\AA\ for both the Keck and HET spectra.  Note that second-order
contamination is present in the HET spectra at wavelengths greater than
7700~\AA.
The spectra are described in detail in \S A.1. }
\label{spectra1}
\end{figure}



\begin{figure}
\epsscale{1.2}
\figurenum{9b}

\plotone{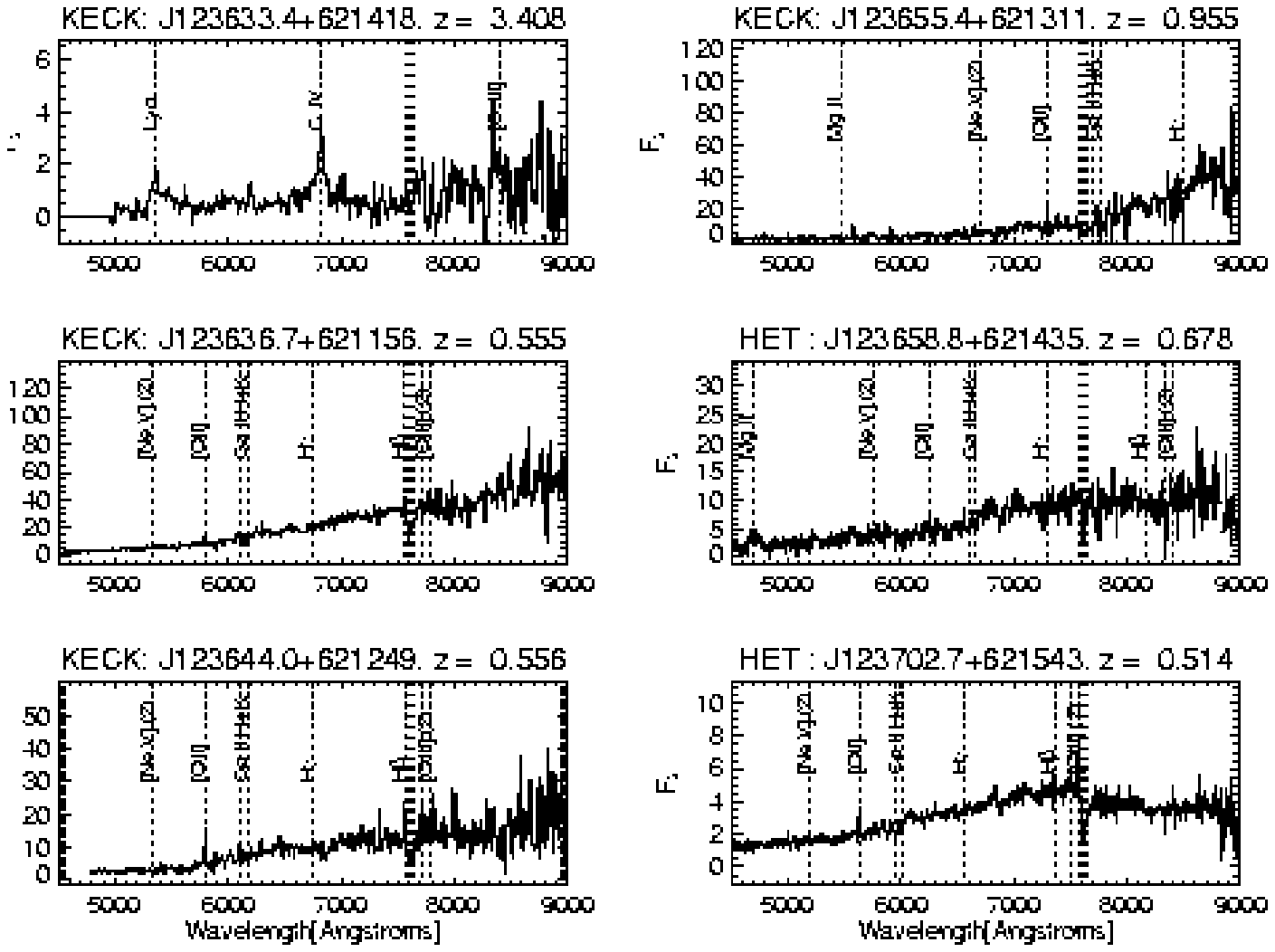}

%
\label{spectra2}
\end{figure}



\begin{figure}
\epsscale{1.2}
\figurenum{9c}

\plotone{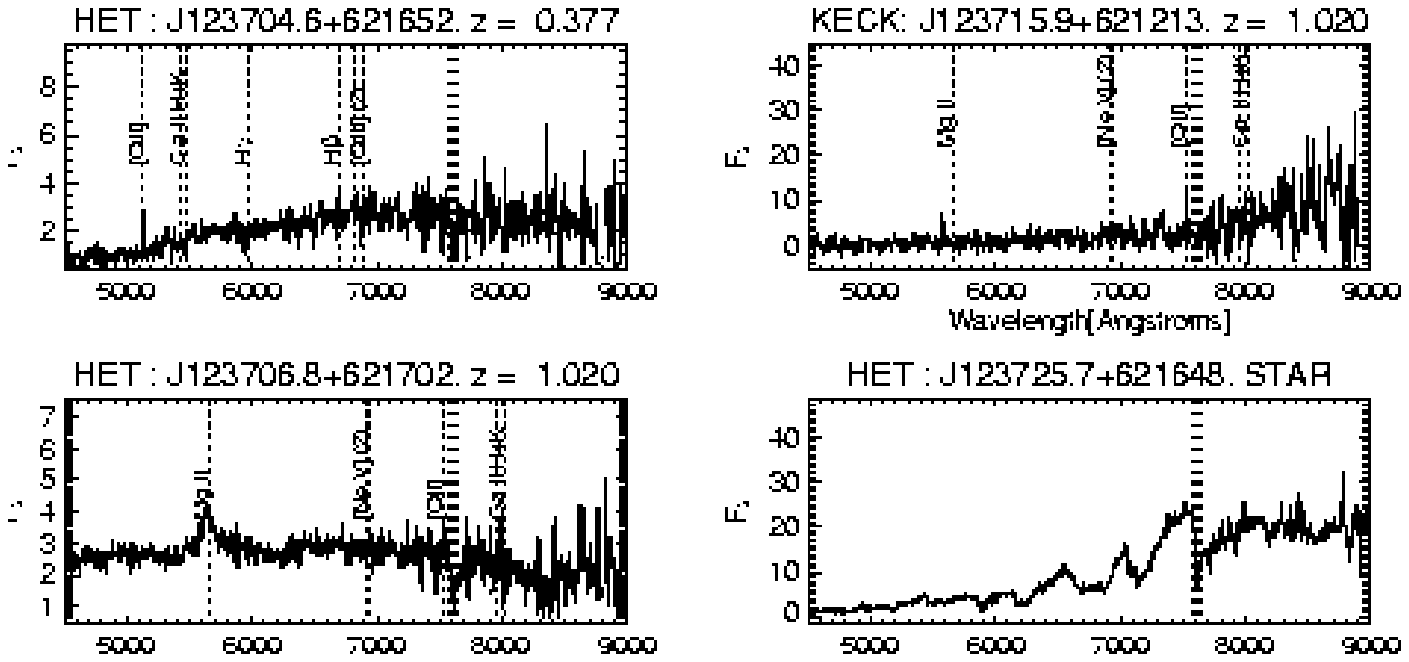}

%
\label{spectra3}
\end{figure}


\begin{figure}
\epsscale{0.95}
\figurenum{10}
\plotone{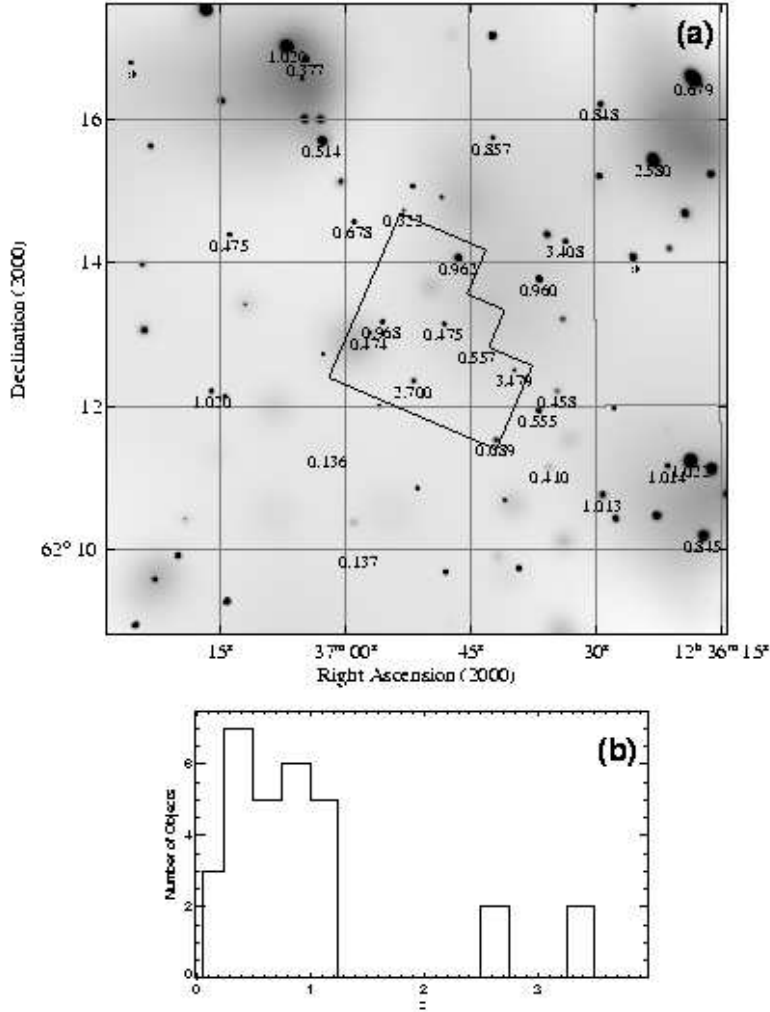}
\vspace{-1.0in}
\caption{Redshifts of \chandra\ X-ray sources.  (a) 
Adaptively smoothed full-band image of the Caltech area with
the available redshifts superposed.  Redshifts are given immediately
below the corresponding X-ray sources, and all redshifts are
spectroscopic except for that of \hbox{CXOHDFN}~J123651.7+621221 (marked at
$z=2.700$) where the redshift is photometric and uncertain (see H00).
The redshift of the object marked at $z=3.479$ 
is determined using both photometry and spectroscopy (see \S \ref{xrayhdfn}).
The two X-ray sources with asterisks beneath them are stars (of type M4).  The
adaptive smoothing is at the 3$\sigma$ level using the code of Ebeling,
White, \& Rangarajan (2001).  (b) Redshift distribution of the 30 
extragalactic \chandra\ sources with measured redshifts.
\label{zmap}}
\end{figure}


\begin{figure}
\vspace{+1.0in}
\epsscale{0.80}
\figurenum{11}
\vspace{-1.0in}

\plotone{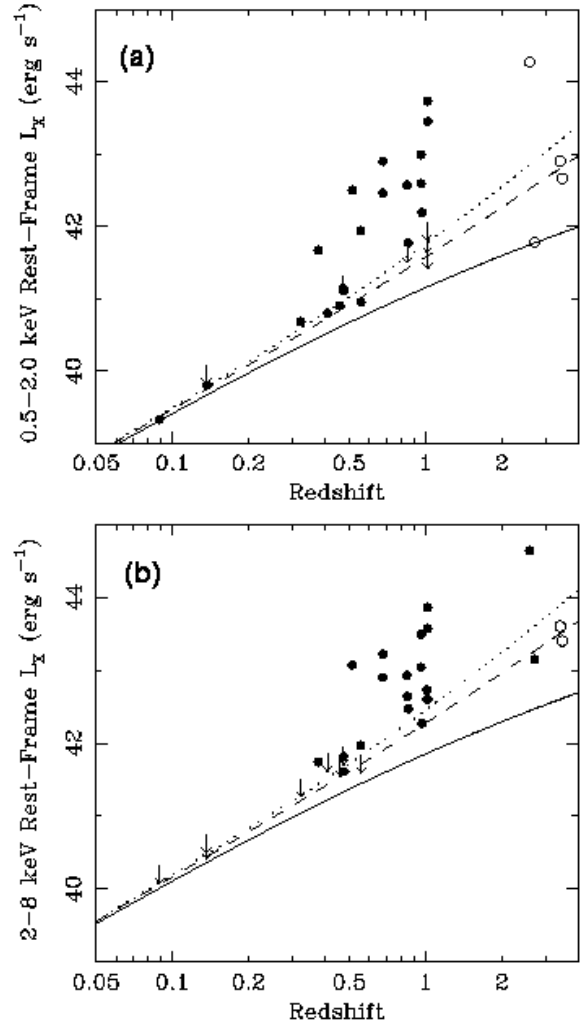}
\caption{Rest-frame luminosity in the (a) soft band and (b) hard band
versus redshift for our \chandra\ sources with spectroscopic
identifications. The solid, dashed and dotted curves are our detection
limits for photon indices of $\Gamma=0.0$, $\Gamma=1.4$ and
$\Gamma=2.0$, respectively. Note that we have used a range of photon indices
in the calculation of the luminosities. Open symbols are used for four soft-band
sources at high redshift due to the fact that we are directly
measuring little or no rest-frame soft-band emission from
these sources; the derived rest-frame soft-band luminosities
are extrapolated from data at higher rest-frame energies. Open
symbols are used for two hard-band sources at high redshift
where the rest-frame hard-band luminosities were calculated
based on observed-frame soft-band data.
\label{Lz}}
\end{figure}


\begin{figure}
\epsscale{1.00}
\figurenum{12}
\plotone{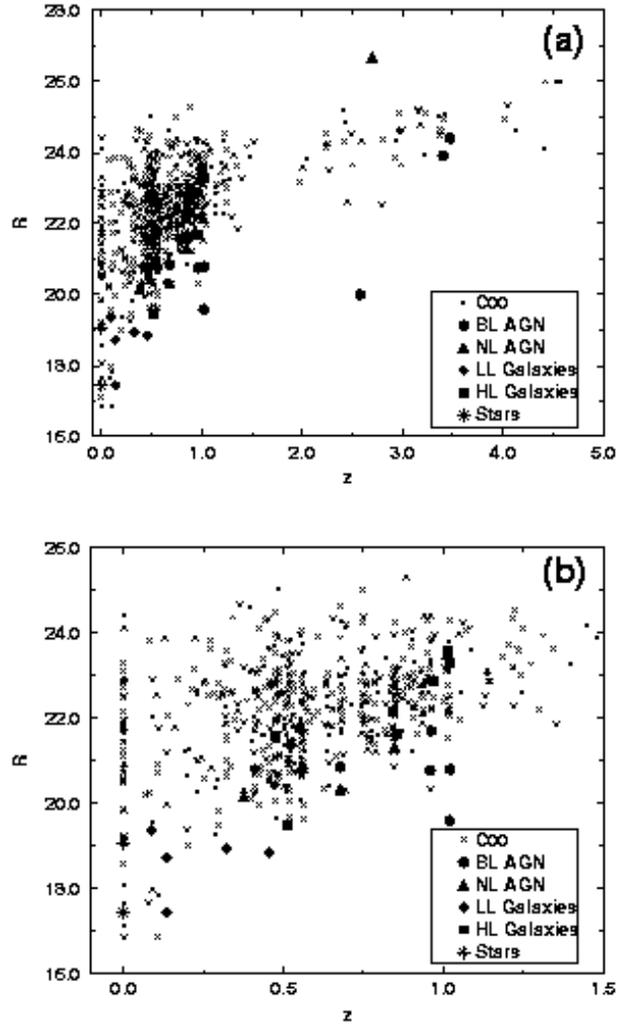}
\vspace{-1.5in}
\caption{${\cal R}$ magnitude versus redshift for all \chandra\ sources having 
spectroscopic identifications, plotted over the redshift catalog of C00.
${\cal R}$ (Hogg00; see \S \ref{optphotom}) is plotted here for the X-ray sources
in order to make direct comparison with C00 
who also use ${\cal R}$. Panel (a) shows the total sample of 32 \chandra\ 
sources with optical identifications, while
panel (b) shows the same data for $0 < z < 1.5$, where most of the \chandra\
objects reside.  All of the C00 sources at $z=0$ are spectroscopically
identified stars.
\label{Rz}}
\end{figure}


\begin{figure}
\epsscale{1.0}
\figurenum{13}

\plotone{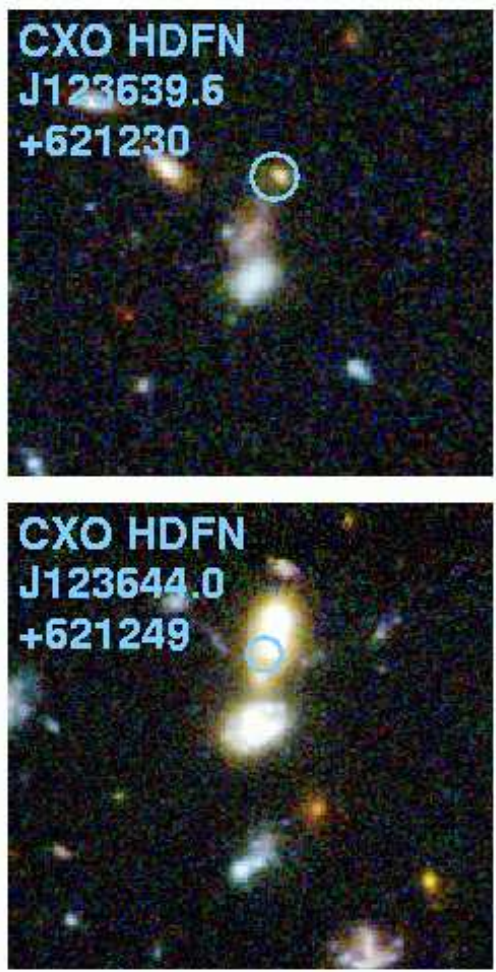}
\vspace{-1.5in}

\caption{Two newly discovered X-ray sources in the \hbox{HDF-N}. \hbox{CXOHDFN}~J123639.6+621230 
is a broad-line AGN at $z=3.479$.
 The two objects to the south of \hbox{CXOHDFN}~J123639.6+621230 do
not have spectroscopic redshifts, but their photometric redshifts of $z=0.00$
and $z=1.16$ (Fern\'andez-Soto et~al. 1999) are sufficiently different to
argue against interaction.   \hbox{CXOHDFN}~J123644.0+621249 is a $z=0.557$ 
emission-line galaxy that also probably contains an AGN (see \S 6.1).  The
object to the the south of \hbox{CXOHDFN}~J123644.0+621249 is at $z=0.555$;
 these two galaxies are likely interacting
(see R98 for discussion).
The blue circles are centered at the best \chandra\ positions and have diameters 
of $1^{\prime \prime}$.
\label{HDFzoom}}
\end{figure}



\begin{figure}
\epsscale{0.8}
\figurenum{14}

\plotone{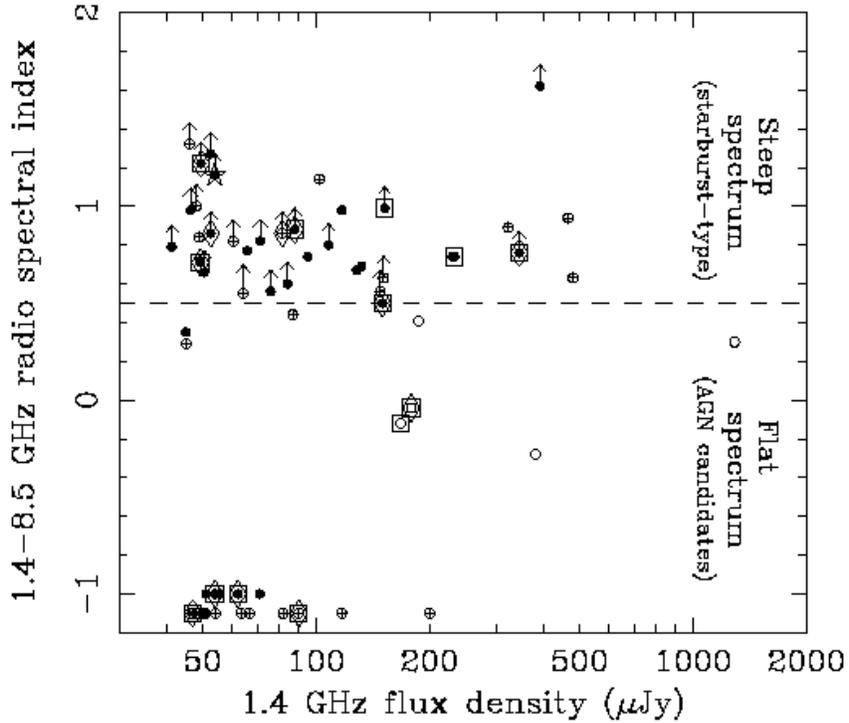}
\vspace{-1.0in}

\caption{
1.4--8.5~GHz spectral index versus flux density at 1.4~GHz for 
radio-classified starburst-type systems (solid dots), AGN candidates 
(open circles), and systems of unknown type (circles with crosses inside) 
detected at 1.4~GHz in the Caltech area (R00). Markers with squares (diamonds) around 
them indicate that the corresponding source has been detected by 
\chandra\ in the soft band (hard band). 
One 1.4~GHz source, J123711.9+621325, was only detected in the full-band data 
(see Table~\ref{radiotable}) and is marked with a star around it.
The horizontal dashed line 
drawn at a 1.4--8.5~GHz 
spectral index of 0.5 denotes the typical value used to separate 
starburst-type systems and AGN candidates (e.g., \S6 of R00). 
Symbols located at or below $-1$ along the ordinate do not 
have measurements of their 1.4--8.5~GHz spectral indices. 
Note that the X-ray detection fractions, in both the soft and hard
bands, are comparable for both the starburst-type systems and AGN candidates.
}
\label{sbagn}
\end{figure}


\begin{figure}
\epsscale{1.0}
\figurenum{15}

\plotone{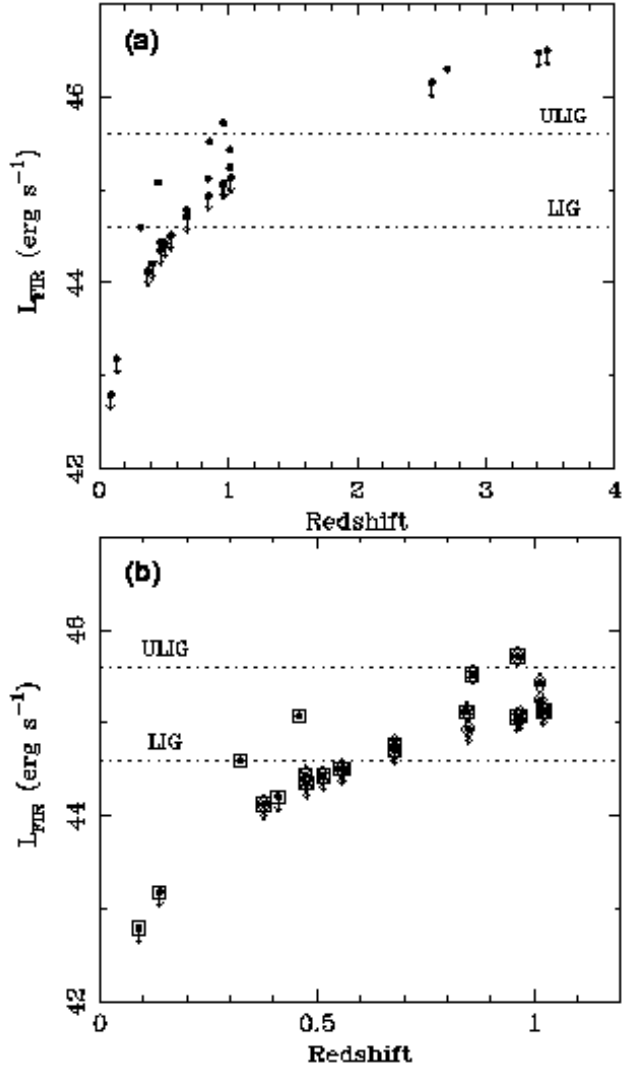}
\vspace{-1.4in}
\caption{
Far-infrared (FIR) luminosity, as estimated from the FIR-radio 
correlation, versus redshift for (a) all our \chandra\ sources 
with redshifts and (b) only those sources with $z=$~\hbox{0--1.2}. 
When sources are not detected at 1.4~GHz, we calculate upper 
limits following \S3 of R00. The two horizontal dotted lines 
show the luminosity thresholds for luminous 
far-infrared galaxies ($L_{\rm FIR}>10^{11}$~$L_\odot$)
and ultraluminous far-infrared galaxies ($L_{\rm FIR}>10^{12}$~$L_\odot$). 
In panel (b) markers with squares (diamonds) around them indicate
that the corresponding source has been detected by \chandra\ 
in the soft band (hard band); we show a restricted redshift  
range in panel (b) to minimize symbol crowding. The two 
sources in panel (a) with $z=$~2--3 are detected in both the 
soft and hard bands, and the two sources with $z>3$ are only 
detected in the soft band.\label{FIR} }
\end{figure}

\begin{figure}
\epsscale{0.9}
\figurenum{16}
\plotone{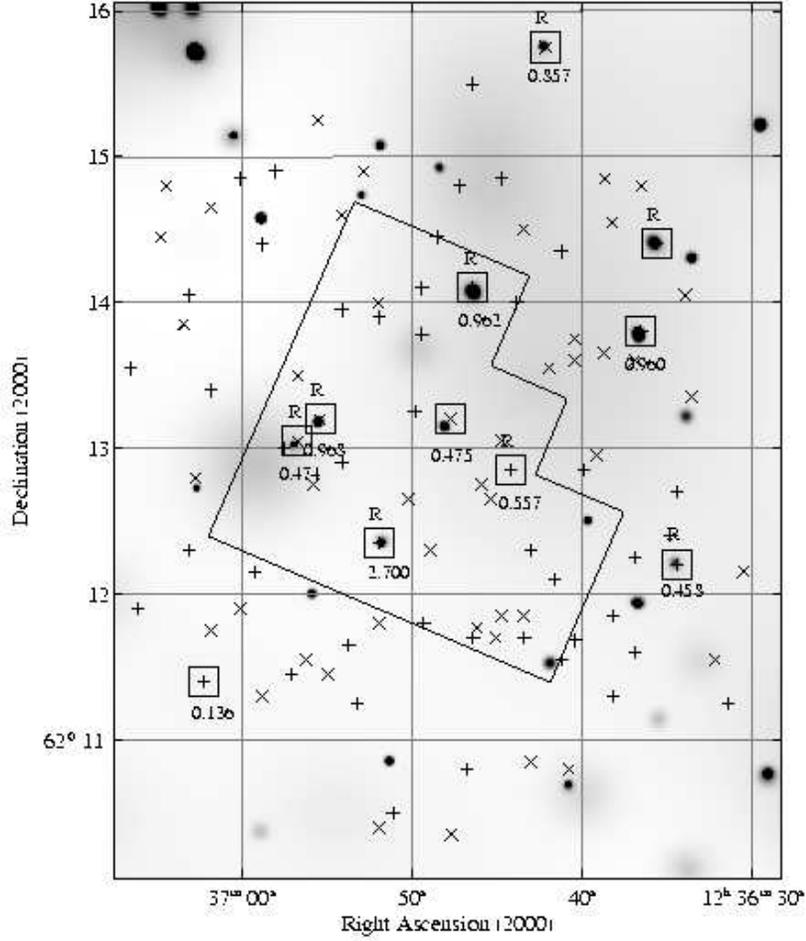}
\vspace{-0.5in}
\caption{
Adaptively smoothed full-band \chandra\ image showing 
the A99 main \iso-CAM sources (plus signs) and 
the A99 supplementary \iso-CAM sources (multiplication signs);
the entire \iso-CAM survey area is shown in this plot. 
\iso-CAM sources with \chandra\ matches are boxed, and 
when available redshifts are given for these matches
below the boxes. Matches with detections at 8.5~GHz 
are marked with ``R'' above the boxes. Note that six of 
the eight \chandra\ sources in the \hbox{HDF-N} have \iso-CAM
matches and that most of these matches are also detected at 8.5~GHz.
 The adaptive smoothing is at the $3\sigma$ level using the code of
Ebeling, White, \& Rangarajan (2001).  \label{ISOCAM_figure}}
\end{figure}



\begin{figure}
\epsscale{1.1}
\figurenum{17}

\plottwo{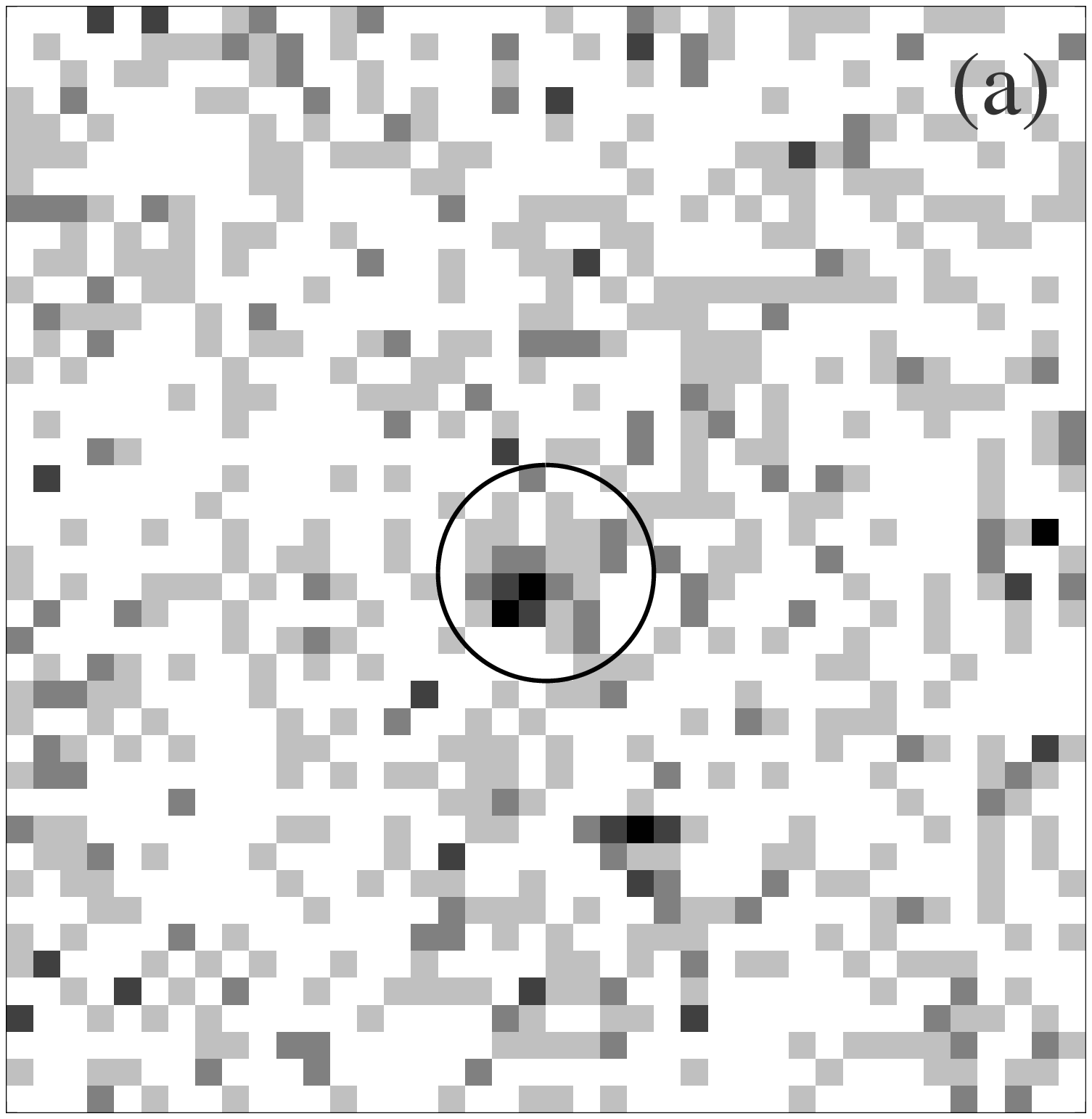}{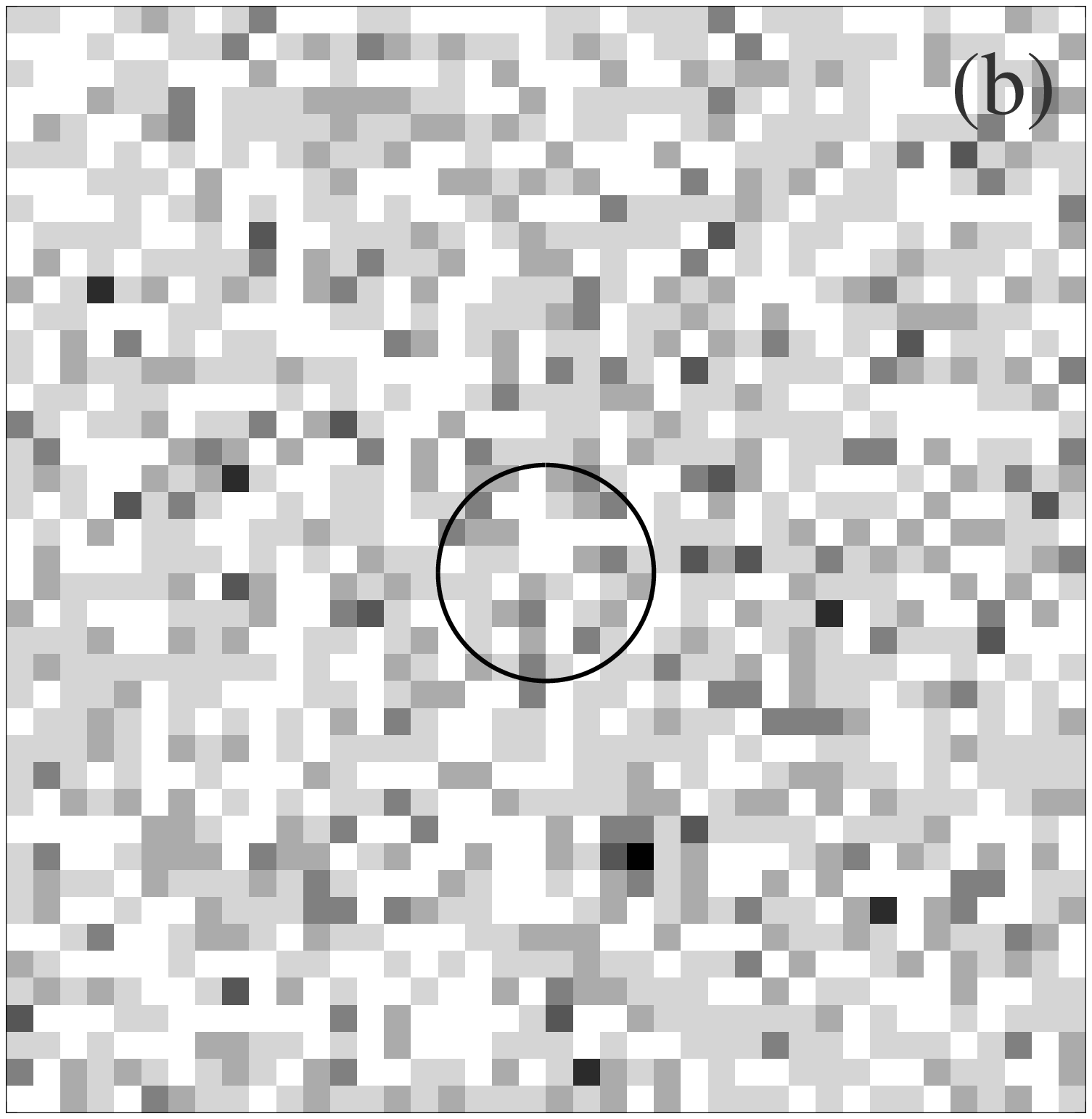}

\caption{
Stacked \chandra\ images of the 29 X-ray weak VROs in the (a) soft band  
and (b) hard band. These images have effective exposure times
of 6.02~Ms each, are $20^{\prime\prime}$ on a side, and have $0.5^{\prime \prime}$ pixels. The circles 
in the centers of the images are centered at the nominal stacking
position and are $2^{\prime\prime}$ in radius. Note the X-ray source 
in the stacked soft-band image that is not visible in the stacked
hard-band image; this source is found by {\sc wavdetect} in the
soft band when {\sc wavdetect} is run with a probability 
threshold of $1\times 10^{-7}$. The detection of the stacked X-ray 
weak VROs only in the soft band contrasts with the hard X-ray 
emission of the four VROs individually detected by \chandra. 
}
\label{VROpostagestamp}
\end{figure}


\end{document}